\documentclass[%
pre,reprint,
 amsmath,amssymb,
 aps,
]{revtex4-1}

\usepackage{graphicx}
\usepackage{dcolumn}
\usepackage{bm}
\usepackage{textcomp}
\usepackage[pdftex,colorlinks=true,bookmarks=false,citecolor=blue,urlcolor=blue]{hyperref}
\usepackage{cancel}

\allowdisplaybreaks
\newcommand{\mrm}[1]{\mathrm{#1}}

\newcommand{\E}{\mathrm{E}} 
\newcommand{\rmi}{\mathrm{i}} 
\newcommand{\var}{\mathrm{var}} 
\newcommand{\tr}{\mathrm{tr}} 
\newcommand{\diag}{\mathrm{diag}} 

\newcommand{\Fvec}{\mathbf{F}}	
\newcommand{\Evec}{\mathbf{E}}	
\newcommand{\Hvec}{\mathbf{H}}	
\newcommand{\Bvec}{\mathbf{B}}	
\newcommand{\Jvec}{\mathbf{J}}	

\newcommand{\pvec}{\mathbf{p}}	
\newcommand{\uvec}{\mathbf{u}}  

\newcommand{\rvec}{\mathbf{r}}
\newcommand{\vvec}{\mathbf{v}}  
\newcommand{\evec}{\mathbf{e}}	

\newcommand{\gvec}{\mathbf{g}}

\newcommand{\zeroth}{$0^{\text{th}}\,$}
\newcommand{\first}{$1^{\text{st}}\,$}
\newcommand{\second}{$2^{\text{nd}}\,$}
\newcommand{\third}{$3^{\text{rd}}\,$}

\DeclareSymbolFont{lettersA}{U}{txmia}{m}{it}
\DeclareMathSymbol{\real}{\mathord}{lettersA}{"92} 
\DeclareMathSymbol{\cplx}{\mathord}{lettersA}{"83} 

\begin{document}


\title{Theory of terahertz emission from femtosecond-laser-induced micro-plasmas}

\author{I. Thiele}
\email{illia-thiele@web.de}
\author{R. Nuter}
\author{B. Bousquet}
\author{V. Tikhonchuk}
\author{S. Skupin}
\affiliation{Univ.~Bordeaux - CNRS - CEA, Centre Lasers Intenses et Applications, UMR 5107, 33405 Talence, France}

\author{X. Davoine}
\author{L. Gremillet}
\author{L. Berg\'e}
\affiliation{CEA, DAM, DIF, 91297 Arpajon, France} 

\date{\today}

\begin{abstract}
We present a theoretical investigation of terahertz (THz) generation in laser-induced gas plasmas. The work is strongly motivated by recent experimental results on micro-plasmas, but our general findings are not limited to such a configuration. The electrons and ions are created by tunnel-ionization of neutral atoms, and the resulting plasma is heated by collisions. Electrons are driven by electromagnetic, convective and diffusive sources and produce a macroscopic current which is responsible for THz emission. The model naturally includes both, ionization current and transition-Cherenkov mechanisms for THz emission, which are usually investigated separately in the literature. The latter mechanism is shown to dominate for single-color multi-cycle lasers pulses, where the observed THz radiation originates from longitudinal electron currents. However, we find that the often discussed oscillations at the plasma frequency do not contribute to the THz emission spectrum. In order to predict the scaling of the conversion efficiency with pulse energy and focusing conditions, we propose a simplified description that is in excellent agreement with rigorous particle-in-cell simulations.  
\end{abstract}

\maketitle


\section{\label{sec:Int}Introduction}

Many applications in spectroscopy and sensing require sources in the terahertz (THz) region \cite{0034-4885-70-8-R02,Tonouchi,Liu,Kampfrath,Tuniz}. Compact small bandwidth THz sources have been established based on optical rectification in nonlinear crystals \cite{Stepanov:08,PhysRevLett.112.213901}. In large electron accelerator facilities, broadband THz radiation is produced via coherent transition radiation~\cite{PhysRevSTAB.12.030705,1.4790427}. THz emission from interaction of ionizing intense laser pulses with gases is a promising approach for more compact broadband THz sources. Recently, various theoretical and experimental efforts have been made in this direction~\cite{Cook,Kim2007,PhysRevE.69.066415,PhysRevLett.98.235002,Kim,PhysRevE.78.046406,PhysRevLett.105.053903,Balakin:10,PhysRevE.82.056409,PhysRevE.81.026407,1367-2630-13-12-123029,Debayle:14,Buccheri:15,PhysRevLett.116.063902,PhysRevA.93.033844,PedrosArticle}.

Several mechanisms responsible for THz radiation in ionized gases have been proposed. Most of them are plasma based. A crucial role plays the actual form of the driving laser pulse, which can be single or multi-color. For single-color driving pulses, THz emission can be caused by excitation of plasma currents via ponderomotive or radiation pressure sources~\cite{PhysRevE.69.066415}. This idea has been applied to explain THz emission for femtosecond (fs) filaments in~\cite{PhysRevLett.98.235002,ISI:000253084000007}, and is usually referred to as transition-Cherenkov (TC) mechanism: The ponderomotive force of the driving laser pulse produces a longitudinal current structure, which propagates approximately with the speed of light. The interference of radiation from distinct points along the propagation axis leads to a conical emission. The name transition-Cherenkov mechanism comes from this characteristic radiation profile. 
While Cherenkov radiation usually requires the source moving at superluminal velocity, this is not necessary for an emission zone of a finite length.


For multi-color driving pulses, in particular the two-color configuration relying on mixing the fundamental harmonic (FH) and the second harmonic (SH) has been studied~\cite{Cook,Kim2007,Kim,PhysRevLett.116.063902,PhysRevLett.105.053903,1367-2630-13-12-123029,Balakin:10,PhysRevE.81.026407}.
Originally, four-wave mixing (FWM) rectification via third-order nonlinearity of the neutral atoms has been suggested as the THz generating mechanism in this case~\cite{Cook}. However later, it has been shown that contributions from FWM are much weaker than those from mechanisms based on excitation of the laser-induced plasma~\cite{PhysRevLett.116.063902,PhysRevLett.105.053903,1367-2630-13-12-123029}.
In particular, the ionization current (IC) mechanism proposed in \cite{Kim2007} has been accepted as the major contributor to THz radiation from such a two-color pump, and this mechanism may also contribute for single-color pulses~\cite{Debayle:14}: The extrema of the first and the second half-laser-cycle create two bunches of charge due to tunnel ionization. Each of these gets accelerated in the laser field and produces a current. Temporal asymmetry in the driving pulse can render the superposition of these currents to be not completely destructive. In this case, the non-vanishing net current can lead to emission of radiation, in particular in the THz domain. The temporal asymmetry of the driving field can be achieved by, e.g., admixing a second laser color or choosing a very short laser pulse.

Recently, a promising approach towards further miniaturization of the THz source has been investigated experimentally in~\cite{Buccheri:15}: A pulsed single-color fs laser is focused strongly into a gas (e.g., air or argon). Intensities of $10^{14}$--$10^{16}$~W/cm$^2$ in the focal region can be reached with \textmu J driving pulses focused down to Abbe's diffraction limit. In the focal region, the neutral gas is ionized and a few micrometer thick and few tens of micrometer long micro-plasma is created. The excitation of the plasma by the ionizing laser pulse leads to THz radiation that can be measured in the far field. The major goal of this paper is to investigate the THz radiation from such a micro-plasma theoretically. While for two-color laser pulses the dominance of the IC mechanism is established, for single-color laser pulses the prevailing mechanism is still under discussion and depends on both laser and gas properties. Moreover, a unified theoretical description of these mechanisms is  missing, as well as simplified models which would allow to identify general scaling laws.
Throughout this paper we assume the laser pulses to interact with argon gas at ambient pressure. The first stage ionization potential of argon is $I_\mrm{p}^{\mrm{Ar}}=15.8$~eV, which is close to the ionization potential $I_\mrm{p}^{\mrm{N}_2}=15.6$~eV of the nitrogen molecule as the main component of air~\cite{NIST}. 

The paper is organized in the following way. In Sec.~\ref{sec:Theo}, we derive a model by means of multiple scale analysis which describes both IC and TC mechanisms in a consistent and straight forward way. In Sec.~\ref{sec:Comp}, the model is analyzed for the one-dimensional (1D) case to understand the main processes occurring in the laser gas interaction: ionization, heating, collisions, and in particular excitation of plasma waves at THz frequencies. We also estimate the laser pulse parameters where either the IC mechanism or the TC mechanism dominates. In Sec.~\ref{sec:Dist}, we extend our analysis to the two and three dimensional (2D, 3D) cases. Symmetry properties of the system are studied and the important role of non-radiative plasma wave excitations is discussed. 
In Sec.~\ref{sec:Tera}, we finally provide a simplified 2D/3D model for THz radiation in the single-color case. By means of this simplified approach, the scaling of the THz conversion efficiency with various pulse parameters is discussed.
All our results are benchmarked by rigorous particle-in-cell (PIC) simulations using the codes OCEAN~\cite{PhysRevE.87.043109} and CALDER~\cite{Nut11}, and we report excellent agreement.

\section{\label{sec:Theo}Model for THz emission}

In the following, we briefly present our model for THz emission from fs-laser-generated gas plasmas. The actual derivation makes use of multiple scale analysis and is detailed in App.~\ref{app:Der}. In the main text, we focus more on discussing the resulting set of equations. The starting point is the non-relativistic Vlasov equation describing the distribution function of electrons $f_\mrm{e}(\rvec,\vvec,t)$ depending on position $\rvec$, velocity $\vvec$ and time $t$
\begin{equation}
 \partial_t f_\mrm{e} + \vvec\cdot\nabla_\rvec f_\mrm{e} + \frac{\Fvec}{m_\mrm{e}}\cdot\nabla_\vvec f_\mrm{e} = S \delta(\vvec) + C\,\mbox{,}
 \label{eq:Vlasov_e}  
\end{equation}
where $m_\mrm{e}$ is the electron mass. The electrons are pushed by the electromagnetic force
\begin{equation}
	\Fvec = q_\mrm{e}\left[\Evec(\rvec,t) + \vvec\times\Bvec(\rvec,t)\right]\mbox{,}	
	\label{eq:Force_elmag}
\end{equation}
with the electric field $\Evec$ and magnetic field $\Bvec$,  and $q_\mrm{e}$ is the electron charge. 
The ionization of atoms is taken into account by the source term $S$. We assume that each electron is born with zero velocity, and that ions do not move on timescales relevant to our problem. For example, for tunnel ionization and singly-charged ions only, the source term simply reads $S=W[\Evec]n_{\textrm{n}}$, where $W[\Evec]$ is the tunnel ionization rate and $n_{\textrm{n}}$ the density of neutral atoms. For intensities of up to $10^{16}$~W/cm$^2$ as considered in this paper, single ionization is not sufficient, and we include multiple ionization as explained in App.~\ref{app:ion}.
Collisions are taken into account via the term $C$ which depends on the properties of the plasma. 
Since for the driving pulses considered here almost all atoms in the interaction region are quickly ionized, we neglect electron neutral collisions, and in particular impact or collisional ionization. Then, collisions are all elastic, and the total momentum of electrons can change via collisions with ions only. Our PIC simulations accounting for electron-ion collisions show that anisotropy of the electron distribution function is negligible due to fast thermalization, and thus employing a scalar electron-ion collision frequency $\nu_\mrm{ei}$ is sufficient [cf.\ Eq.~(\ref{eq:nu_ei_NRL_2}) below]. Details can be found in App.~\ref{app:Der}. Finally, introducing the macroscopic current density $\Jvec$ as the \first moment of the electron distribution function $f_\mrm{e}$,
\begin{equation}
	\Jvec = q_\mrm{e}\int \vvec f_\mrm{e}d\,^3\vvec\,\mbox{,}
	\label{eq:J}
\end{equation}
allows for coupling to the macroscopic Maxwell equations. Like in the PIC codes we use to benchmark our theory, we do not account for linear or nonlinear polarization of the neutral atoms or ions. Moreover, in our model we neglect losses in the electromagnetic fields due to ionization. Both simplifications are justified by the small size of the interaction region considered in this paper.  

The \zeroth, \first and \second moments of Eq.~(\ref{eq:Vlasov_e}) leads to continuity equation, Euler equation and free electron energy balance, respectively. A brief review can be found in App.~\ref{app_chap:Vlasov2THz}. We perform a multiple scale expansion on these three equations employing a small parameter $\epsilon$. For instance, electric and magnetic field as well as the macroscopic current density are then expressed as
\begin{equation}
	\Evec = \sum\limits_{i=1}^\infty\epsilon^i \Evec_i,\quad
	\Bvec = \sum\limits_{i=1}^\infty\epsilon^i \Bvec_i,\quad
	\Jvec = \sum\limits_{i=1}^\infty\epsilon^i \Jvec_i\,\mbox{.}	
	\label{eq:mult_scale_exp}
\end{equation}
The parameter $\epsilon$ represents the ratio of respective orders of the fields, for instance, the ratio of $\Jvec_2$ over $\Jvec_1$. As discussed in App.~\ref{sec:Mult_scale}, for underdense gas plasmas this ratio can be estimated to be smaller than $|q_e E_\mrm{L} / m_\mrm{e} \omega_\mrm{L}c|$. Here, $E_\mrm{L}$ is the laser electric field amplitude, $c$ is the vacuum speed of light, $\omega_\mrm{L}=2\pi c/\lambda_\mrm{L}$ is the laser frequency and $\lambda_\mrm{L}$ is the vacuum laser wavelength. Following this estimation, the multiple scale expansion is valid for laser intensities below $2\times 10^{16}$~W/cm$^2$ at $\lambda_\mrm{L}=0.8$~\textmu m. We will furthermore verify a posteriori the ratio of different orders of the fields at the end of Sec.~\ref{sec:Tera}. 

It turns out that in order to get meaningful results out of the multiple scale expansion, the source term $S$ has to be of order $\epsilon^0$. Then, the continuity equation immediately dictates
\begin{equation}
 \partial_t n_0 = S\mbox{,} \label{eq:cont_0}
\end{equation}
where $n_0$ is the \zeroth order electron density. For practical purposes, $n_0$ can be seen as the macroscopic free electron density. Considering the momenta equations at scales $\epsilon^1$ and $\epsilon^2$ gives, after some algebra, the following set of equations (see App.~\ref{sec:Mult_scale} for details): 
\begin{align}
	&\epsilon^1:\quad & \partial_t \Jvec_1 + \nu_\mrm{ei}\Jvec_1 &= \frac{q_\mrm{e}^2}{m_\mrm{e}} n_0 \Evec_1\label{eq:cont_1}\\
	&\epsilon^2:\quad & \partial_t\Jvec_2 + \nu_\mrm{ei}\Jvec_2 &=\frac{q_\mrm{e}^2}{m_\mrm{e}} n_0\Evec_2 + \bm\iota_2\,\mbox{,}\label{eq:cont_2}
\end{align}
where
\begin{equation}
\begin{split}
	\bm\iota_2 & = - \frac{n_0}{2 q_\mrm{e}}\nabla{\left|\frac{\Jvec_1}{n_0}\right|}^2 - \frac{\Jvec_1}{q_\mrm{e}}\times\nabla\times \! \int\limits_{-\infty}^t\!\frac{\Jvec_1}{n_0} \left(\nu_\mrm{ei} + \frac{\partial_{t'}n_0}{n_0}\right)dt' \\
	& \quad - \frac{\left(\nu_\mrm{ei} + \partial_t\right)}{q_\mrm{e} n_0}\left(\Jvec_1\int\limits_{-\infty}^t\nabla\cdot\Jvec_1\,dt'\right)-\frac{2q_\mrm{e}}{3m_\mrm{e}}\nabla \left(n_0E_{\mrm{th}}\right)
	\label{eq:iota_2}
\end{split}
\end{equation}
and $E_\mrm{th}$ is the lowest order thermal energy. The electron-ion collision frequency $\nu_\mrm{ei}$ is of order $\epsilon^0$, like the ionization source $S$. Following \cite{Huba2013}, we assume
\begin{equation}
	\nu_\mrm{ei}[\mrm{s}^{-1}] = \frac{3.9 \times 10^{-6}\sum\limits_Z Z^2 n_\mrm{ion}^{(Z)}[\mrm{cm}^{-3}]\lambda_\mrm{ei}}{\left(E_{\mrm{th}}[\mrm{eV}] + E_{\mrm{kin}}[\mrm{eV}]\right)^{3/2}}  \,\mbox{,}
	\label{eq:nu_ei_NRL_2}
\end{equation}
where $Z$ is the ion charge for ions with density $n_\mrm{ion}^{\mrm{(Z)}}$ and $\lambda_\mrm{ei}$ is the Coulomb logarithm. For our choice of scaling, the lowest order of the electron thermal energy $E_{\mrm{th}}$ appears at order $\epsilon^2$ and is given by (see App.~\ref{app:Der})
\begin{equation}
	\partial_t\left(n_0 E_{\mrm{th}}\right) = E_{\mrm{kin}}\left(2n_0\nu_\mrm{ei}+\partial_t n_0\right)\,\mbox{,}\label{eq:heating}
\end{equation}
and the \second order kinetic energy reads
\begin{align}
	E_{\mrm{kin}} &= \frac{m_\mrm{e}}{2}\left|\frac{\Jvec_1}{n_0 q_\mrm{e}}\right|^2\label{eq:Kin_E}\,\mbox{.}
\end{align}
Thus, the knowledge of $\Jvec_1$ is sufficient to compute $E_{\mrm{th}}$ and $\nu_\mathrm{ei}$. A similar reasoning is possible for the ionization source $S$, for which a system of rate equations, as detailed in App.~(\ref{app:ion}), has to be solved. 
In general, the ionization rates involved depend on the total electric field $\Evec$, which is unknown. In the spirit of the multiple scale expansion, one has to take $\Evec$ up to the highest known order. However, in practice the \first order electric field $\Evec_1$ is sufficient to compute the ionization source $S$. Finally, we have to plug our multiple scale expansion into Maxwell's equations, and linearity implies
\begin{align}
	\nabla\times\Evec_i &= - \partial_t \Bvec_i
	\label{eq:Far_1}\\
	\nabla\times\Bvec_i &= \frac{1}{c^2}\partial_t \Evec_i + \mu_0 \Jvec_i
	\label{eq:Amp_1}
\end{align}
for all orders $\epsilon^i$ with $i=1,2,\ldots$\,\,.

The above system of Eqs.~(\ref{eq:cont_0})--(\ref{eq:Amp_1}) is complete. Before going on and discussing the solutions in detail, a few remarks are in order. Firstly, the electric field $\Evec_1$ contains the laser field $\Evec_\mrm{L}$ fixing the boundary conditions, and for $S\equiv0$ we recover just vacuum propagation at order $\epsilon^1$. Secondly, in Eqs.~(\ref{eq:cont_0}),~(\ref{eq:cont_1}), and (\ref{eq:nu_ei_NRL_2})--(\ref{eq:Amp_1}) all quantities up to order $\epsilon^1$ are treated independently from higher orders. This \first order set of equations already describes the IC mechanism~\cite{Kim,PhysRevE.78.046406,PhysRevLett.105.053903,1367-2630-13-12-123029,Debayle:14,Debayle:14}. The THz radiation due to the IC mechanism can be computed as soon as $\Jvec_1$ is known by using Jefimenko's equation~\cite{jackson}. Thirdly, the current $\Jvec_1$ allows to compute the nonlinear source term $\iota_2$ driving the current $\Jvec_2$ in Eq.~(\ref{eq:cont_2}). The source term $\iota_2$ contains ponderomotive, radiation pressure, convection and diffusion sources that are discussed in more detail in Sec.~(\ref{sec:Comp}). Thus, the TC mechanism appears at order $\epsilon^2$. Formally, it would be possible to extend the multiple scale approach to even higher orders. However, at least for driving pulse and plasma configurations investigated here, \second order solutions show already excellent agreement with rigorous PIC simulations.

\section{\label{sec:Comp}Comparing mechanisms of THz excitation}

In the following section, the excitation of plasma currents in the THz range is analyzed for various laser pulse durations and intensities. To this end, we restrict ourselves to a 1D configuration, where translational invariance is assumed in the $x$ and $y$ directions while the laser pulse propagates along $z$. The laser pulse propagates through vacuum for $z\leq0$ and enters the gas at $z=0$. The incoming linear polarized laser pulse is prescribed as
\begin{equation}
	\Evec_{\mrm{L}}(t,z=0) = E_{\mrm{L}}^0 \sin\!\left(\omega_\mrm{L}t\right) \exp\!\left(-t^2/t_0^2\right) \evec_x \,\mbox{,}
\end{equation}
where $t_0$ characterizes the pulse duration, $E_{\mrm{L}}^0$ is the electric field amplitude, and $\evec_x$ the unit vector in $x$ direction. The corresponding intensity can be calculated as $I_\mrm{L}^0 = \epsilon_0 c \left(E_\mrm{L}^0\right)^2 \!/ 2$ where $\epsilon_0$ is the vacuum permittivity.

One advantage of our multiple scale model is that the orders $\epsilon^1$ and $\epsilon^2$, i.e., Eq.~(\ref{eq:cont_1}) and Eq.~(\ref{eq:cont_2}), can be analyzed separately. Let us start with order $\epsilon^1$.
The current $\Jvec_1$ is driven by the electric field $\Evec_1$ [c.f. Eq.~(\ref{eq:cont_1})]. As suggested in previous works \cite{Kim, PhysRevLett.105.053903, 1367-2630-13-12-123029}, we can approximate the electric field $\Evec_1$ by the laser field $\Evec_\mrm{L}$ when computing $\Jvec_1$. By doing so, electrons are treated as test particles driven by the laser electric field. As a consequence, radiation emitted by the current $\Jvec_1$ does not affect the electric field which drives $\Jvec_1$. However, such back-coupling has an important impact in particular in the THz frequency range~\cite{Cabrera-Granado2015}, and therefore this approximation can only serve as a very rough estimation for $\Jvec_1$ at THz frequencies. For the main spectral components of $\Jvec_1$ however, namely at the laser frequency $\omega_\mrm{L}$, this approximation works very well. This is justified by the short propagation distances ($\sim10$~\textmu m) and underdense plasmas we are interested in. In the following we will compute the laser field $\Evec_\mrm{L}$ for vacuum propagation, and thus neglect plasma dispersion and nonlinear propagation effects. We therefore approximate Eq.~(\ref{eq:cont_1}) for the \first order current as
\begin{align}
	&\partial_\tau \Jvec_1 + \nu_\mrm{ei} \Jvec_1 \approx \bm\iota_1 \,\mbox{,} \label{eq:J_L}
\end{align}
where the incident laser field contributes to the first-order nonlinear source term 
\begin{align}
	\bm\iota_1 = \frac{q_\mrm{e}^2 n_0}{m_\mrm{e}} \Evec_\mrm{L}\label{eq:iota_1}\,\mbox{.}
\end{align}
The current density $\Jvec_1$ is transverse, as the electric field $\Evec_\mrm{L}$.
For technical convenience, we switched to the co-moving pulse frame by introducing the new time variable $\tau=t-z/c$. The collision frequency $\nu_\mrm{ei}$ has to be computed from Eqs.~(\ref{eq:nu_ei_NRL_2})--(\ref{eq:Kin_E}), and $n_0$ follows from Eq.~(\ref{eq:cont_0}).

\begin{figure}
  \centering
  \includegraphics[width=0.8\columnwidth]{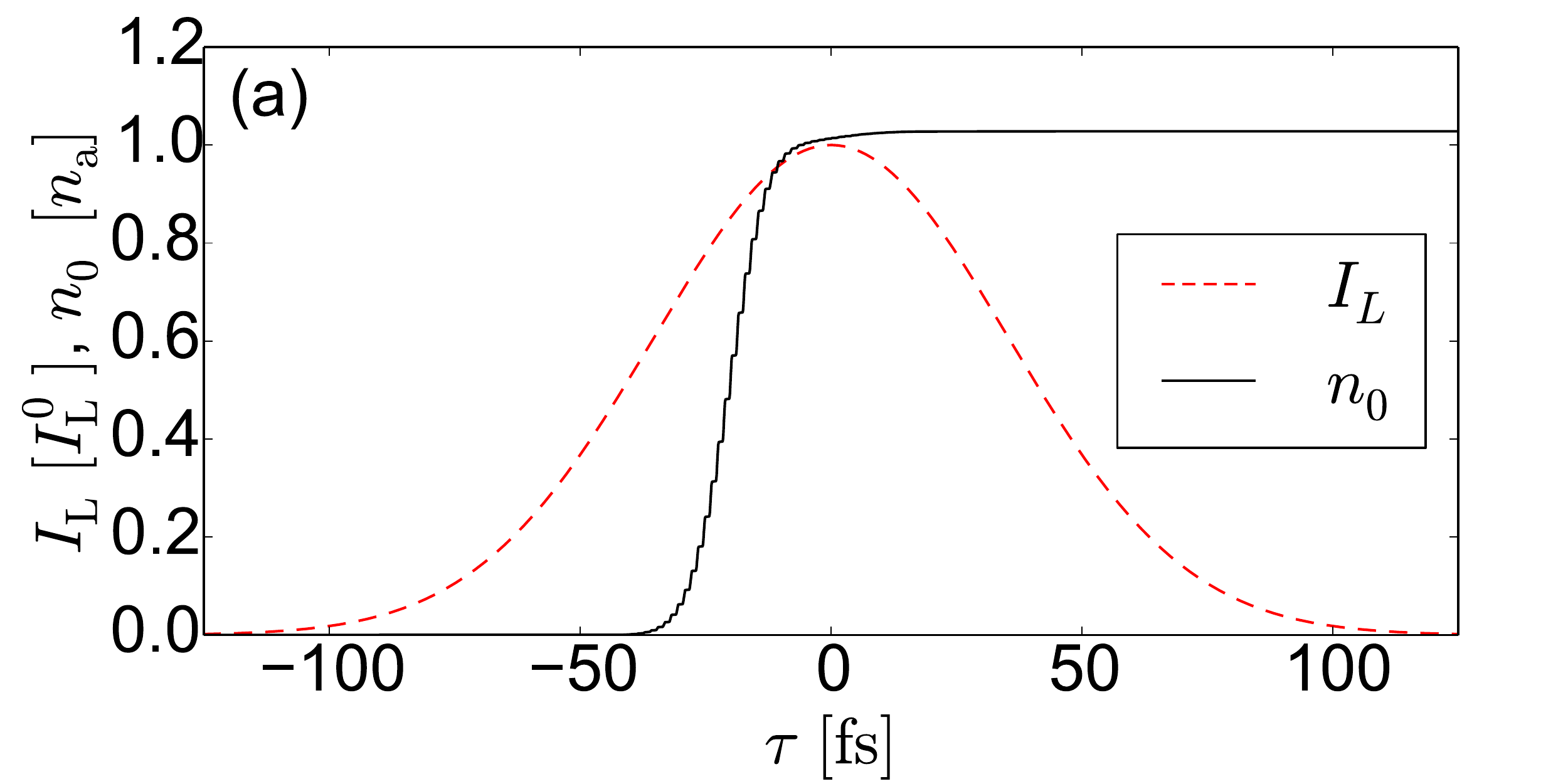}
  \includegraphics[width=0.8\columnwidth]{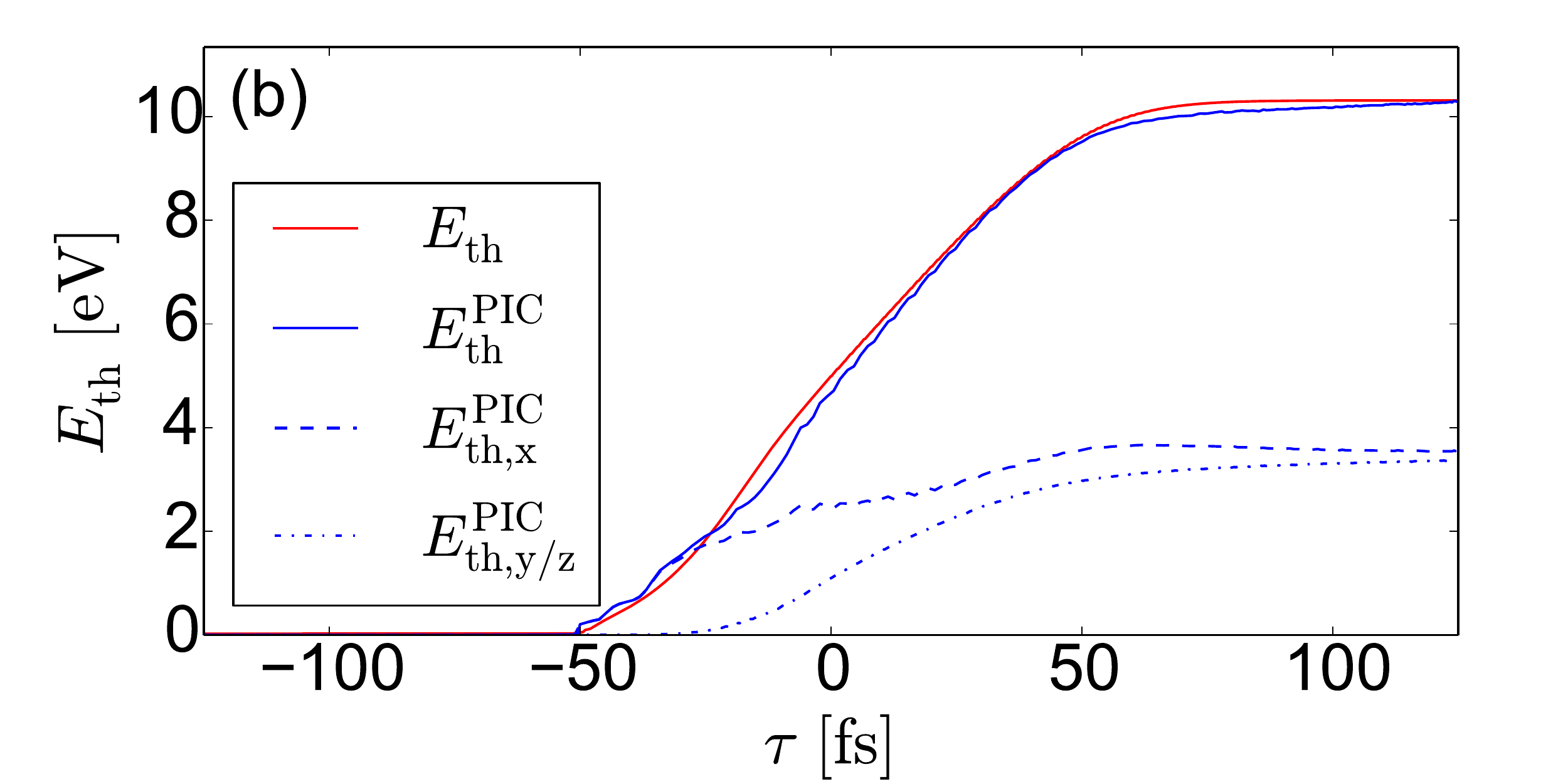}
  \includegraphics[width=0.8\columnwidth]{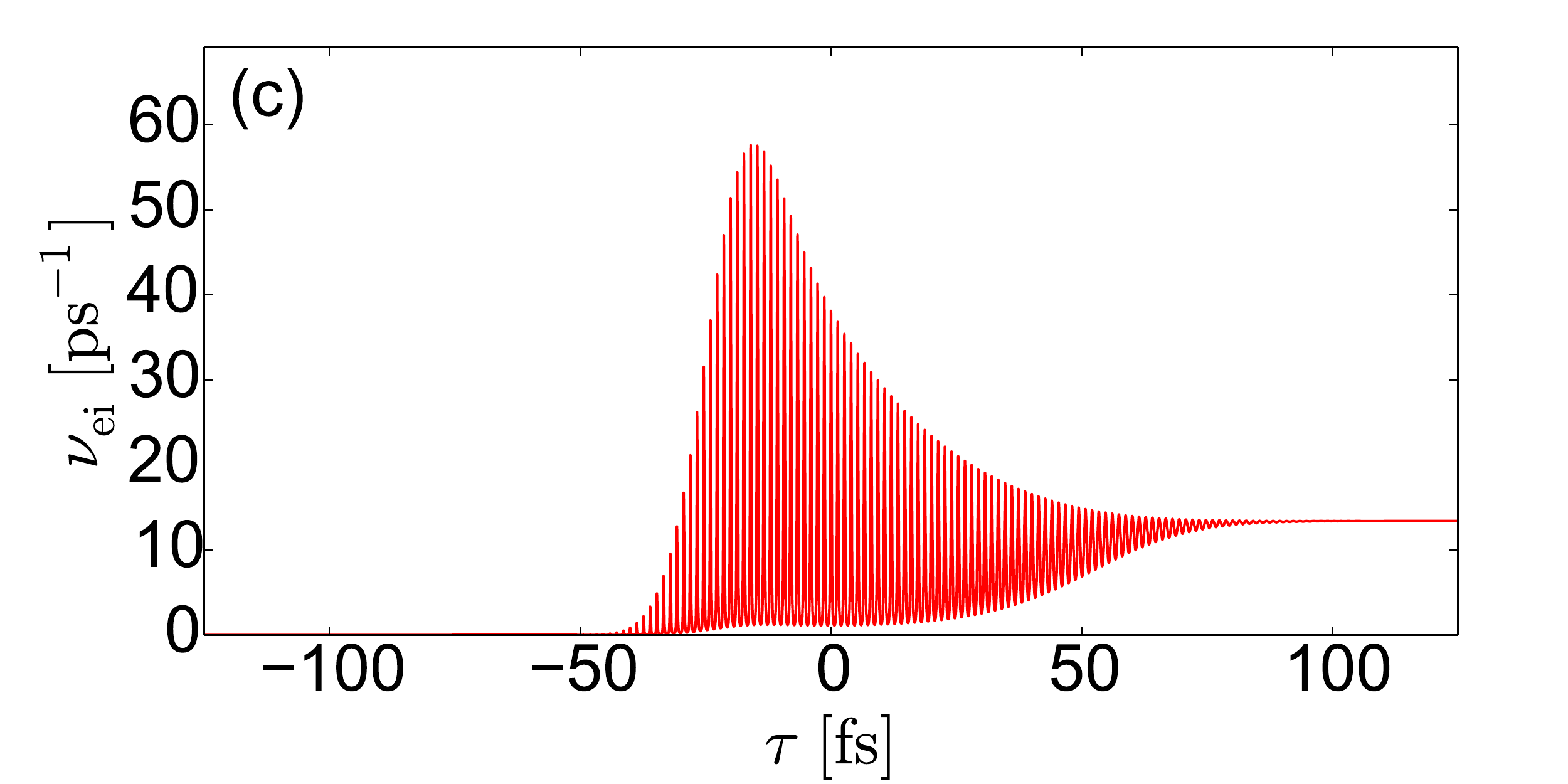}
  \caption{Example of a $t_0=50\,\mrm{fs}$, $I_{\mrm{L}}^0=4\times10^{14}\,\mrm{W/cm}^2$ laser pulse at $\lambda_\mrm{L} = 800$\,nm in argon gas with initial atom density $n_\mrm{a} = 3\times10^{19}\,\mrm{cm}^{-3}$ in 1D configuration. Because we neglect laser propagation effects, the problem depends on the co-moving time $\tau=t-z/c$ only. In (a) the laser intensity $I_\mrm{L}$ (red dashed line) and the resulting electron density $n_0$ (black solid line) according to our model are shown. Figure (b) presents the thermal energy $E_{\mrm{th}}$ as captured by the model (red line) for $\lambda_\mrm{ei}=3.5$, in excellent agreement with the thermal energy $E_\mrm{th}^\mrm{PIC}$ obtained from a PIC simulation (solid blue line, see text for details). In (c) the collision frequency $\nu_\mrm{ei}$ (red line) according to Eq.~(\ref{eq:nu_ei_NRL_2}) is shown.}
  \label{fig:1D_example_1}
\end{figure}

Let us have a look at a first illustrative example showing some basic processes, namely the ionization and laser heating, captured by the $\epsilon^1$ model.
We consider a laser pulse with $t_0=50\,\mrm{fs}$, $I_{\mrm{L}}^0=4\times10^{14}\,\mrm{W/cm}^2$, $\lambda_\mrm{L} = 800$\,nm in argon gas with initial atom density $n_\mrm{a} = 3\times10^{19}\,\mrm{cm}^{-3}$, corresponding to about 1\,bar pressure. 
The laser pulse profile is shown in Fig.~\ref{fig:1D_example_1}(a) (red dashed line). The gas atoms are getting ionized and the electron density $n_0$ in Fig.~\ref{fig:1D_example_1}(a) (black solid line) is growing step-wise near time points corresponding to extrema of $\Evec_\mrm{L}$. In this particular case the final electron density $n_0$ reaches the initial gas density $n_\mrm{a}$, thus the atoms undergo complete single ionization. The electron kinetic energy $E_{\mrm{kin}}$ reaches about $48\,\mrm{eV}$, and oscillates at $2\omega_\mrm{L}$ (not shown). According to our model, the thermal energy $E_{\mrm{th}}$ of the electrons increases up to $10\,\mrm{eV}$ [red solid line in Fig.~\ref{fig:1D_example_1}(b)]. As shown by Eq.~(\ref{eq:heating}), the heating of the electrons is driven by two mechanisms. Firstly, we have the contribution of electron-ion collisions $\propto \nu_\mrm{ei}$. Secondly, the ionization term $\propto \partial_t n_0$ increases the thermal energy as well, consistent with results published in~\cite{PhysRevA.46.2077}: Electrons which are born at a time point in the laser cycle displaced from the peak electric field acquire a dephasing energy. This mechanism is important for fs-laser pulses but becomes negligible for longer pulses ($>100\,\mrm{fs}$), where the heating from electron-ion collisions dominates. The evolution of the collision frequency $\nu_\mrm{ei}$ according to Eq.~(\ref{eq:nu_ei_NRL_2}) is shown in Fig.~\ref{fig:1D_example_1}(c) (red line). It features a maximum near the peak intensity of the driving pulse and decreases finally to $13\,\mrm{ps}^{-1}$ corresponding to a collision time of 77\,fs. Oscillations at $2\omega_\mrm{L}$ appear due to the dependency on $E_{\mrm{kin}}$.

The prediction of the thermal energy by our model is now confronted with a 1D PIC simulation accounting for electron-ion and electron-electron collisions by means of the code CALDER~\cite{Collisions}. 
The input laser pulse [same as shown in Fig.~\ref{fig:1D_example_1}(a)] propagates over 10\,\textmu m in argon, without noticeable deformation. Thus, the thermal energies extracted from this simulation depend only on the retarded time $\tau$ as well.
The PIC electron thermal energy $E_{\mrm{th}}^\mrm{PIC}$ shown in Fig.~\ref{fig:1D_example_1}(b) (solid blue line) is in excellent agreement with the model.
As already mentioned above, we expect contributions from two different heating processes, which can be visualized in the PIC results. The dashed blue line in Fig.~\ref{fig:1D_example_1}(b) shows the thermal energy $E_{\mrm{th},x}^\mrm{PIC}$ in the motion of the electrons along the $x$-axis, which is the laser polarization direction. The dash-dotted blue line shows the thermal energy $E_{\mrm{th},y/z}^\mrm{PIC}$ contained in each of the other two degrees of freedom. Thus, the momentum distribution of the electrons in the PIC simulation is anisotropic. 
The reason for this anisotropy is the heating by the second term on r.h.s in Eq.~(\ref{eq:heating}): The corresponding dephasing energy (see above) is acquired solely along the laser polarization direction leading to a momentum spread of the electron distribution function along $x$ only.
In contrast, heating by electron-ion collisions is isotropic. 
However, the phase space quickly thermalizes in the PIC simulation due to electron-electron and electron-ion collisions. This fact justifies the assumption of instantaneous thermalization in our model. 

Before discussing THz emission from the $\epsilon^1$ model, let us proceed with the calculation of $\Jvec_2$. In 1D configuration, the current at order $\epsilon^2$ is driven by the purely longitudinal source term $\bm\iota_2 = \iota_{2,z}\evec_z$, which according to Eq.~(\ref{eq:iota_2}) contains four contributions
\begin{equation}
	\iota_{2,z} = \iota^\mrm{pond}_{2,z} +  \iota^\mrm{ion}_{2,z} +  \iota^\mrm{col}_{2,z} + \iota^\mrm{heat}_{2,z} \,\mbox{.}
\end{equation}
In the co-moving pulse frame, these contributions read (see App.~\ref{app:iota} for details)
\begin{equation}
\begin{aligned}
     \iota^\mrm{pond}_{2,z} &= 
\frac{n_0}{2q_\mrm{e}c}\partial_\tau{\left|\frac{\Jvec_{1}}{n_0}\right|}^2 \,\mbox{,} &
     \iota^\mrm{ion}_{2,z} &= \frac{\left(\partial_\tau n_0\right)}{q_\mrm{e}c} {\left|\frac{\Jvec_{1}}{n_0}\right|}^2  \,\mbox{,} \\
     \iota^\mrm{col}_{2,z} &= 
\frac{n_0\nu_\mrm{ei}}{q_\mrm{e}c}{\left|\frac{\Jvec_{1}}{n_0}\right|}^2\,\mbox{,} &
     \iota^\mrm{heat}_{2,z} &= 
\frac{2\partial_\tau \! \left(n_0 E_{\mrm{th}}\right)}{3m_\mrm{e}q_\mrm{e}c}\,\mbox{.}
	\label{eq:iota_2z_cont}
\end{aligned}
\end{equation}
The first term is the ponderomotive source $\iota^\mrm{pond}_{2,z}$. The second source term $\iota^\mrm{ion}_{2,z}$ is a direct consequence of the ionization, and is absent in preformed plasmas. The third source term $\iota^\mrm{col}_{2,z}$ takes into account the radiation pressure. Finally, the fourth source term $\iota^\mrm{heat}_{2,z}$ is caused by  diffusion or pressure of the electrons. We note that $\iota^\mrm{pond}_{2,z}$, $\iota^\mrm{ion}_{2,z}$, and $\iota^\mrm{col}_{2,z}$ have already been derived in~\cite{PhysRevE.69.066415}. 
In a 1D configuration, the longitudinal component of $\nabla\times\Bvec$ vanishes and Eq.~(\ref{eq:Amp_1}) dictates that the longitudinal electric field $E_{2,z}$ is connected to the longitudinal current $J_{2,z}$ via
\begin{equation}
	J_{2,z} = -\epsilon_0 \partial_{\tau} E_{2,z}\,\mbox{.} \label{eq:jz1d}
\end{equation}
Thus, we can substitute $J_{2,z}$ in Eq.~(\ref{eq:cont_2}) and end up with the following equation for the longitudinal field $E_{2,z}$:
\begin{equation}
	\partial_{\tau\tau} E_{2,z} + \nu_\mrm{ei}\partial_{\tau} E_{2,z} + \left(\frac{q_\mrm{e}^2n_0}{m_\mrm{e} \epsilon_0}\right) E_{2,z} = -\frac{\iota_{2,z}}{\epsilon_0}\,\mbox{.}
	\label{eq:osc_E_2NL}
\end{equation}
All quantities involved, in particular $\iota_{2,z}$, can be computed from the solution to the \first order problem.

\begin{figure}
  \centering
  \includegraphics[width=0.8\columnwidth]{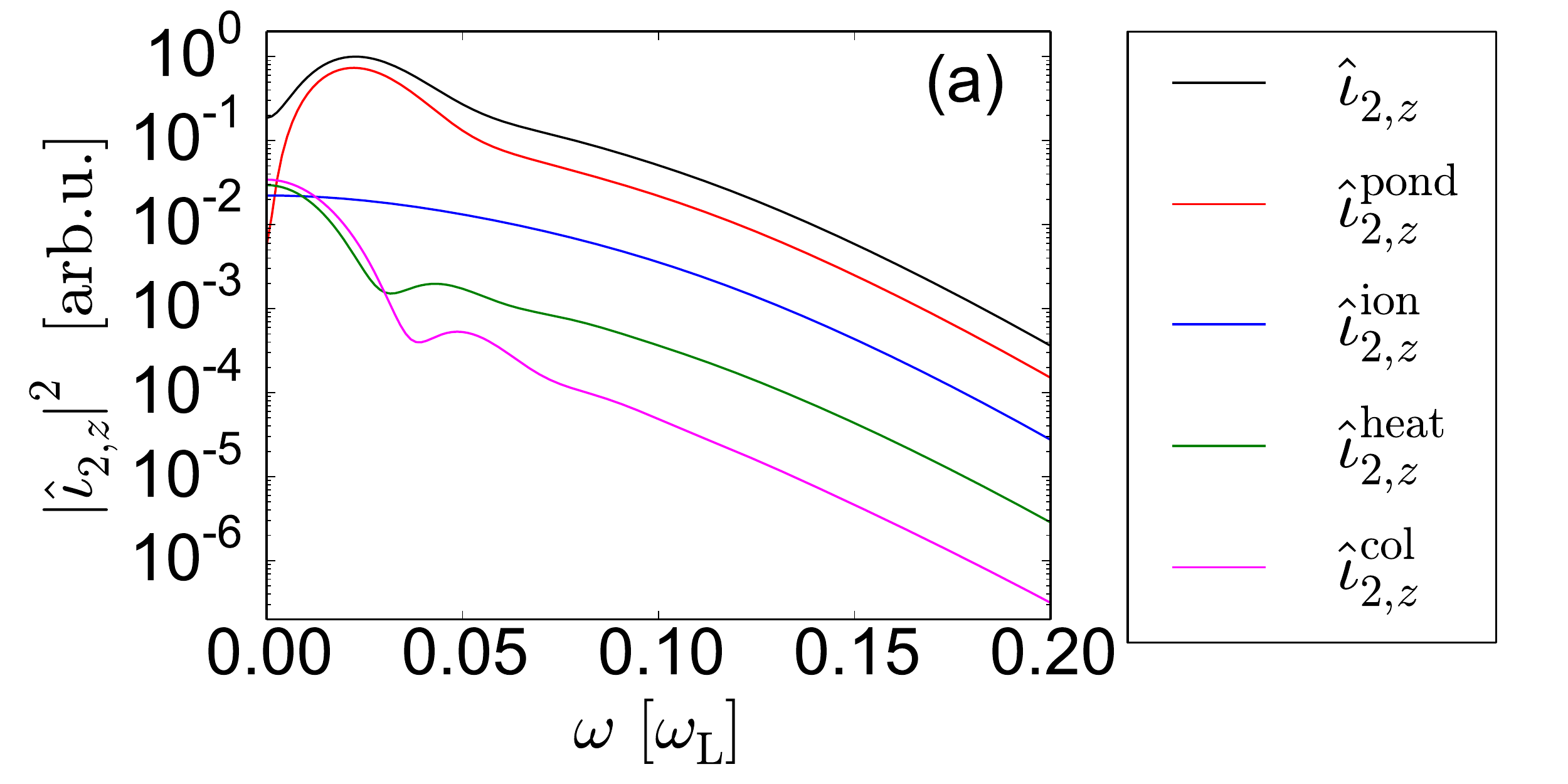}
  \includegraphics[width=0.8\columnwidth]{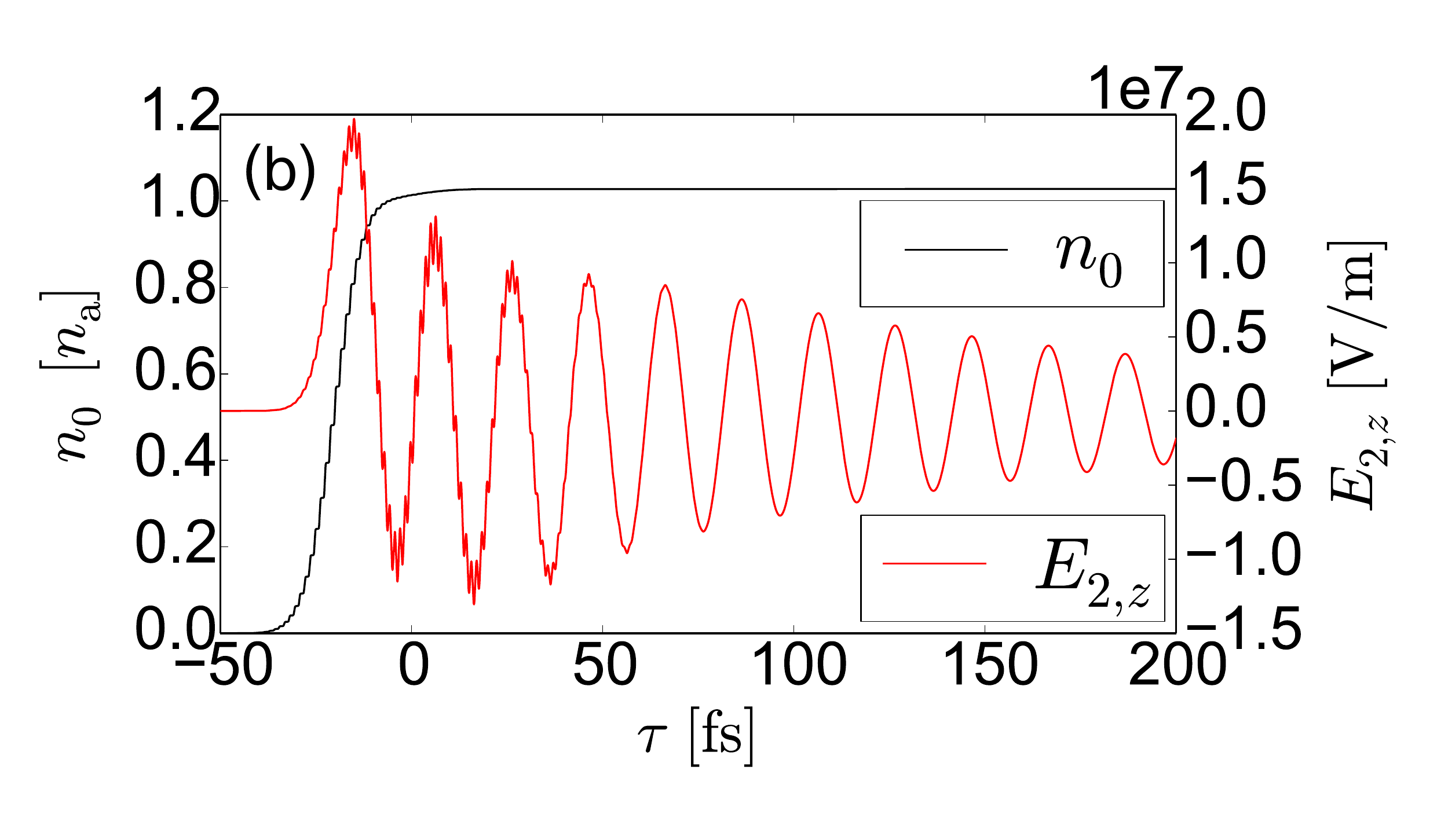}
  \includegraphics[width=0.8\columnwidth]{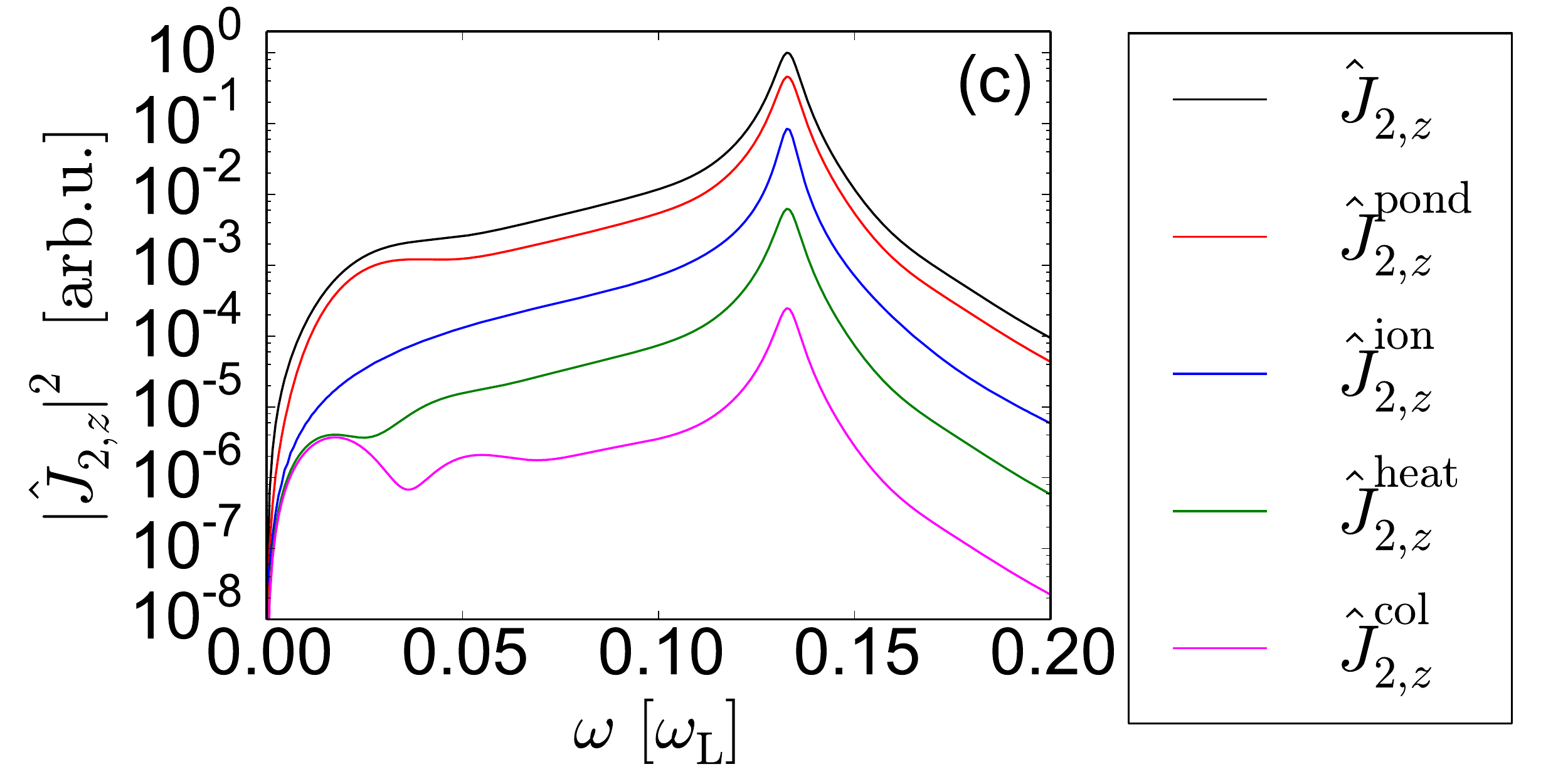}
  \caption{In (a) the low-frequency power spectra of the second-order source term and its constituents as specified in the legend are shown. The longitudinal electric field $E_{2,z}$ is presented in (b) together with the electron density $n_0$. In (c) the power spectra of the longitudinal currents corresponding to the source terms in (a) are plotted. Driving laser parameters are the same as in Fig.~\ref{fig:1D_example_1}.}
  \label{fig:1D_example_2}
\end{figure}

\begin{figure*}
  \centering
  \includegraphics[width=1.55\columnwidth]{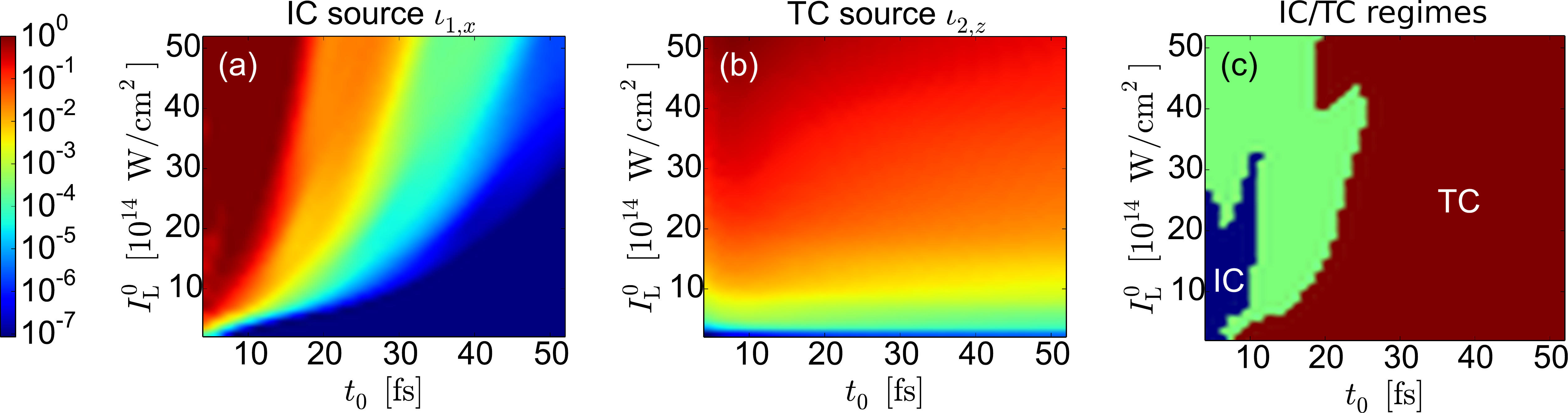}
  \caption{In (a) the up to $0.2\omega_\mrm{L}$ ($\nu\leq75\,\mrm{THz}$) integrated power spectrum of the source term $\iota_{1,x}$ (IC source) is shown as a function of laser pulse duration $t_0$ and intensity $I_\mrm{L}^0$. The same data for $\iota_{2,z}$ (TC source) are shown in (b). These two terms are compared in panel (c): In the blue region the IC source term $\iota_{1,x}$ dominates by at least one order of magnitude, in the red region the same is true for the TC source term $\iota_{2,z}$, and in the green region $\iota_{1,x}$ and $\iota_{2,z}$ are both important. The computations are performed for an argon gas with the initial atom density $n_\mrm{a} = 3\times10^{19}\,\mrm{cm}^{-3}$.}
  \label{fig:IC_TC_comp}
\end{figure*}

\begin{figure}[b]
  \centering
  \includegraphics[width=\columnwidth]{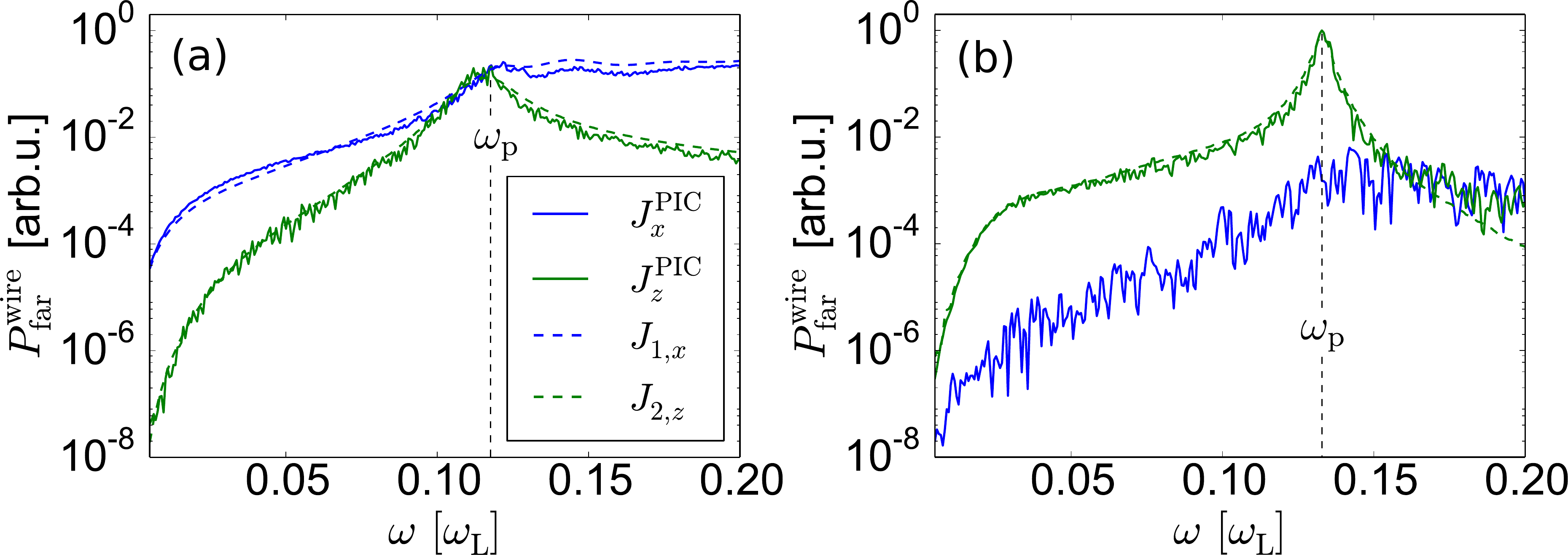}
  \caption{Hypothetical far field spectra integrated over all angles computed by assuming an infinitely thin 10\,\textmu m long plasma wire (see text) are shown for $I_\mrm{L}^0 = 4\times10^{14}$~W/cm$^2$, $t_0=5$~fs~(a) and $t_0=50$~fs~(b). Power spectra are calculated from current densities obtained by 1D PIC simulations and the model according to the legend in (a).}
  \label{fig:IC_TC_comp_PIC}
\end{figure}

Let us now come back to the case study of a 50-fs pulse from above. The low-frequency power spectra of the \second order source term $\iota_{2,z}$ and its four constituents defined in Eq.~(\ref{eq:iota_2z_cont}) are presented in Fig.~\ref{fig:1D_example_2}(a). 
Obviously, $\iota_{2,z}$ (black line) is dominated by the ponderomotive source $\iota^\mrm{pond}_{2,z}$ (red line). Other contributions are at least one order of magnitude smaller for this driving  pulse. The peak excitation happens around $0.022\omega_\mrm{L}$ (i.e., $\nu\approx8.25\,\mrm{THz}$). In comparison, the power spectrum of $\iota_{2,z}$ at the plasma frequency $\omega_\mrm{p}\approx\sqrt{(q_\mrm{e}^2 n_\mrm{a}/m_\mrm{e} \epsilon_0)}\approx0.13\omega_\mrm{L}$ (i.e., $\nu\approx50\,\mrm{THz}$) is almost two orders of magnitude smaller~\footnote{For our 50-fs example pulse the final electron density is $n_0(t\rightarrow\infty) \approx n_\mrm{a}$}. Nevertheless, longitudinal plasma oscillations at $\omega_\mrm{p}$ are excited in $E_{2,z}$ when the electron density $n_0$ builds up, as can be seen in Fig.~\ref{fig:1D_example_2}(b). This excitation is also visible in the spectrum of the current $J_{2,z}$ shown in Fig.~\ref{fig:1D_example_2}(c).
It is interesting to note that for our example the observed strong excitation at the plasma frequency $\omega_\mrm{p}$ is intimately linked to ionization. Shooting the same driving pulse into a preformed plasma with constant density $n_0 \equiv n_\mrm{a}$ triggers almost no oscillations at $\omega_\mrm{p}$ (not shown). For constant $n_0$, the power spectrum of the second-order source term $\iota_{2,z}$ is more narrow, and in particular its value at $\omega_\mrm{p}$ is more than two orders of magnitude lower. Only very short driving pulses fulfill the resonance condition $t_0\lesssim\pi/\omega_\mrm{p}$ and significantly excite plasma oscillations in a preformed plasma.

In our discussion of plasma currents in the THz spectral range above we completely ignored the first-order current $\Jvec_{1}$. The reason for this is simply that for our 50-fs single-color driving pulse, $J_{1,x}$ has no significant THz component. In the spectral range shown in  Fig.~\ref{fig:1D_example_2}(c), the power spectrum of $J_{1,x}$ is more than ten orders of magnitude lower than that of $J_{2,z}$. Thus, in our example the IC mechanism is not present and THz emission results from the TC mechanism only. However, this may change for other driving pulse parameters, even in single-color configuration.
In the following, the IC ($\Jvec_1$) and TC ($\Jvec_2$) mechanisms are compared for laser intensities $I_\mrm{L}^0=2-50\times10^{14}\,\mrm{W/cm}^2$ and pulse durations $t_0=4-50\,\mrm{fs}$.
Figures~\ref{fig:IC_TC_comp}(a,b) show the power spectra of $\iota_{1,x}$ and $\iota_{2,z}$ (IC resp.\ TC source) integrated up to $0.2\omega_\mrm{L}$ ($\nu\leq75\,\mrm{THz}$). Figure~\ref{fig:IC_TC_comp}(a) reveals that the IC mechanism requires short and intense pulses to play a role, in agreement with \cite{Debayle:14}. By contrast, the source term $\iota_{2,z}$ of the TC mechanism varies only weakly in the considered parameter range [see Fig.~\ref{fig:IC_TC_comp}(b)]. Finally, Fig.~\ref{fig:IC_TC_comp}(c) displays the parameter regions where one of the source terms dominates by at least one order of magnitude. We can conclude that the IC mechanism is important for very short pulses only, whereas the TC mechanism is the key player for sufficiently long pulses.

We now want to cross-check the predictions of Fig.~\ref{fig:IC_TC_comp} by means of 1D PIC simulations. To this end, we consider two pulse configurations: A few-cycle pulse with $t_0=5\,\mrm{fs}$, $I_\mrm{L}^0 = 4\times10^{14}\,\mrm{W/cm}^2$ to illustrate the IC dominated regime and the multi-cycle pulse with $t_0=50\,\mrm{fs}$, $I_\mrm{L}^0 = 4\times10^{14}\,\mrm{W/cm}^2$ already used above as an example for the TC dominated regime. Following~\cite{PhysRevLett.98.235002,ISI:000253084000007}, we use the 1D results for the current density $\Jvec$ and compute a hypothetical far-field spectrum $P_\mrm{far}^\mrm{wire}$ by assuming the plasma having a transverse shape of an infinitely thin 10\,\textmu m long wire and by means of Jefimenko's equations~\cite{jackson}. Simulation (solid lines) and model results (dashed lines) are presented in Fig.~\ref{fig:IC_TC_comp_PIC}. For the IC mechanism (blue lines) we use the transverse current $J_x^\mrm{PIC}$ from the PIC simulation and $J_{1,x}$ from the full $\epsilon^1$ model, without approximating $\Evec_1$ by $\Evec_\mrm{L}$. For the TC mechanism (green lines) we use the longitudinal current $J_z^\mrm{PIC}$ from the PIC simulation and $J_{2,z}$ from the $\epsilon^2$ model. Obviously, PIC simulations (solid lines) and the model (dashed lines) are in an excellent agreement. The PIC results confirm that the IC mechanism (blue lines) dominates the TC mechanism (green lines) for the short pump pulse (a), and vice versa for the longer pulse (b). In the latter case, $P_\mrm{far}^\mrm{wire}$ produced by $J_x^\mrm{PIC}$ is even dominated by the noise of the PIC simulation and the model gives a four orders of magnitude lower signal (not shown here), far below the signal from the TC mechanism.
 
It is important to note that the IC mechanism requires the full treatment of the equations at order $\epsilon^1$. In contrast, assuming $\Evec_1\approx\Evec_\mrm{L}$ causes almost no loss of accuracy when evaluating the $\epsilon^2$ order of the model: The computation of the nonlinear source $\iota_{2,z}$ for Fig.~\ref{fig:IC_TC_comp_PIC}(b) was performed approximating $\Jvec_1$ according to Eq.~(\ref{eq:J_L}), and gives already perfect agreement with the PIC simulation. 
For the few-cycle pulse in Fig.~\ref{fig:IC_TC_comp_PIC}(a), laser absorption due to ionization and electron heating becomes notable: The electric field amplitude decreases during the propagation through the 10 \textmu m long gas plasma by 3\,\%, and the final electron density at $z=10$~\textmu m is about 11\,\% smaller than at  $z=0$~\textmu m (not shown here). Therefore, in Fig.~\ref{fig:IC_TC_comp_PIC}(a) a full treatment of the model up to order $\epsilon^2$ was necessary to obtain perfect agreement with PIC results.

It is quite tempting to conclude from hypothetical far field spectra obtained from 1D results as shown in Fig.~\ref{fig:IC_TC_comp_PIC} on actual THz emission from a real 3D plasma as produced in experiments. While such reasoning can be found in the literature, e.g.\ in~\cite{PhysRevLett.98.235002}, it is generally incorrect. 
On the one hand, we assume translation invariance in the transverse directions when computing the 1D current, on the other hand, we impose later a thin transverse shape of the plasma wire when computing the hypothetical far field. 
As we will see in the next section, realistic THz emission spectra differ very strongly from Fig.~\ref{fig:IC_TC_comp_PIC}. The reason for this discrepancy is that not all plasma currents lead to emission of radiation, and in particular oscillations at the plasma frequency may not radiate~\cite{PhysRevLett.89.209301}.

\section{\label{sec:Dist}Radiating and non-radiating excitations} 

In the previous section we have analyzed plasma currents generated by an intense fs laser pulse in a gas, with particular emphasize on excitations in THz spectral range. The question we want to answer now is how the radiation produced by these currents looks like. 
As already indicated in the concluding remarks of Sec.~\ref{sec:Comp}, one has to be careful with longitudinal excitations at the plasma frequency, which may not contribute to the radiation spectrum  \cite{PhysRevLett.89.209301}. The reason for this will be elaborated in the following, however, a simple physical picture gives already some insight: The plasma oscillations at $\omega_p$, whenever they are eigen-oscillations of the system, would continue forever (for $\nu_\mrm{ei}=0$) and their energy would stay inside the plasma. Thus, they can not contribute to radiation, or otherwise energy conservation is violated. 

In this section, we focus on the TC mechanism and thus emission from the \second order current $\Jvec_2$. As we have seen above, this mechanism is expected to dominate the THz emission from micro-plasmas created by multi-cycle single-color laser pulses as used in~\cite{Buccheri:15}. At the end of the section we briefly comment on the IC mechanism. Our starting point is thus Eq.~(\ref{eq:cont_2}). Using Maxwell's equations (\ref{eq:Far_1}),~(\ref{eq:Amp_1}) we find
\begin{equation}
\begin{split}
	\partial_{tt} \Evec_2 + \nu_\mrm{ei}\partial_t \Evec_2 + \frac{q_\mrm{e}^2n_0}{m_\mrm{e} \epsilon_0} \Evec_2 & \\
	+ \, c^2 \nabla\times\nabla\times \Evec_2  + \nu_\mrm{ei} c^2 \! \int\limits_{-\infty}^t \! \nabla\times\nabla\times \Evec_2 \, dt' & = - \frac{\bm{\iota_2}}{\epsilon_0}
	\label{eq:E_wave}
\end{split}
\end{equation}
for the field $\Evec_{2}$. This equation is nothing else but the 3D version of Eq.~(\ref{eq:osc_E_2NL}). In contrast to the 1D case studied above (where $\bm{\iota_2}$ is purely longitudinal and the $z$-component of $\nabla\times\Evec_2$ vanishes), in 3D all components of $\Evec_2$ are non-zero and coupled. Moreover, focusing dynamics of the driving laser pulse render a transformation to the co-moving pulse frame useless. 

In order to identify the part of $\Evec_2$ which actually contributes to the far field, we note that according to Eq.~(\ref{eq:Far_1}) a curl-free field ($\nabla\times\Evec_2=0$) does not lead to radiation in the far-field, because no electro-magnetic wave is produced ($\partial_t\Bvec_2=0$). 
By using the Helmholtz decomposition theorem, we can decompose $\Evec_2=\Evec_{2,\mrm{d}}+\Evec_{2,\mrm{r}}$ into a curl-free field $\Evec_{2,\mrm{d}}$ with $\nabla\times\Evec_{2,\mrm{d}}=0$ and a divergence-free field $\Evec_{2,\mrm{r}}$ with $\nabla\cdot\Evec_{2,\mrm{r}}=0$. 
In general, both fields are coupled in Eq.~(\ref{eq:E_wave}) by the terms $\propto n_0$ and $\propto\nu_\mrm{ei}$. By taking the curl of Eq.~(\ref{eq:E_wave}) we find that $\Evec_{2,\mrm{r}}$ can decouple from $\Evec_{2,\mrm{d}}$ if
\begin{align}
	\Evec_{2,\mrm{d}}\times\nabla n_0&=0\,\mbox{,}  & \partial_t \Evec_{2,\mrm{d}}\times\nabla \nu_\mrm{ei}&=0\,\mbox{.}  \label{eq:E_curl_free}
\end{align}

\begin{figure}
  \centering
  \includegraphics[width=\columnwidth]{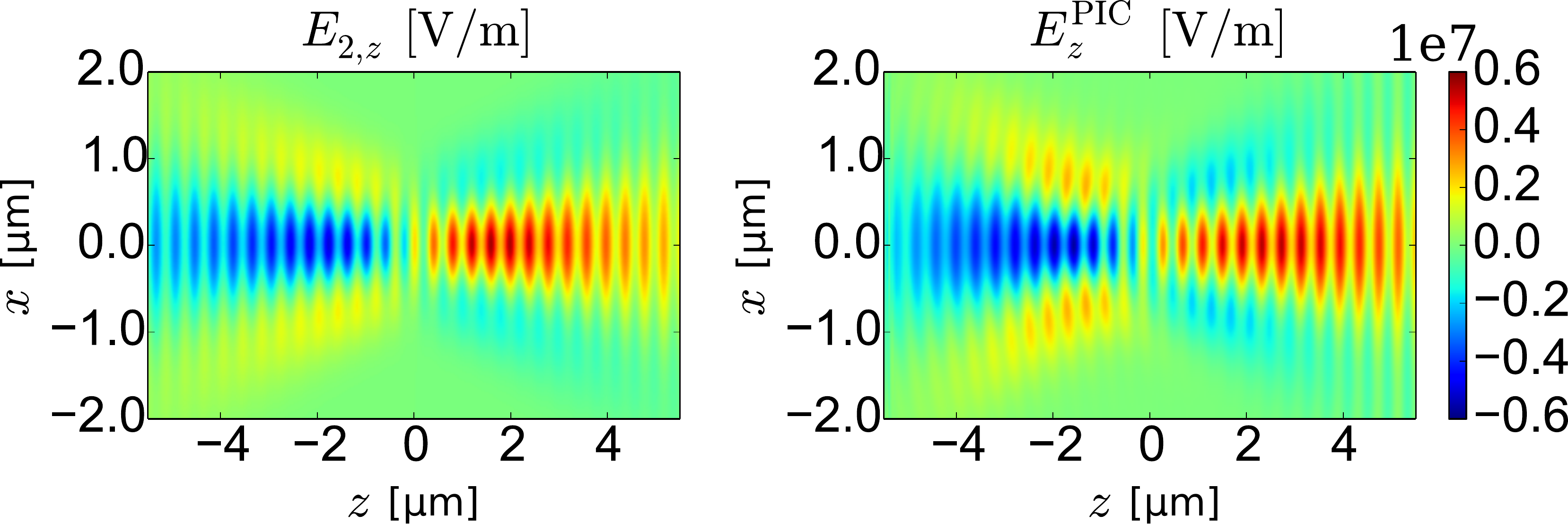}
  \caption{Snapshot of the longitudinal electric field (a) $E_{2,z}$ from our model and (b) $E_z^\mrm{PIC}$ from a corresponding 2D PIC simulation at the time moment when the laser pulse is at focus. The $y$-polarized Gaussian laser pulse ($t_0=50$\,fs, $I_{\mrm{max}}=4\times10^{14}\,\mrm{W/cm}^2$) is focused to $w_0=0.8$\,\textmu m into a uniform preformed plasma ($n_0=3\times10^{19}\,\mrm{cm}^{-3}$).}
  \label{fig:E_z_preformed}
\end{figure}

Let us have a look at a simple but illustrative example for the occurrence of such non-radiative curl-free electric fields: 
In a preformed collisionless plasma ($n_0=\mrm{const.}$, $\nu_\mrm{ei}=0$), the curl-free and divergence-free fields are decoupled (see also \cite{jackson}). 
We found in the previous section that the ponderomotive source $\bm{\iota}_2^\mrm{pond}=-n_0/2q_\mrm{e}\nabla|\Jvec_1/n_0|^2$ gives the dominant contribution.
Because $\bm{\iota}_2^\mrm{pond}$ is obviously curl-free, the solution to Eq.~(\ref{eq:E_wave}) is also curl-free and the wave equation reduces to a simple oscillator equation
\begin{equation}
	\partial_{tt} \Evec_2 + \frac{q_\mrm{e}^2n_0}{m_\mrm{e} \epsilon_0} \Evec_2 = - \frac{n_0}{2 q_\mrm{e}}\nabla{\left\|\frac{\Jvec_1}{n_0}\right\|}_2^2\,\mbox{.}
\end{equation}
We now consider a 2D (translational invariance in $y$-direction) 50-fs Gaussian pulse, linearly polarized in $y$-direction, and strongly focused into a uniform preformed plasma ($n_0=3\times10^{19}\,\mrm{cm}^{-3}$). The peak intensity at focus is $I_{\mrm{max}}=4\times10^{14}\,\mrm{W/cm}^2$, and the transform-limited beam width is $0.8$\,\textmu m.
This particular 2D configuration has the advantage that in the PIC simulation the electric field of the driving laser appears in the $E_y$ component only, and the longitudinal component $E_z$ is produced by the plasma only. Thus a direct confrontation of $E_{2,z}$ from the model with $E_{z}^{\mrm{PIC}}$ is possible.
When evaluating the model, the ponderomotive source and the laser field are approximated (see App.~\ref{app:iota_pond_par}). Nevertheless, a temporal snapshot of the longitudinal electric field at focus (see Fig.~\ref{fig:E_z_preformed}) shows excellent agreement between analytical model (a) and 2D PIC simulation (b). A low-frequency field and a second harmonic (SH) field are clearly visible as a fast and slow modulation pattern along $z$. Both fields are non-radiating according to our previous argumentation. 
This is confirmed by inspecting the magnetic field component $B_y$ in the PIC simulation, which is found to be at background noise level (not shown).

For laser-induced plasmas, we have a finite plasma volume with spatially (and temporally) varying electron density $(\nabla n_0 \ne 0)$. 
For the sake of simplicity, we will look for curl-free solutions of Eq.~(\ref{eq:E_wave}) for $\bm\iota_2=0$ only, i.e., after the driving pulse has passed. Then, $n_0$ is constant in time and the general solution in the collisionless case has the form (see App.~\ref{app:static})
\begin{equation}
	\Evec_{2,\mrm{d}} = A(n_0) \cos\!\left[\sqrt{\frac{q_\mrm{e}^2n_0}{m_\mrm{e} \epsilon_0}}t+\phi(n_0)\right] \nabla n_0\,\mbox{,}
	\label{eq:stat_osc}
\end{equation}
where $A$ and $\phi$ are scalar functions depending on the electron density $n_0$. 
The solution $\Evec_{2,\mrm{d}}$ oscillates at the local plasma frequency $\omega_\mrm{p}(\rvec)=\sqrt{q_\mrm{e}^2n_0(\rvec)/m_\mrm{e} \epsilon_0}$, and the electric field vector is always parallel to $\nabla n_0$, thus $\Evec_{2,\mrm{d}}$ is decoupled from radiating fields $\Evec_{2,\mrm{r}}$ [c.f. Eq.~(\ref{eq:E_curl_free})].

\begin{figure}[t]
  \centering
  \includegraphics[width=\columnwidth]{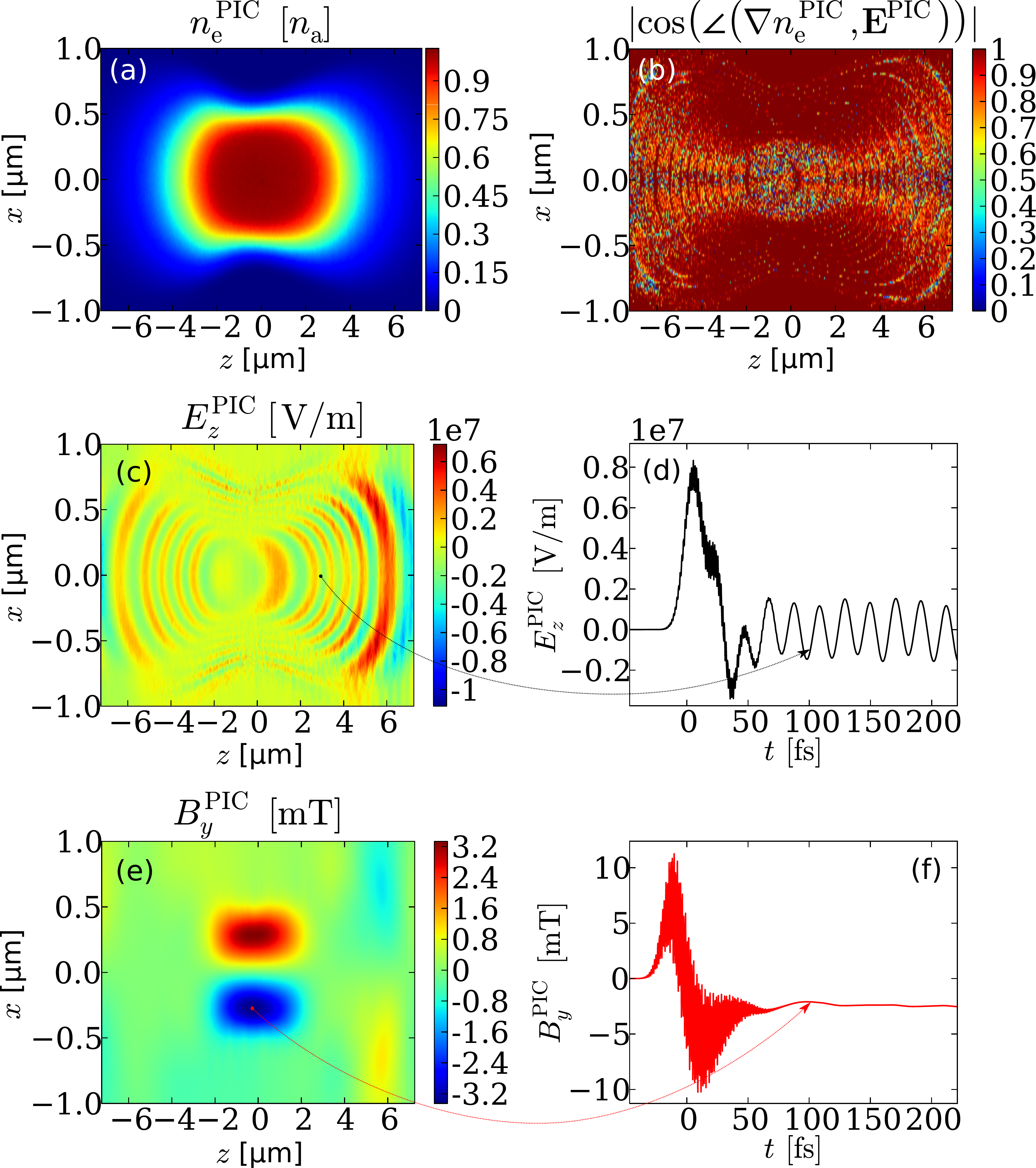}
  \caption{The same laser pulse as in Fig.~\ref{fig:E_z_preformed} is focused into argon gas at ambient pressure. A snapshot of the generated plasma (a) and the absolute value of the cosine of the angle between $\Evec^\mrm{PIC}$ and $\nabla n^\mrm{PIC}$ (b) after the pulse has passed the focus are shown. The corresponding snapshot of the electric field component $\E_z^\mrm{PIC}$ is depicted in (c). The exemplary time trace of $\E_z^\mrm{PIC}$ in (d) features oscillations at the local plasma frequency, in agreement with Eq.~(\ref{eq:stat_osc}). The snapshot of $B_y^\mrm{PIC}$ in (e) shows the static field which is present in the interaction region after the laser pulse has passed (see corresponding time trace in (f)). All temporal snapshots in (a,b,c,e) are taken about 100\,fs after the pulse has passed the focus. Recording positions of the time traces shown in (d,f) are indicated by the respective arrows.}
  \label{fig:plasma_Ar}
\end{figure}

\begin{figure}[t]
  \centering
  \includegraphics[width=0.9\columnwidth]{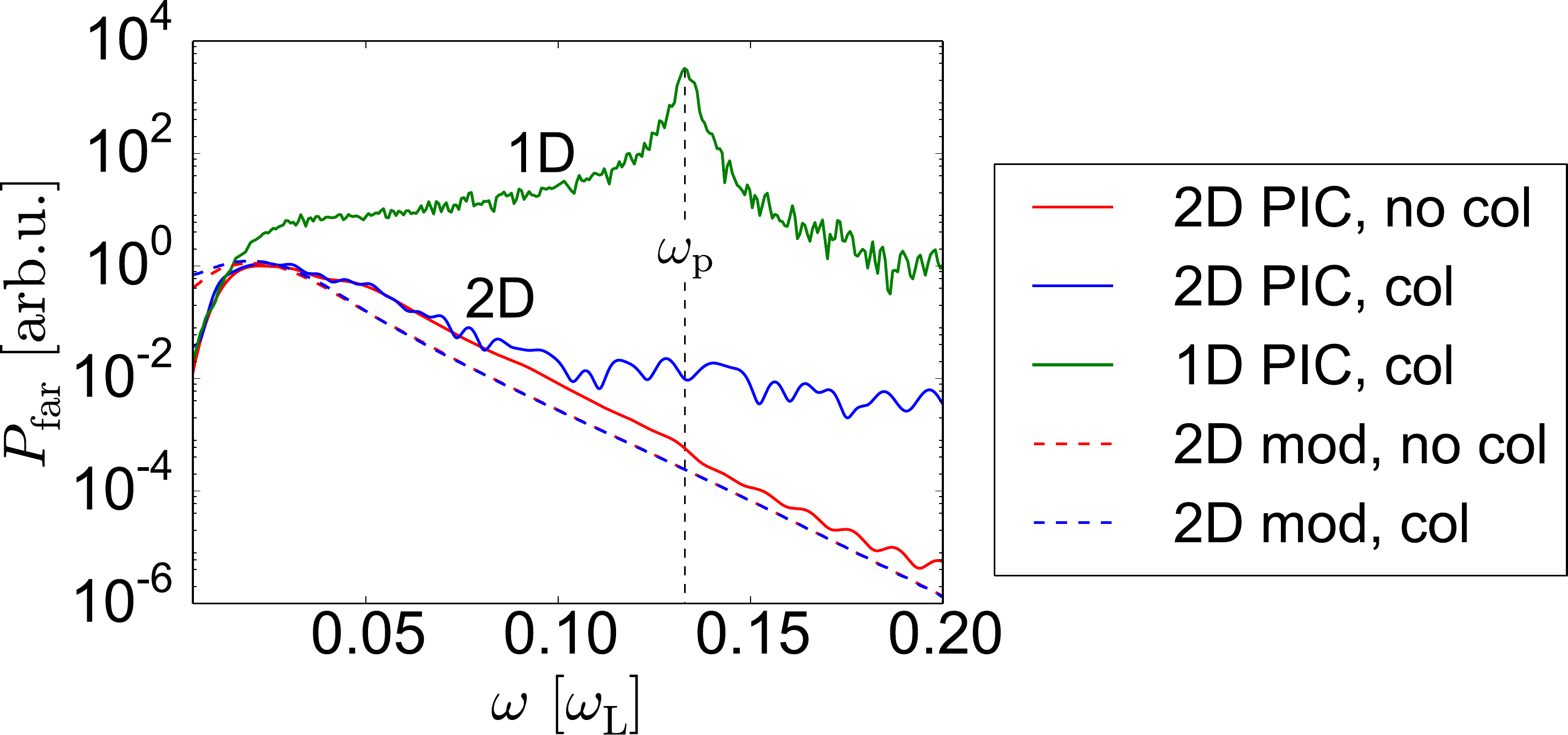}
  \caption{The same laser pulse as in Figs.~\ref{fig:E_z_preformed},~\ref{fig:plasma_Ar} is focused into argon gas at ambient pressure. The angle-integrated far field spectra obtained from 2D PIC simulation, model and 1D wire model are presented according to the legend.}
  \label{fig:P_far_2D}
\end{figure}

Equation~(\ref{eq:stat_osc}) presents the solution for a non-radiating eigen-oscillation at the plasma frequency in 2D/3D configuration. We now want to show that such fields are really excited in laser-induced micro-plasmas and shoot the same $y$-polarized laser as in the previous example of a preformed plasma in argon gas with $n_\mrm{a}=3\times10^{19}\,\mrm{cm}^{-3}$. 
As above, we use the 2D geometry with translational invariance in $y$-direction in the PIC simulation and neglect collisions.
The resulting electron density profile after the laser pulse has passed through the interaction region is shown in Fig.~\ref{fig:plasma_Ar}(a): A 10\,\textmu m long and 1\,\textmu m wide plasma with fully singly ionized argon at focus. In order to check whether the electric field in the PIC simulation after the laser pulse has passed is of the form Eq.~(\ref{eq:stat_osc}), we compute the absolute value of the cosine of the angle between $\Evec^\mrm{PIC}$ and $\nabla n_\mrm{e}^\mrm{PIC}$~\footnote{If the absolute of $\Evec^\mrm{PIC}$ or $\nabla n_\mrm{e}^\mrm{PIC}$ is smaller than 1\% of its average value in the whole box, we set the value to unity since the angle between zero-vectors cannot be defined.} and present the result in Fig.~\ref{fig:plasma_Ar}(b).
Obviously, after the laser has left the interaction region the two vectors are (anti-)parallel almost everywhere. Moreover, Figs.~\ref{fig:plasma_Ar}(c,d) confirm that after the laser has left the interaction region, the electric field $\Evec^\mrm{PIC}$ oscillates at the local plasma frequency (here shown for the $z$-component only). However, as expected from our previous reasoning, these oscillations occur inside the plasma only, the angle-integrated far field spectrum in Fig.~\ref{fig:P_far_2D} (solid red line) exhibits no feature at the plasma frequency, in direct contradiction to the results from the 1D wire model discussed in the previous section (solid green line). Moreover, 2D PIC simulations with collisions (solid blue line) coincide perfectly with PIC simulations without collisions, up to the noise level around $10^{-2}$. In particular, collisions do not enhance the far-field amplitude at the plasma frequency.

As previously explained, in our simulation no radiation is emitted due to plasma oscillations after the excitation by the laser pulse. During the laser pulse, the non curl-free $\bm\iota_2^\mrm{pond}$ term (as $ \nabla n_0 \ne 0$) is able to generate a potentially radiating field $\Evec_{2,r}$. The investigation of this TC radiation will be presented in the next section and explains the spectrum observed in Fig.~\ref{fig:P_far_2D}.

Besides the non-radiating excitation $\Evec_{2,\mrm{d}}$, there exists a second non-radiating, magneto-static excitation.
The corresponding field $\Bvec_{2,\mrm{m}}$ is linked to a non-radiating, static current $\Jvec_{2,\mrm{m}}$ via
\begin{equation}
	\Delta \Bvec_{2,\mrm{m}} = - \mu_0 \nabla \times \Jvec_{2,\mrm{m}}\,\mbox{.}
	\label{eq:magnetostat}
\end{equation}
Our PIC simulation above confirms the existence of such a magneto-static excitation as well: The $y$-component of the magnetic field $\Bvec^\mrm{PIC}$ is constant in time after the laser has passed the interaction region. In Fig.~\ref{fig:plasma_Ar}(e) a snapshot of this static magnetic field component is shown, together with an exemplary time trace in Fig.~\ref{fig:plasma_Ar}(f).

The total non-radiative current created in the micro-plasma can thus be written as
\begin{equation}
	\Jvec_2^\mrm{nonrad } = \Jvec_{2,\mrm{m}} - \epsilon_0\partial_t \Evec_\mrm{2,d}\,\mbox{.} \label{eq:jnonrad}
\end{equation}
Interestingly, the second (curl-free) term in Eq.~(\ref{eq:jnonrad}) is exactly what we used in the previous 1D model [c.f.\ Eq.~(\ref{eq:jz1d})]. 
In order to compute the hypothetical far field spectra in Fig.~\ref{fig:IC_TC_comp_PIC}, this (curl-free) current density has been then multiplied by a narrow transverse electron distribution in order to represent a thin wire. It is obvious that such operation destroys the ''curl-free'' property of the current and thus introduces an artificial radiation. To judge if a current is radiating its transverse spatial dependence is of great importance. Therefore, 1D modeling is not suitable for the description of THz emission from plasma currents and we perform 2D/3D modeling for the radiating fields in the next section.

The PIC simulation results presented above show that non-radiating plasma oscillations following Eq.~(\ref{eq:stat_osc}) can be excited in the particular 2D configuration. Finally, we want to show that non-radiating plasma oscillations can in principle appear whenever the TC mechanism is active. To this end, let us consider an axisymmetric laser pulse ($|\Evec_\mrm{L}(\rvec_\perp,z,t)|=|\Evec_\mrm{L}(-\rvec_\perp,z,t)|$, e.g., a Gaussian or vortex beam profile. For such a driving pulse, the electron density $n_0$ and collision frequency $\nu_\mrm{ei}$ feature the same symmetry. Thus, the left-hand-side of Eq.~(\ref{eq:E_wave}) conserves the following two symmetries:
$\Evec_2^\mrm{s}$ with symmetric transverse field
\begin{equation}
\begin{split}
	\Evec_{2,\perp}^\mrm{s}(\rvec_\perp,z,t) &= \Evec_{2,\perp}^\mrm{s}(-\rvec_\perp,z,t) \\
	E_{2,z}^\mrm{s}(\rvec_\perp,z,t) &= -E_{2,z}^\mrm{s}(-\rvec_\perp,z,t)\,\mbox{,}\label{eq:sym_s}
\end{split}
\end{equation}
and $\Evec_2^\mrm{a}$ with anti-symmetric transverse field
\begin{align}
\begin{split}
	\Evec_{2,\perp}^\mrm{a}(\rvec_\perp,z,t) &= -\Evec_{2,\perp}^\mrm{a}(-\rvec_\perp,z,t)\\
	E_{2,z}^\mrm{a}(\rvec_\perp,z,t) &= E_{2,z}^\mrm{a}(-\rvec_\perp,z,t)\,\mbox{.}\label{eq:sym_a}
\end{split}
\end{align}
Hence, if we decompose the source term $\bm{\iota}_2=\bm{\iota}^\mrm{s}+\bm{\iota}^\mrm{a}$ in the spirit of Eqs.~(\ref{eq:sym_s}) and Eqs.~(\ref{eq:sym_a}),
the symmetric solution $\Evec_2^\mrm{s}$ is solely driven by $\bm{\iota}^\mrm{s}$, and the anti-symmetric solution $\Evec_2^\mrm{a}$ is solely driven by $\bm{\iota}^\mrm{a}$.
It is easy to verify that the non-radiating curl-free solution $\Evec_{2,\mrm{d}}$ given in Eq.~(\ref{eq:stat_osc}) is anti-symmetric. Thus, $\Evec_{2,\mrm{d}}$ can be excited only if the source has an anti-symmetric part. Because the source term $\bm\iota_2$ given in Eq.~(\ref{eq:iota_2}) contains always an anti-symmetric part $\bm{\iota}^\mrm{a}$, non-radiating plasma oscillations can in principle appear whenever the TC mechanism is active.

So what about the IC mechanism we completely ignored in the previous discussion?
With the simple substitution $\tilde{\Evec}_{1} = \Evec_1 - \Evec_\mrm{L}$, which has already been used in~\cite{Debayle:14}, we can bring Eq.~(\ref{eq:cont_1}) in the same form
as Eq.~(\ref{eq:E_wave}), where $\Evec_2 \rightarrow \tilde{\Evec}_{1}$ and $\bm{\iota}_2\rightarrow\bm{\iota}_1$ as defined in Eq.~(\ref{eq:iota_1}).
Then, by analyzing symmetry properties completely analogous to above, it turns out that for driving pulses with a Gaussian beam profile, $\bm\iota_1$ is symmetric, and thus cannot excite a non-radiative solution $\Evec_\mrm{1,d}$ of the form in Eq.~(\ref{eq:stat_osc}). However, anti-symmetric beams as a singly charged vortex or a radially polarized doughnut could in principle produce non-radiating plasma oscillation even for the IC mechanism. 

\section{\label{sec:Tera}Terahertz radiation from laser-induced micro-plasmas}

We have seen in the previous section that one has to be careful when concluding from plasma excitations on THz radiation in the far field.
Plasma oscillations can be excited at the local plasma frequency, which do not emit radiation. Using a 1D plasma model in order to predict THz emission spectra from fs-laser gas interaction, as proposed in~\cite{PhysRevLett.98.235002} and later used, e.g., in \cite{ISI:000253084000007,Buccheri:15}, may give incorrect results:
As we have shown in Fig.~\ref{fig:plasma_Ar}(e), THz emission spectra obtained from such model deviate strongly from those obtained from PIC simulations.
Thus, in order to understand the THz emission spectra, 2D or even 3D models are inevitable. 

\begin{figure}[t]
  \centering
  \includegraphics[width=0.8\columnwidth]{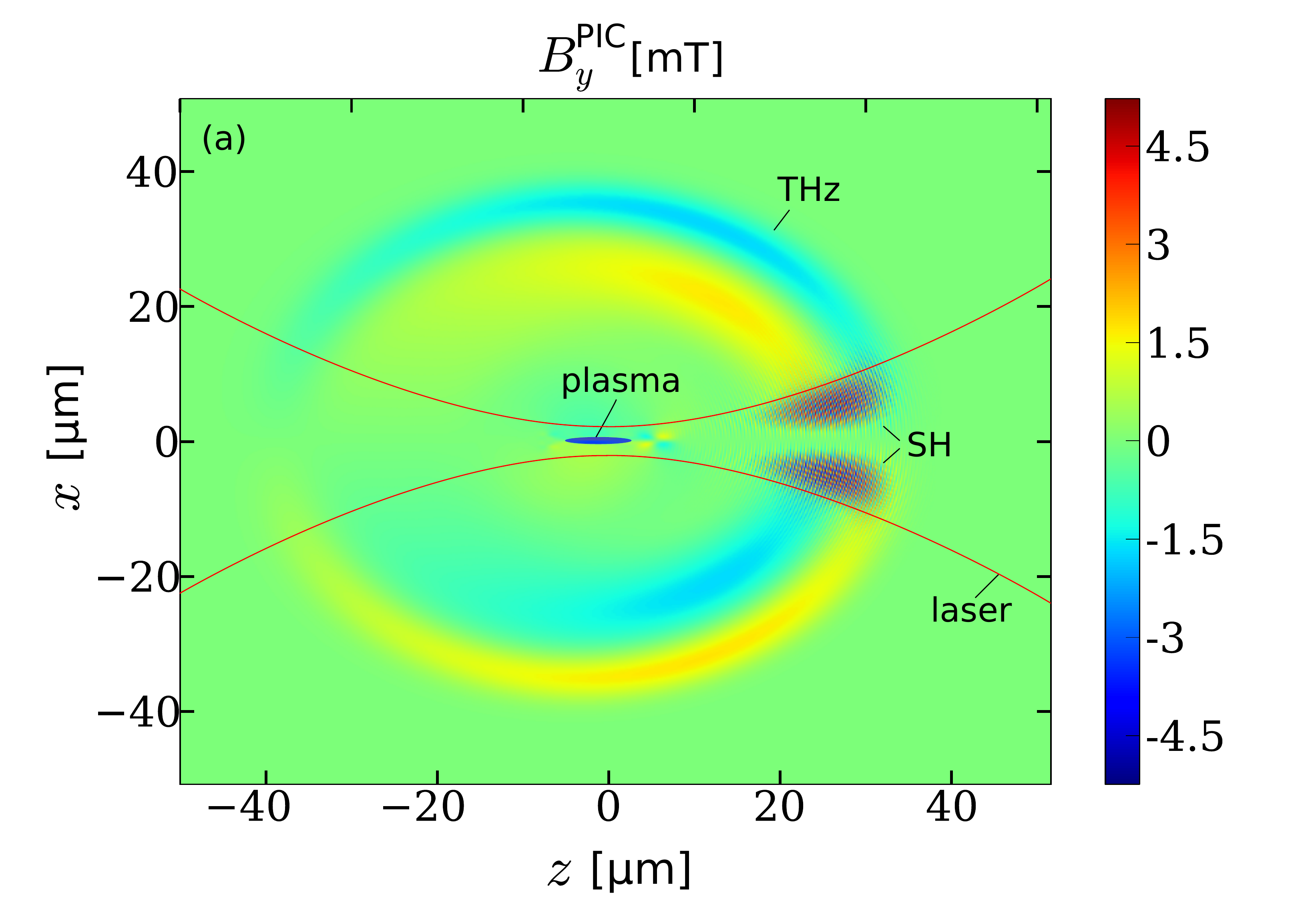}
  \caption{Snapshot of the magnetic field $B_y^\mrm{PIC}$ from the same PIC simulation as shown in Fig.~\ref{fig:plasma_Ar}. The figure is a zoom-out of Fig.~\ref{fig:plasma_Ar}(e),
  so the emitted THz and SH waves are visible (denoted as 'SH' and 'THz'). The mean width of the focused laser is indicated by the red lines, and the position of the generated plasma is marked as a blue oval.}
  \label{fig:2D_example}
\end{figure}

Throughout this section, we consider the 50-fs $y$-polarized laser pulse configuration already employed in Figs.~\ref{fig:plasma_Ar} and \ref{fig:P_far_2D} as an example. For the sake of computational costs, we restrict our PIC simulations to this 2D configuration, and treat the experimentally relevant 3D geometry in the framework of the model only.
We have already seen that for our example the TC mechanism dominates and the current $\Jvec_2$ is mainly driven ponderomotively, i.e., in the $xz$-plane. Thus, the radiation driven by the ponderomotive source is fully described by the magnetic field component $B_{2,y}$, while $B_{2,x}=B_{2,z}=0$. On the other hand, the driving laser pulse is $y$-polarized and hence $B_{\mrm{L},y}=0$. This natural separation, which is a special feature of the chosen 2D geometry, is very handy when it comes to analyzing the PIC simulation results.
In Fig.~\ref{fig:2D_example} a snapshot of $B_y^\mrm{PIC}$ from the PIC simulation is presented. The beam envelope of the focused laser is shown schematically as red lines. 
The snapshot is taken about 100\,fs after the driving pulse has passed the focus, and the extension of the created plasma is sketched as a blue oval. 
In fact, zooming in on the region where the plasma is created would reproduce Fig.~\ref{fig:plasma_Ar}(e). In the larger frame of Fig.~\ref{fig:2D_example} emitted THz and also SH waves propagating forward inside a cone are clearly visible. No radiation is emitted on-axis along $z$ in agreement with PIC simulations in \cite{PedrosArticle} for a simple reason: For Gaussian beams the source $\bm{\iota}_2$ is symmetric in the sense of Eq.~(\ref{eq:sym_s}), and the beam excites the symmetric solution $\Evec_2^\mrm{s}$ only. This implies an anti-symmetric $\Bvec_{2}$ and thus $B_{2,y}(\rvec_\perp = 0)=0$, which forbids on-axis radiation. 

While solving the full model up to order $\epsilon^2$, i.e., Eqs.~(\ref{eq:cont_0})--(\ref{eq:Amp_1}), is already much cheaper in terms of computational costs compared to full PIC simulations, it is still too heavy for quick estimations and parameter scans, in particular in 3D.
Therefore, we propose in the following further simplifications for the computation of the THz emission in the far field. Firstly, we will approximate the source term $\iota_2$ by the ponderomotive source as we already did in the previous sections (see App.~\ref{app:iota_pond_par}). Secondly, we will not solve the full wave equation~(\ref{eq:E_wave}) to obtain $\Evec_2$,
but neglect the term proportional to $n_0$. Then, for the collision-less case ($\nu_\mrm{ei}=0$), the simplified equation reads
\begin{align}
	\partial_{tt} \Evec_2 + \xcancel{\frac{q_\mrm{e}^2n_0}{m_\mrm{e} \epsilon_0} \Evec_2} + c^2 \nabla\times\nabla\times \Evec_2 = - \frac{\bm{\iota_2}}{\epsilon_0}\,\mbox{.}
	\label{eq:E_wave_simp}
\end{align}
With this approximation, the curl-free part $\Evec_{2,\mrm{d}}$ and divergence-free part $\Evec_{2,\mrm{r}}$ of the solution $\Evec_2$ decouple, and we disregard the following three effects: 
\begin{itemize}
	\item[1.] Oscillations at the plasma frequency in the curl-free $\Evec_{2,\mrm{d}}$ defined by Eq.~(\ref{eq:stat_osc}) are neglected.
	\item[2.] In particular for the divergence free $\Evec_{2,\mrm{r}}$, both dispersion and absorption are neglected.
	\item[3.] The coupling between $\Evec_{2,\mrm{d}}$ and $\Evec_{2,\mrm{r}}$ is neglected.
\end{itemize}
The first point is not problematic for the description of the THz far-field spectra, because such plasma oscillations do not lead to radiation as they are curl-free. The second point is critical and has to be accounted for, in particular for larger and more dense plasmas as we will see below. The third point cannot be avoided and may be important for certain plasma shapes: The coupling can lead to resonant excitation as in plasma wave-guides, similar to what is frequently exploited in the case of meta-materials, see e.g.~\cite{PhysRevB.88.205125}. However, for laser-induced micro-plasmas we never observed such wave-guiding effects in the PIC simulations. 

It is interesting to note that using Eq.~(\ref{eq:E_wave_simp}) instead of Eq.~(\ref{eq:E_wave}) is equivalent to solving the simplified current equation $\partial_t \Jvec_2 = \bm\iota_2$, or 
\begin{equation}
\partial_t \Jvec_2 +\nu_\mrm{ei}\Jvec_2 = \bm\iota_2 \,\mbox{,}\label{eq:j_simp}
\end{equation}
when taking into account collisions. In this light, the approximation is not exactly new and has already been successfully applied for the current $\Jvec_1$ and the IC mechanism~\cite{Kim, PhysRevLett.105.053903, 1367-2630-13-12-123029}. By using Eq.~(\ref{eq:j_simp}), we can easily compute the current $\Jvec_2$ from $\bm\iota_2$, and then $\Evec_2$ and $\Bvec_2$ in the far field by means of Jefimenko's equations~\cite{jackson}, or equivalently via the results presented in App.~\ref{app:rad_far}. For Gaussian beam profiles and transversely narrow plasmas, it is moreover possible to show that only the longitudinal component of the current $J_{2,z}$ contributes to the far field and thus to
the THz power spectrum (see App.~\ref{app:rad_thin}).

\begin{figure}
  \centering
  \includegraphics[width=\columnwidth]{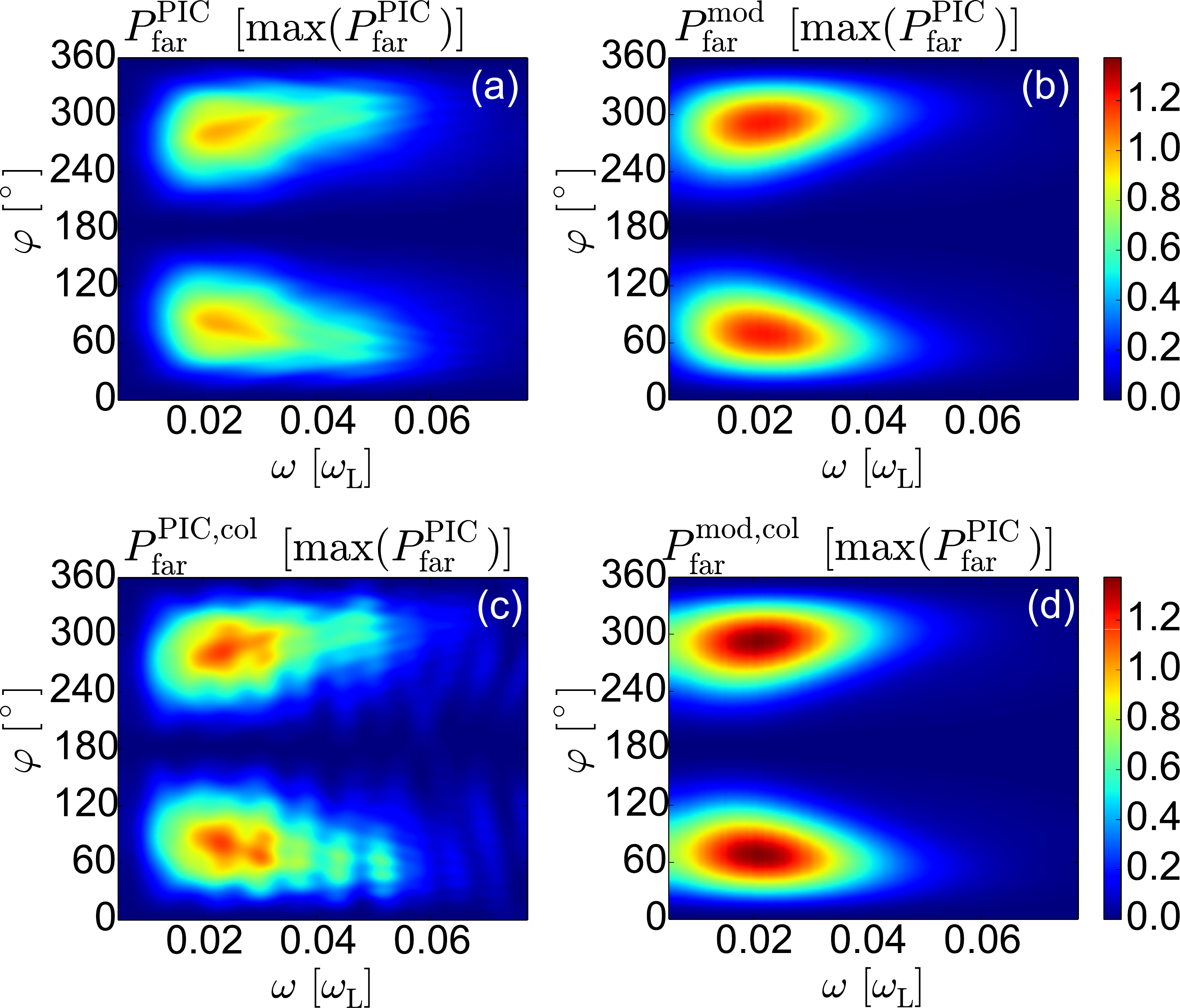}
  \caption{Far field THz power spectra as a function of frequency $\omega_\mrm{THz}$ and detection angle $\varphi$ for the laser pulse of Figs.~\ref{fig:plasma_Ar},~\ref{fig:P_far_2D} and \ref{fig:2D_example}. In (a) the result of the PIC simulation and in (b) those of the simplified model (see text) are presented. In (c) and (d) analogous results accounting for collisions are shown. The color scale allows for quantitative comparison of the amplitudes, which are normalized to $\max P_\mrm{far}^\mrm{PIC}$.}
  \label{fig:2D_model}
\end{figure}

\begin{figure*}
   \centering
   \includegraphics[width=0.63\columnwidth]{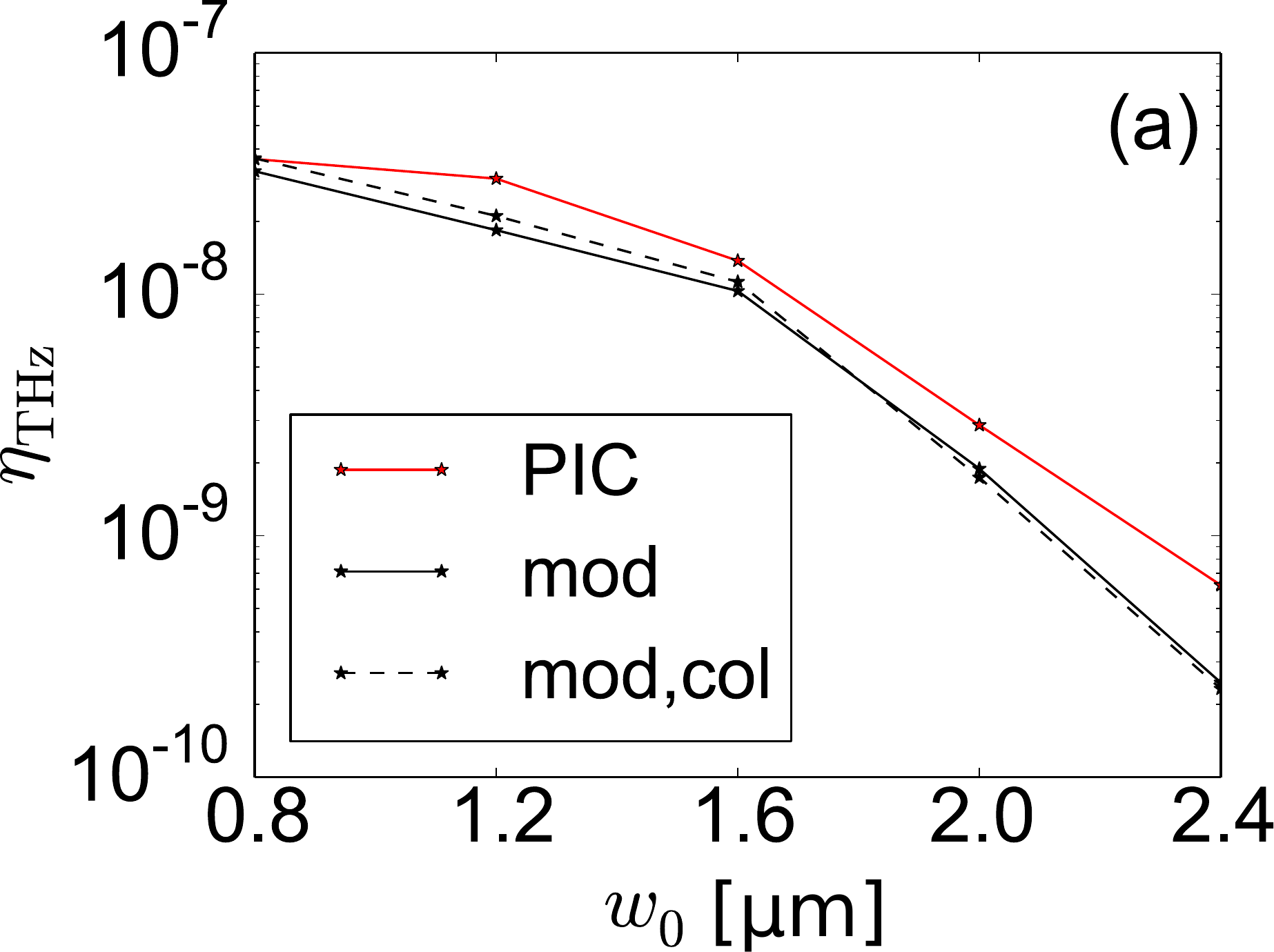}~
   \includegraphics[width=0.64\columnwidth]{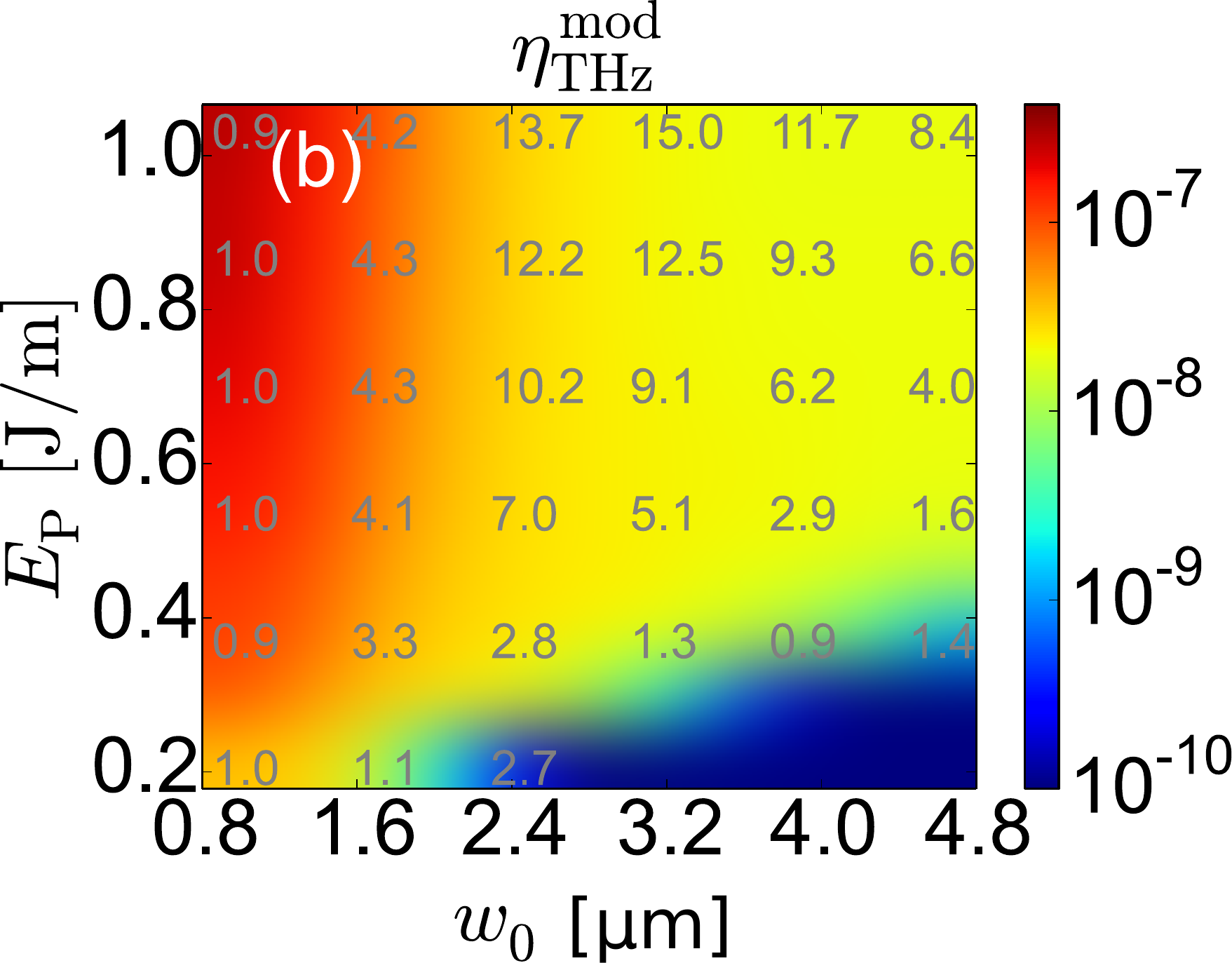}~
   \includegraphics[width=0.6\columnwidth]{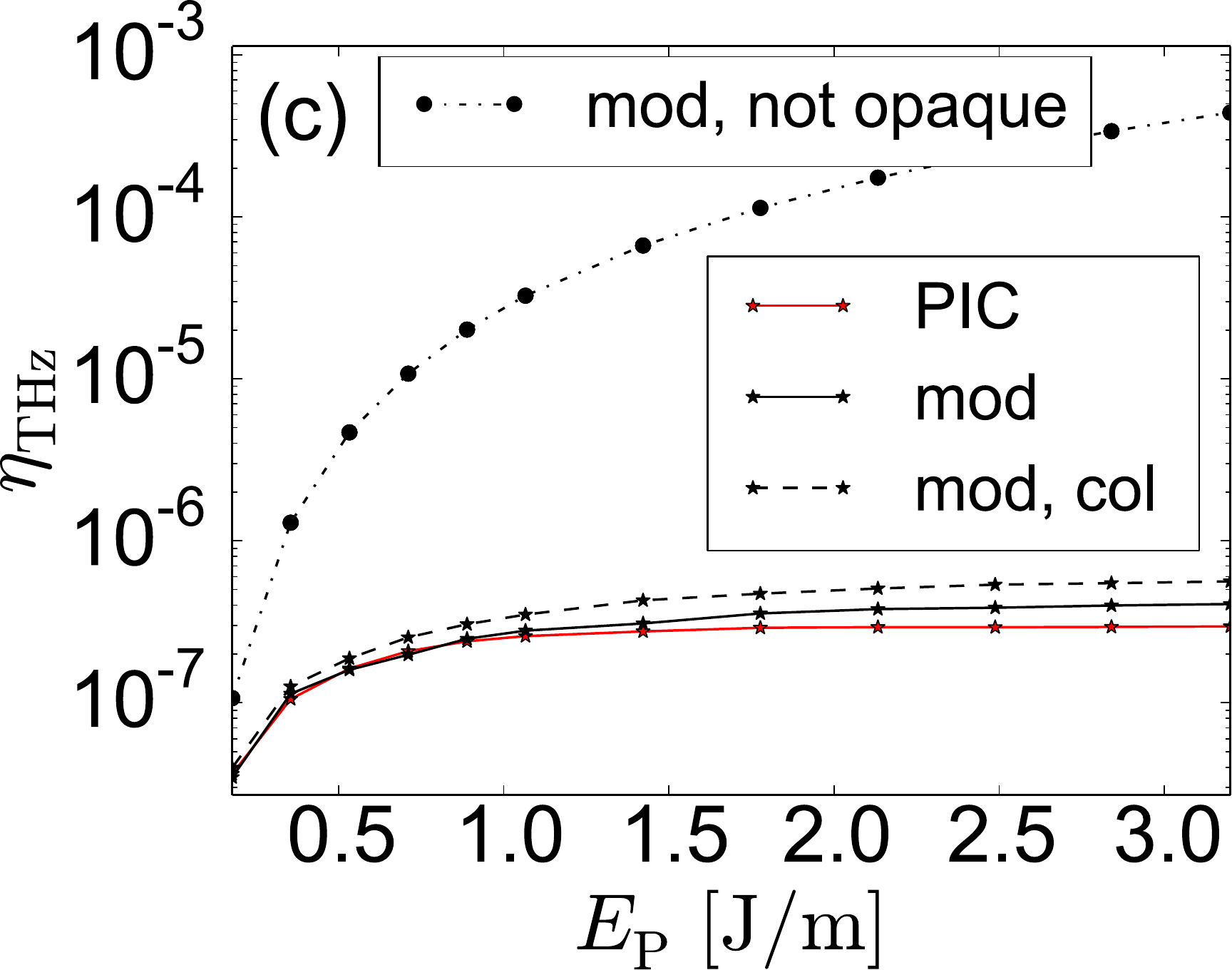}
   \caption{Scaling of the conversion efficiency $\eta_\mrm{THz}$ with 
focal spot-size and pulse energy, for fixed laser wavelength 
$\lambda_\mrm{L}=0.8$\,\textmu m, pulse duration $t_0=50\,\mrm{fs}$ and 
argon gas density $n_\mrm{a}=3\times10^{19}\,\mrm{cm}^{-3}$.  In (a), 
PIC results (solid red line) and the simplified model (solid black line) 
for a laser pulse energy of $E_\mrm{p}=0.18\,\mrm{J/m}$ are shown. The 
black dashed line shows model results accounting for collisions. In (b), 
the model is evaluated in the $(w_0,E_\mrm{p})$-plane. Ratios of PIC and model conversion efficiencies are indicated in 
gray. In (c), $\eta_\mrm{THz}$ as a function of the pulse energy is 
shown for tight focusing ($w_0=0.8$\,\textmu m). Results from PIC 
simulations (solid red line) and the simplified model with and without collisions 
(dashed and solid black line, resp.) are in good agreement. The black 
dashed-dotted line shows model results when the opaqueness of the plasma 
is ignored (see text).}
   \label{fig:2D_model_eta}
\end{figure*}

Before evaluating our simplified model and confront it with PIC results, we have to take into account the second point above, namely the incorrect treatment of THz dispersion.
For frequencies below the plasma frequency $\omega_\mrm{p}$ the plasma becomes opaque. For given $\omega_\mrm{p}$ and $\nu_\mrm{ei}$, it is possible to compute the penetration depth of the electromagnetic field as 
\begin{equation}
	s_\mrm{p}=\frac{c}{2\omega_\mrm{THz}}\Im\sqrt{\frac{\omega_\mrm{THz}^2+\rmi\nu_\mrm{ei}\omega_\mrm{THz}}{\omega_\mrm{THz}^2+\rmi\nu_\mrm{ei}\omega_\mrm{THz}-\omega_\mrm{p}^2}}\,\mbox{,} \label{eq:penetration}
\end{equation} 
where the symbol $\Im$ denotes the imaginary part of a complex quantity. For singly ionized argon gas, the penetration depth $s_\mrm{p}$ is about 0.5\,\textmu m for $\omega_\mrm{THz} \approx 0.02\omega_\mrm{L}\ll 0.13\omega_\mrm{L}\approx\omega_\mrm{p}$. This is about half the thickness of the plasma in the example of Fig.~\ref{fig:plasma_Ar}(a). For driving pulse configurations which produce larger plasmas or cause multiple ionization, the penetration depth may become significantly smaller than the plasma width. Then, the plasma emits mainly from a thin layer at its surface, where radiation at frequencies below $\omega_\mrm{p}$ can still exit due to optical tunneling. The frequency dependent thickness of this layer is related to the penetration depth in Eq.~(\ref{eq:penetration}). In order to mimic this effect in our simplified model, we do not take into account contributions from the whole plasma when calculating the far field THz power spectrum as described above. Instead, we only take contributions from the current density $\Jvec_2$ in a thin layer at the plasma surface, i.e., from positions $\rvec$ and frequencies $\omega_\mrm{THZ}$ with distance to the transparent outer area less than $1.2s_\mrm{p}$. The empirical factor 1.2 was chosen by equaling the THz pulse energy obtained with model and PIC simulation in Fig.~\ref{fig:2D_model}, and is kept constant for the rest of the paper. Of course, this approach implies a strong simplification of the situation, however, as we will see below, it leads to reasonable agreement with PIC simulations with respect to the spectral and angular distribution of the THz emission.

In Fig.~\ref{fig:2D_model}(a), we present the angle resolved far field power spectrum $P_\mrm{far}^\mrm{PIC}$ for the PIC simulation of Figs.~\ref{fig:plasma_Ar},~\ref{fig:P_far_2D} and \ref{fig:2D_example}. The exact definition for $P_\mrm{far}$ we use is provided in App.~\ref{app:spectrum}. For comparison, Fig.~\ref{fig:2D_model}(b) shows the power spectrum $P_\mrm{far}^\mrm{mod}$ obtained from our simplified model. We find good qualitative and even quantitative agreement, both spectra feature a broad peak around $0.02\omega_\mrm{L}$ (i.e., $7.5\,\mrm{THz}$). Simulation and simplified model predict the strongest radiation under an angle of $\varphi\approx70^\circ$ with respect to the optical axis. Because the length of the plasma is about 10\,\textmu m only, THz emission due to the TC mechanism is expected at such large angles~\cite{PhysRevLett.98.235002,ISI:000253084000007}. Comparing the power spectrum $P_\mrm{far}^\mrm{PIC}$ with results from PIC simulation accounting for collisions, as shown in Fig.~\ref{fig:2D_model}(c), confirms that collisions are of minor importance for the THz emission. 

Let us finally discuss scaling properties of the THz conversion efficiency 
\begin{equation}
	\eta_\mrm{THz} = \frac{E_\mrm{THz}}{E_\mrm{p}}\,\mbox{,}
\end{equation}
where $E_\mrm{THz}$ is the THz pulse energy [or energy density in 2D, see App.~(\ref{app:spectrum}) for details] containing frequencies below $0.2\omega_\mrm{L}$ and $E_\mrm{p}$ is the energy (or energy density) of the incoming laser pulse. In the following, we will use the term ''energy'' also in the 2D case for the sake of readability, even if we mean ''energy density''. 
Moreover, we fix the laser wavelength $\lambda_\mrm{L}=0.8$ \textmu m, pulse duration $t_0=50\,\mrm{fs}$ and neutral argon gas density $n_\mrm{a}=3\times10^{19}\,\mrm{cm}^{-3}$. 
The conversion efficiency $\eta_\mrm{THz}$ for $E_\mrm{p}=0.18\,\mrm{J/m}$ as a function of the focal beam width $w_0$ is presented in Fig.~\ref{fig:2D_model_eta}(a). Our simplified model is in good agreement with the PIC simulations. It turns out that strong focusing leads to the highest conversion efficiency $\eta_\mrm{THz}$ for the chosen pulse energy. Strong focusing is also preferable for higher pulse energies as shown by the $(w_0,E_\mrm{p})$-parameter scan in Fig.~\ref{fig:2D_model_eta}(b). The results of the parameter scan are presented for the model. The ratio between the PIC and the model conversion efficiencies is indicated by the gray numbers, showing reasonable agreement within one order of magnitude.

For all focusing conditions in Fig.~\ref{fig:2D_model_eta}(b), the conversion efficiency $\eta_\mrm{THz}$ first increases with the driving laser pulse energy, and then saturates for higher energies around $10^{-6}-10^{-7}$. This behavior, which translates into linear growth of the THz energy with the laser energy, is a direct consequence of the opaqueness of the plasma for THz radiation. As we have argued above, due to this opaqueness the plasma radiates from a thin layer only when the plasma diameter is sufficiently large. 
The thickness of this layer does not change much when increasing $E_\mrm{p}$, and is of the order of the penetration depth $s_\mrm{p}$ given in Eq.~(\ref{eq:penetration}).
We show in App.~(\ref{app:plasma_extend}) that in the 2D geometry the plasma diameter $d_\mrm{plasma}^\mrm{2D}\propto E_\mrm{p}$ grows linearly with the laser pulse energy $E_\mrm{p}$. 
In 3D, we find $d_\mrm{plasma}^\mrm{3D}\propto \sqrt{E_\mrm{p}}$. Thus, the length/surface (2D/3D) of the radiating layer increases as $d_\mrm{plasma}^\mrm{2D}\propto E_\mrm{p}$ in 2D or $(d_\mrm{plasma}^\mrm{3D})^2 \propto E_\mrm{p}$ in 3D. In both cases, the volume of the radiating layer grows therefore linearly in $E_\mrm{p}$. 
This simple consideration explains why the conversion efficiency $\eta_\mrm{THz}$ is expected to saturate at higher pulse energies, and the THz energy $E_\mrm{THz}$ increases only linearly with $E_\mrm{p}$. 

Figure~\ref{fig:2D_model_eta}(c) shows a line-out of Fig.~\ref{fig:2D_model_eta}(b) for tight focusing ($w_0=0.8$\,\textmu m) and corroborates this explanation. Both PIC results (red solid line) and model are in good agreement. The dashed-dotted black line corresponds to results from our simplified model when the opaqueness of the plasma is ignored, i.e., the current density $\Jvec_2$ in the whole plasma volume is taken into account. In particular for larger pulse energies, where the plasma is much wider than the penetration depth $s_{\mrm{p}}$, conversion efficiencies are overestimated by several orders of magnitude. Thus, it is crucial to take into account the opacity of the plasma when the plasma width exceeds the penetration depth.

Finally, we want to comment a posteriori on validity of the multiple scale expansion for the configuration studied in this paper with $t_0=50\,\mrm{fs}$, $I_{\mrm{L}}^0=4\times10^{14}\,\mrm{W/cm}^2$, $E_\mrm{L}^0=55\times10^9\,\mrm{V/m}$, $\lambda_\mrm{L} = 800$\,nm in argon gas with initial atom density $n_\mrm{a} = 3\times10^{19}\,\mrm{cm}^{-3}$: In Fig.~\ref{fig:1D_example_2}(b) for 1D and Fig.~\ref{fig:plasma_Ar}(d) for 2D configuration we found the electric field $\Evec_2$ driven by $\bm\iota_2$ of the order $10^7\,\mrm{V/m}$. The ratio of \first and \second order electric field is thus about $2\times10^{-3}<1.4\times10^{-2}=|q_e E^0_\mrm{L} / m_\mrm{e} \omega_\mrm{L}c|$, i.e., the upper bound for the ration established in App.~\ref{sec:Mult_scale}. 

\section{\label{sec:Conc}Summary and Conclusion}

We have analyzed the emission of broadband THz radiation from
femtosecond laser-induced gas plasmas by means of theoretical modeling
and numerical simulations. Our approach is based on a multiple scale
analysis of the non-relativistic Vlasov equation, which allows us to
identify distinct THz generation mechanisms. We
have obtained closed systems of equations describing the ionization
current (IC)~\cite{Kim2007} as well as the transition-Cherenkov
(TC)~\cite{PhysRevLett.98.235002} mechanism. Both mechanisms have been
discussed already in the literature, but usually without a direct comparison. 
Our model accounts for, among others, field ionization, damping of
the current due to collisions, and heating of the electron
plasma. Plasma currents can be excited due to the electric laser field
(IC) as well as ponderomotive, radiation pressure, convective and
diffusive sources (TC). Confrontation of the model with rigorous PIC
simulations shows excellent agreement. The main results of this paper
are as follows. 

For single-color driving pulses in gases at ambient pressure, as used, e.g., in~\cite{Buccheri:15},
the TC mechanisms is dominant for sufficiently long (50 fs) multi-cycle
pulses. The ponderomotive excitation dominates radiation pressure and
other convective and diffusive sources. Strong plasma oscillations are
excited at the plasma frequency, which are damped due to collisions.

Angularly-resolved far-field spectra confirm the angular THz emission
characteristics for the TC mechanism as proposed
in~\cite{PhysRevLett.98.235002}. In particular, for micro-plasmas the
THz radiation is predominantly emitted at large angles $>70^\circ$,
as also observed in~\cite{Buccheri:15}. However, the frequency
dependence of the far field power spectra according to the model derived
in~\cite{PhysRevLett.98.235002} is not correct: Oscillations at the
plasma frequency are present in the micro-plasma but do not contribute to the far field emission spectrum. We analyzed this behavior in detail, showing
that 1D modeling of the plasma currents is not sufficient for
predicting correct THz emission spectra.

Our investigation of the THz conversion efficiency suggests that, as
far as low-energy pulses below the filamentation threshold are
concerned, strong focusing is advantageous. Conversion efficiencies of
$10^{-6}-10^{-7}$ have been observed in PIC simulations as well as in our
model. We find that for higher laser energies, the conversion
efficiency saturates, and the THz energy increases linearly with the
pump energy. We explained this saturation by the opacity of the plasma
at THz frequencies: For higher laser pulse energies, the plasma volume becomes larger and radiates from a thin layer at its surface only. For strong focusing conditions, the plasma surface increases linearly with the
laser pulse energy. 

We believe that our model will be useful for further analysis of THz
emission from fs-laser-induced gas plasmas. Besides single-color
driving pulses, it also allows the treatment of multi-color driving
pulses, for which the IC mechanism is expected to play a key role.

\section*{Acknowledgements}
Numerical simulations were performed using computing resources at
M\'esocentre de Calcul Intensif Aquitaine (MCIA), Grand Equipement
National pour le Calcul Intensif (GENCI, Grants No.~2015-056129 and
No.~2016-057594), and Partnership for Advanced Computing in Europe
AISBL (PRACE, Grant No.~2014112576). This study was supported by ANR
(Projet ALTESSE).

\appendix

\section{\label{app:Der}Derivation of the model}
\subsection{\label{app_chap:Vlasov2THz}Moments of the Vlasov equation}

Our starting point are Maxwell's equations and the non-relativistic Vlasov equation, i.e., Eqs.~(\ref{eq:Vlasov_e}) and (\ref{eq:Force_elmag}) of the main text, governing the distribution function of electrons $f_\mrm{e}(\rvec,\vvec,t)$. We assume ions to be fixed, so the Vlasov equation for ions with charge $Z$ simply reads
\begin{equation}
	\partial_t f_\mrm{ion}^{(Z)}(\rvec,t) = S^{(Z)} \delta(\vvec)~\mbox{.}
	\label{eq:Vlasov_i_app}
\end{equation}
The ionization of atoms is taken into account by the source terms $S^{(Z)}$, which are related to the corresponding term $S$ in Eq.~(\ref{eq:Vlasov_e}) via 
$\sum_{Z} Z S^{(Z)} = S$ (see Appendix~\ref{app:ion}). Elastic collisions in Eq.~(\ref{eq:Vlasov_e}) are described by the term $C[f_\mrm{e},f_\mrm{ion}^{(Z)}]$, which depends on the state of the plasma. $C$ has the property to conserve the density and the total energy of the particles. The dominating effect of collisions is to change the direction of the electron momenta. The momenta of non-relativistic electrons 'a' and 'b' after an electron-electron collision event are $\pvec_\mrm{a}'=\pvec_\mrm{b}$ and $\pvec_\mrm{b}'=\pvec_\mrm{a}$ where $\pvec_\mrm{a}$ and $\pvec_\mrm{b}$ are the momenta before the collision event. Thus, electron-electron collisions do not lead to a change of the overall momentum. The total momentum of the electrons can change under collisions with ions only. 

As already mentioned in the main text we neglect electron-neutral collisions, mainly because for the driving pulses considered here almost all atoms in the interaction region are quickly ionized. We checked this assumption by including the electron-neutral collision frequency $\nu_\mrm{en}\approx5 \times 10^{-15}\mrm{cm}^2 \times n_\mrm{n}\sqrt{(E_\mrm{kin}+E_\mrm{th})/m_\mrm{e}}$~\cite{Huba2013} in the model below. Here $n_n$ is the neutral density, and $E_\mrm{kin}$ and $E_\mrm{th}$ are kinetic and thermal electron energy, respectively. 
As expected, electron-neutral collisions contribute to the heating of the electron plasma at the beginning of the ionization process, but quickly become negligible because of the depletion of neutral atoms. For driving pulse configurations considered in this paper, neglecting electron-neutral collisions gives a total error below 30\,\% in the final thermal energy and collision frequency. 
Because it turns out that collisions play a minor role for THz generation in laser-induced micro-plasmas anyway, and moreover electron-neutral collisions are not implemented in both of our PIC codes yet, we will neglect electron-neutral collisions throughout this paper. 

In the following, we will re-derive the \zeroth, \first, and \second velocity momenta of the distribution function $f_\mrm{e}$ leading to continuity and Euler equation, respectively, as well as 
the equation for the total energy of the electrons.

\paragraph{Continuity equation}
Let us compute the \zeroth velocity moment of Eq.~(\ref{eq:Vlasov_e}). The first term on the left hand side provides
\begin{equation}
	\int\limits \partial_t f_\mrm{e} \,d^3\vvec  = \partial_t \int\limits f_\mrm{e} \,d^3\vvec = \partial_t n_\mrm{e}\,\mbox{,}
\end{equation}
where we introduced the electron density
\begin{equation}
	n_\mrm{e} = \int f_\mrm{e}d\,^3\vvec\,\mbox{.}
	\label{eq:densiy_e}
\end{equation}
Using integration by part for the second term gives
\begin{equation}
	\int \vvec\cdot\nabla_\rvec f_\mrm{e}d\,^3\vvec  = \nabla_\rvec\cdot\int \vvec f_\mrm{e}d\,^3\vvec= \nabla_\rvec\cdot(n_\mrm{e}\uvec)\,\mrm{,}
\end{equation}
where we defined the (electron) fluid velocity as 
\begin{equation}
	\uvec = \frac{1}{n_\mrm{e}}\int \vvec f_\mrm{e}d\,^3\vvec\,\mbox{.}
	\label{eq:velocity_e}
\end{equation}
Because ions are fixed, the total current density reads
\begin{equation}
	\Jvec = q_\mrm{e} n_\mrm{e} \uvec\,\mbox{.}
\end{equation}
The \zeroth moment of the force term vanishes because 
\begin{equation}
	\int \Fvec\cdot\nabla_{\vvec}f_\mrm{e}d\,^3\vvec = - \int f_\mrm{e} \nabla_{\vvec}\cdot\Fvec d\,^3\vvec = 0\,\mbox{,} \label{eq:force_zero}
\end{equation}
where we used integration by parts and $f_\mrm{e}(|\vvec|=\infty)=0$ as well as $\nabla_{\vvec}\cdot\Fvec = q_\mrm{e}\nabla_{\vvec}\cdot\left(\Evec + \vvec\times\Bvec\right) = 0$. Moreover, it is easy to see that
\begin{equation}
	\int S\delta(\vvec)d\,^3\vvec = S\,\mbox{.}
\end{equation}
Finally, the \zeroth moment of the term $C$ has to vanish because elastic collisions conserve the particle density.
Adding up all the terms leads to the continuity equation
\begin{equation}
	\partial_t n_\mrm{e} + \nabla_{\rvec}\cdot\left(n_\mrm{e}\uvec\right) = S~\mbox{,}
	\label{eq:continuity_e}
\end{equation}
and analogous for the ions by using Eq.~(\ref{eq:Vlasov_i_app})
\begin{equation}
	\partial_t n^{(Z)}_\mrm{ion} = S^{(Z)}~\mbox{,}
	\label{eq:cont_e}
\end{equation}
where we introduced ion densities as $n^{(Z)}_\mrm{ion}=\int\! f_\mrm{ion}^{(Z)} d\,^3\vvec$.
The total charge density $\rho$ is then given by
\begin{equation}
	\rho = q_\mrm{e}\left(n_\mrm{e}-\sum\limits_Z Z n^{(Z)}_\mrm{ion}\right)\,\mbox{.}
\end{equation}

\paragraph{Euler equation}
In analogy to above, we now compute the \first velocity moment of all terms in Eq.~(\ref{eq:Vlasov_e}):
\begin{equation}
	\int\vvec\partial_t f_\mrm{e}d^3\,\vvec = \partial_t\left(n_\mrm{e}\uvec\right)\,\mbox{.}
\end{equation}
We rewrite the moment of the second term as
\begin{equation}
	\int \vvec \left(\vvec\cdot\nabla_\rvec f_\mrm{e}\right) d^3\vvec = \nabla_\rvec \cdot \int f_\mrm{e} \vvec \otimes \vvec d^3\vvec \,\mbox{,}
\end{equation}
where $\otimes$ denotes the usual outer product, and the divergence operator applied to a matrix-valued function yields the divergence for each row of the matrix.
By using the expression for the velocity spread of the electrons 
\begin{equation}
	\var\,\vvec = \frac{1}{n_\mrm{e}}\int f_\mrm{e} (\vvec-\uvec)\otimes(\vvec-\uvec)\,d^3\vvec
\end{equation}
as well as electron density Eq.~(\ref{eq:densiy_e}) and fluid velocity Eq.~(\ref{eq:velocity_e}), one can show that
\begin{equation}
	\int \vvec \left(\vvec\cdot\nabla_\rvec f_\mrm{e}\right) d^3\vvec = \nabla_\rvec \cdot \left(n_\mrm{e}\uvec\otimes\uvec + n_\mrm{e} \var\,\vvec\right)\,\mbox{.}
\end{equation}
We assume instantaneous thermalization of the plasma, which renders $\var\,\vvec$ proportional to the identity matrix:
\begin{equation}
	\var\,\vvec = \diag(1,1,1) \frac{1}{3 n_\mrm{e}} \int \left|\vvec-\uvec\right|^2 f_\mrm{e}\,d^3\vvec \,\mbox{.} \label{eq:instther}
\end{equation}
Again via integration by parts, we find 
\begin{equation}
\begin{split}
	&\quad\int \vvec\left(\frac{\Fvec}{m_\mrm{e}}\cdot\nabla_\vvec f_\mrm{e}\right)d^3\vvec = -\int f_\mrm{e} \nabla_\vvec \cdot \left(\vvec\otimes\frac{\Fvec}{m_\mrm{e}}\right)d^3\vvec \\
	&= -\int f_\mrm{e} \left(\frac{\Fvec}{m_\mrm{e}}+\vvec \frac{\nabla_\vvec \cdot\Fvec}{m_\mrm{e}}\right)d^3\vvec =-\frac{q_\mrm{e} n_\mrm{e}}{m_\mrm{e}}\left(\Evec + \uvec\times\Bvec\right)
\end{split}
\end{equation}
for the \first moment of the force term.
The contribution from the ionization source term $S$ vanishes.
In order to handle the \first velocity moment of the collision term, we introduce the electron-ion collision frequency $\nu_\mrm{ei}$ via
\begin{equation}
	\int\vvec C \,d^3\vvec = - n_\mrm{e}\nu_\mrm{ei}\uvec \,\mbox{.} \label{eq:nucoll}
\end{equation}
This term is responsible for the damping of the electron current, as will be clear later. Note that in Eq.~(\ref{eq:nucoll}) we make again the assumption of instantaneous thermalization and thus 
electron-ion collisions are isotropic, i.e., a scalar collision frequency $\nu_\mrm{ei}$ is sufficient.

In summary, the \first moment of Eq.~(\ref{eq:Vlasov_e}) is giving Euler's equation 
\begin{equation}
\begin{split}
	& \quad \partial_t\left(n_\mrm{e}\uvec\right)  + \nabla_\rvec \cdot \left(n_\mrm{e}\uvec\otimes\uvec + n_\mrm{e} \var\,\vvec\right) \\
	& = \frac{q_\mrm{e}n_\mrm{e}}{m_\mrm{e}}\left(\Evec+\uvec\times\Bvec\right) - n_\mrm{e}\nu_\mrm{ei}\uvec\,\mbox{.}
	\label{eq:Euler_e}
\end{split}
\end{equation}

\paragraph{Energy equation}
Finally, we consider the energy density of electrons defined as
\begin{equation}
	\mathcal{E} = \frac{m_\mrm{e}}{2}\int \left|\vvec\right|^2 f_\mrm{e}\,d^3\vvec\,\mbox{.}
\end{equation}
Again, we compute the \second velocity momenta of all terms in Eq.~(\ref{eq:Vlasov_e}) first:
\begin{equation}
	\int \left|\vvec\right|^2 \partial_t f_\mrm{e}\,d^3\vvec = \frac{2}{m_\mrm{e}}\partial_t\mathcal{E}\,\mbox{.}
\end{equation}
Next we have
\begin{equation}
	\int\left|\vvec\right|^2\,\vvec\cdot\nabla_\rvec f_\mrm{e}\,d^3\vvec = \nabla_\rvec \cdot \int\left|\vvec\right|^2\vvec f_\mrm{e}\,d^3\vvec\,\mbox{,}
\end{equation}
which is already of \third order in the velocity and therefore, as we will see below, not relevant for our multiple scale analysis in the next subsection. 
Furthermore, 
\begin{align}
	&\quad\int\left|\vvec\right|^2\frac{\Fvec}{m_\mrm{e}}\cdot\nabla_\vvec f_\mrm{e}\,d^3\vvec = - \int f_\mrm{e} \nabla_\vvec \cdot \left(\left|\vvec\right|^2\frac{\Fvec}{m_\mrm{e}} \right) d^3\vvec \nonumber \\
	&= - 2\int f_\mrm{e} \, \vvec \cdot \frac{\Fvec}{m_\mrm{e}} \, d^3\vvec = - \frac{2q_\mrm{e}n_\mrm{e}}{m_\mrm{e}} \uvec\cdot\Evec\,\mbox{,}
\end{align}
where we used integration by parts,  Eq.~(\ref{eq:force_zero}), and $\vvec\cdot\left(\vvec\times\Bvec\right)=0$.
The ionization source $S$ gives no contribution, and the assumption of elastic collisions dictates
\begin{equation}
	\int\left|\vvec\right|^2C\,d^3\vvec = 0 \mbox{.}
	\label{eq:coll_elast}
\end{equation}
Using these results and the Euler equation~(\ref{eq:Euler_e}), the free electron energy density is governed by
\begin{equation}
\begin{split}
	&\quad \partial_t\mathcal{E} + \frac{m_\mrm{e}}{2} \nabla_\rvec \cdot \int\left|\vvec\right|^2\vvec f_\mrm{e}\,d^3\vvec -m_\mrm{e} n_\mrm{e}\nu_\mrm{ei}\left|\uvec\right|^2 \\
	& = m_\mrm{e} \uvec \cdot \left[\partial_t\left(n_\mrm{e}\uvec\right) + \nabla_\rvec \cdot \left(n_\mrm{e}\uvec\otimes\uvec + n_\mrm{e} \var\,\vvec\right)\right]\,\mbox{.}
	\label{eq:energy}
\end{split}
\end{equation}
It is possible to recast the electron energy density in terms of $\uvec$ and $\var\,\vvec$:
\begin{align}
	\int \left|\vvec\right|^2 f_\mrm{e}\,d^3\vvec & =\int \left|\uvec\right|^2 f_\mrm{e}\,d^3\vvec+\int \left|\vvec-\uvec\right|^2 f_\mrm{e}\,d^3\vvec \nonumber \\ 
	& = n_\mrm{e} \left|\uvec\right|^2+ \tr \left( n_\mrm{e} \var\,\vvec \right) \,\mbox{,}
\end{align}
where $\tr$ denotes the trace of the matrix. Then, we can identify kinetic and thermal energy density as
\begin{equation}
\mathcal{E}_\mrm{kin} = \frac{m_\mrm{e}}{2}n_\mrm{e} \left|\uvec\right|^2\,\mbox{,} \qquad  \mathcal{E}_\mrm{th} = \frac{m_\mrm{e}}{2} \tr \left( n_\mrm{e} \var\,\vvec \right) \,\mbox{,} \label{eq:ekinth}
\end{equation}
respectively.

\subsection{Multiple scale expansion}
\label{sec:Mult_scale}

In the following, Eqs.~(\ref{eq:continuity_e}),~(\ref{eq:Euler_e}) and (\ref{eq:energy}) are simplified by means of a multiple scale analysis. The general idea behind this approach is that electron velocities are small compared to the speed of light, and thus velocity momenta become less important with increasing order. What exactly ''small'' means in this context is discussed at the end of this section. We introduce a scaling parameter $\epsilon\ll 1$, and expand the relevant quantities
\begin{gather}
	n_{e} = \sum\limits_{i=0}^\infty\epsilon^i n_i,\quad
	\uvec = \sum\limits_{i=1}^\infty\epsilon^i \uvec_i,\quad
	\mathcal{E} = \sum\limits_{i=2}^\infty\epsilon^i \mathcal{E}_i, \nonumber \\
	\var\,\vvec = \sum\limits_{i=2}^\infty\epsilon^i \left(\var\,\vvec\right)_i, \label{eq:mult_scale_exp_app} \\
	\int\left|\vvec\right|^2\vvec f_\mrm{e}\,d^3\vvec = \sum\limits_{i=3}^\infty\epsilon^i \left( \int\left|\vvec\right|^2\vvec f_\mrm{e}\,d^3\vvec \right)_i\mbox{.} \nonumber 	
\end{gather}
Each summation in Eq.~(\ref{eq:mult_scale_exp_app}) starts at the order of the respective power in $\vvec$.
Furthermore, we assume that both ionization source $S$ and collision frequency $\nu_\mrm{ei}$ are of order $\epsilon^0$.
These scalings already imply that all macroscopic quantities in Maxwell's equations start at order $\epsilon^1$:
\begin{equation}
	\Evec = \sum\limits_{i=1}^\infty\epsilon^i \Evec_i,~
	\Bvec = \sum\limits_{i=1}^\infty\epsilon^i \Bvec_i,~
	\Jvec = \sum\limits_{i=1}^\infty\epsilon^i \Jvec_i,~
	\rho = \sum\limits_{i=1}^\infty\epsilon^i \rho_i\,\mbox{.}	
	\label{eq:mult_scale_exp1}
\end{equation}
While there certainly is some arbitrariness in our scaling assumptions, they lead to a set of meaningful equations. In particular, 
they insure that the $k^{\mrm{th}}$ moment is driven by terms of order $\epsilon^k$ and higher, only.

Plugging Eqs.~(\ref{eq:mult_scale_exp_app}), (\ref{eq:mult_scale_exp1}) into Eqs.~(\ref{eq:continuity_e}),~(\ref{eq:Euler_e}) and (\ref{eq:energy}) and separating the different orders of $\epsilon$, we find:
\begin{align}
	\epsilon^0:&\quad \partial_t n_0 = S \label{eq:cont_0_app}\\
	\epsilon^1:&\quad \partial_t n_1 + \nabla\cdot\left(n_0\uvec_1\right) = 0 \label{eq:n1} \\
	\epsilon^1:&\quad \partial_t\left(n_0 \uvec_1\right) + n_0\nu_\mrm{ei}\uvec_1 = \frac{q_\mrm{e}n_0}{m_\mrm{e}}\Evec_1\label{eq:Euler_eps1_1}\\
	\epsilon^2:&\quad \partial_t\left(n_0 \uvec_2\right) + \partial_t\left(n_1 \uvec_1\right) + n_0\nu_\mrm{ei}\uvec_2 + n_1\nu_\mrm{ei}\uvec_1 \nonumber \\
	&\quad \quad + \nabla_\rvec \cdot \left[n_0\uvec_1\otimes\uvec_1 + n_0 \left(\var\,\vvec\right)_2\right] \label{eq:Euler_eps2_1} \\
	&= \frac{q_\mrm{e}}{m_\mrm{e}}\left(n_0\Evec_2+n_1\Evec_1+n_0\uvec_1\times\Bvec_1\right) \nonumber \\
	\epsilon^2:&\quad \partial_t\mathcal{E}_2 = m_\mrm{e} \nu_\mrm{ei} n_0 \left|\uvec_{1}\right|^2 + m_\mrm{e} \uvec_{1}\cdot\partial_t\left(n_0 \uvec_{1}\right)\,\mbox{.}
	\label{eq:energy_eps2_1}
\end{align}
According to Eq.~(\ref{eq:ekinth}), the lowest order kinetic energy density reads 
\begin{equation}
\epsilon^2: \quad \mathcal{E}_\mrm{kin,2} = \frac{m_\mrm{e}}{2}n_0 \left|\uvec_1\right|^2 \,\mbox{,} \label{eq:e_kin_app}
\end{equation}
and with Eqs.~(\ref{eq:energy_eps2_1}) we find
\begin{equation}
\epsilon^2: \quad \partial_t\mathcal{E}_\mrm{th,2} 
	=  m_\mrm{e} n_0 \nu_\mrm{ei}\left|\uvec_1\right|^2 + \frac{m_\mrm{e}}{2}\left|\uvec_1\right|^2 \partial_t n_0 \,\mbox{.}
	\label{eq:E_th_2}
\end{equation}
Finally, we can exploit our assumption of instantaneous thermalization in Eq.~(\ref{eq:instther}) and relate $\left(\var\,\vvec\right)_2$
via Eq.~(\ref{eq:ekinth}) to the thermal energy density
\begin{equation}
\epsilon^2: \quad \left(\var\,\vvec\right)_2 = \diag(1,1,1)\,\frac{2 \mathcal{E}_\mrm{th,2}}{3 m_\mrm{e} n_0}\,\mbox{.}
	\label{eq:thermal_anisotropy}
\end{equation}
For the electron-ion collision frequency $\nu_\mrm{ei}$ we take the expression Eq.~(\ref{eq:nu_ei_NRL_2})~\cite{Huba2013}, which depends on the kinetic and thermal energy per electron 
\begin{equation}
	E_{\mrm{th},2} = \frac{\mathcal{E}_\mrm{kin,2}}{n_0}, \qquad
	E_{\mrm{kin},2} = \frac{\mathcal{E}_\mrm{th,2}}{n_0}\,\mbox{.}
\end{equation}

Herewith, our model is complete, and can be recast as given in the main text by replacing fluid velocities by current densities.
Equation~(\ref{eq:cont_0_app}) is Eq.~(\ref{eq:cont_0}) in the main text; with $\Jvec_1=q_\mrm{e}n_0\uvec_1$, Eq.~(\ref{eq:Euler_eps1_1}) gives Eq.~(\ref{eq:cont_1}). Equations~(\ref{eq:cont_2}) and (\ref{eq:iota_2}) follow from Eq.~(\ref{eq:Euler_eps2_1}) with $\Jvec_2=q_\mrm{e}n_0\uvec_2+q_\mrm{e}n_1\uvec_1$, Eqs.~(\ref{eq:n1}), (\ref{eq:thermal_anisotropy}) and Eqs.~(\ref{eq:Far_1}), (\ref{eq:Amp_1}):
\begin{widetext}
\begin{equation}
\begin{split}
        &\quad \frac{q_\mrm{e}^2}{m_\mrm{e}}\left(n_1\Evec_1+n_0\uvec_1\times\Bvec_1\right) - q_\mrm{e} \nabla_\rvec \cdot \left[n_0\uvec_1\otimes\uvec_1 + n_0 \left(\var\,\vvec\right)_2\right] \\
        &= \frac{q_\mrm{e}^2}{m_\mrm{e}} n_1 \Evec_1
        -\frac{q_\mrm{e}}{m_\mrm{e}}\Jvec_1\times\int\limits_{-\infty}^t\left(\nabla_\rvec\times\Evec_1\right)\,dt' 
        - \nabla_\rvec \cdot \left(\uvec_1\otimes\Jvec_1 \right) - 
        \frac{2 q_\mrm{e}}{3 m_\mrm{e} }\nabla_\rvec \mathcal{E}_\mrm{th,2} \\
        &= \frac{n_1}{n_0}\left(\nu_\mrm{ei} + \partial_t\right)\Jvec_1 - \Jvec_1\times\left[\nabla_\rvec\times\int\limits_{-\infty}^t \frac{\left(\nu_\mrm{ei} + \partial_t'\right)}{q_\mrm{e} n_0}\Jvec_1 \,dt'\right] 
        - \uvec_1 \nabla_\rvec \cdot \Jvec_1 
        - \left(\Jvec_1 \cdot \nabla_\rvec \right) \uvec_1
        - \frac{2 q_\mrm{e}}{3 m_\mrm{e} }\nabla_\rvec \mathcal{E}_\mrm{th,2} \\
        &= \frac{\left(\nu_\mrm{ei} + \partial_t\right)}{n_0}\left(n_1\Jvec_1\right) 
        - \Jvec_1 \times \left(\nabla_\rvec \times \uvec \right) - \Jvec_1\times\left[\nabla_\rvec\times
        \int\limits_{-\infty}^t \uvec_1 \left(\nu_\mrm{ei} + \frac{\partial_{t'}n_0}{n_0}\right)dt'\right]
        - \left(\Jvec_1 \cdot \nabla_\rvec \right) \uvec_1
        - \frac{2 q_\mrm{e}}{3 m_\mrm{e} }\nabla_\rvec \mathcal{E}_\mrm{th,2} \\
        &= - \frac{\left(\nu_\mrm{ei} + \partial_t\right)}{q_\mrm{e} n_0}\left(\Jvec_1\int\limits_{-\infty}^t\nabla_\rvec\cdot\Jvec_1\,dt'\right)
        - \frac{n_0}{2 q_\mrm{e}}\nabla_\rvec{\left|\frac{\Jvec_1}{n_0}\right|}^2
        - \frac{\Jvec_1}{q_\mrm{e}}\times\nabla_\rvec\times  \int\limits_{-\infty}^t \frac{\Jvec_1}{n_0} \left(\nu_\mrm{ei} + \frac{\partial_{t'}n_0}{n_0}\right)dt' -\frac{2q_\mrm{e}}{3m_\mrm{e}}\nabla_\rvec \mathcal{E}_\mrm{th,2} = \bm\iota_2
        \label{eq:iota_2_exp_app}
\end{split}
\end{equation}
\end{widetext}
Finally, Eqs.~(\ref{eq:e_kin_app}) and (\ref{eq:E_th_2}) give Eqs.~(\ref{eq:heating}) and (\ref{eq:Kin_E}), where we omitted the index $_{,2}$ in the main text.

To ensure that quantities at order $\epsilon^2$ are small compared to those at order $\epsilon^1$ we have to inspect the driving source terms $\bm\iota_2$ and $\bm\iota_1$ defined in Eq.~(\ref{eq:iota_2}) and Eq.~(\ref{eq:iota_1}) of the main text, respectively. The magnitude of $\bm\iota_1$ can be estimated as
\begin{equation}
	\vert\bm\iota_1\vert\sim\frac{q_\mrm{e}^2 n_0}{m_\mrm{e}} E_\mrm{L}\,\mbox{,}
\end{equation}
where $E_\mrm{L}$ is the field amplitude of the driving laser pulse.
The \first order current $\Jvec_1$ is dominated by the laser field oscillating at $\omega_\mrm{L}$, and an upper bound for its magnitude follows from Eq.~(\ref{eq:cont_1}): 
\begin{equation}
	\vert\Jvec_1\vert\lesssim\frac{q_\mrm{e}^2 n_0 E_\mrm{L}}{m_\mrm{e} \omega_\mrm{L}} \label{eq:bound_J1}\,\mbox{.}
\end{equation}
We further estimate an upper bound for $\mathcal{E}_\mrm{th,2}$ by integrating Eq.~(\ref{eq:E_th_2}) to
\begin{equation}
	\mathcal{E}_\mrm{th,2} \lesssim \frac{m_\mrm{e}|\Jvec_1|^2}{2q_\mrm{e}^2n_0}\left(2\nu_\mrm{ei}t_0+1\right)\,\mbox{,}
	\label{eq:bound_Eth}
\end{equation}
where $t_0$ is the laser pulse duration and we used Eq.~(\ref{eq:e_kin_app}) with $\Jvec_1=q_\mrm{e}n_0\uvec_1$. 
By using Eqs.~(\ref{eq:bound_Eth}) and (\ref{eq:bound_J1}) we can given an upper bound for the magnitude of $\bm{\iota}_2$. To this end,
we replace in Eq.~(\ref{eq:iota_2_exp_app}) all spatial derivatives by $\partial_{x/y/z}\rightarrow2\pi/\lambda_\mrm{L}$ and temporal integration of functions oscillating dominantly at $\omega_\mrm{L}$ by $dt\rightarrow1/\omega_\mrm{L}$. Moreover, we assume that $1/\nu_\mrm{ei}>t_0>t_\mrm{ion}>1/\omega_\mrm{L}$, where $t_\mrm{ion}$ is the typical ionization time defined by $\partial_{t}n_0 \lesssim n_0/t_\mrm{ion}$. Then, after some algebra, we find
\begin{equation}
	|\bm\iota_2| \lesssim \left|\frac{q_\mrm{e}E_\mrm{L}}{m_\mrm{e}\omega_\mrm{L}c}\right| |\bm\iota_1|\,\mbox{.}
	\label{eq:iota_2_est_app}
\end{equation}
Thus, we can conclude that the ratio of $|\bm\iota_2|$ over $|\bm\iota_1|$ and thus of $|\Jvec_2|$ over $|\Jvec_1|$ and $|\Evec_2|$ over $|\Evec_1|$ is typically much smaller than $|q_\mrm{e}E_\mrm{L}/m_\mrm{e}\omega_\mrm{L}c|$, rendering the multiple scale approach valid for non-relativistic laser pulses. 

\section{\label{app:ion}The Ionization Model}
We are working at peak intensities $I_{\mrm{L}}>10^{14}$ W/cm$^2$, thus in tunneling ionization regime for argon. The ionization rate $W^{(Z)}$ in quasi-static approximation creating ions with charge $Z$ is given in~\cite{Ammosov-1986-Tunnel,PhysRevA.64.013409}. The same approach is used in the PIC codes OCEAN~\cite{PhysRevE.87.043109} and CALDER~\cite{Nut11}.
The ion densities $n_\mrm{ion}^{(Z)}$ of the ions with charge $Z$ are of order $\epsilon^0$. These quantities appear in the \zeroth moment of Eq.~(\ref{eq:Vlasov_i_app}) and are determined by the set of equations
\begin{equation}
\begin{split}
	\partial_t n_\mrm{ion}^{(Z)} & = W^{(Z)} n_\mrm{ion}^{(Z-1)} - W^{(Z+1)} n_\mrm{ion}^{(Z)} \\
	\partial_t n_\mrm{ion}^{(0)} & = - W^{(1)} n_\mrm{ion}^{(0)}
	\label{eq:rate_eq_ion_dens}
\end{split}
\end{equation}
for $Z=1,2,3,\ldots$, and the initial neutral density is $n_\mrm{ion}^{(0)}(t=-\infty) = n_\mrm{n}(t=-\infty)=n_\mrm{a}$. 
Because we consider fixed ions, there are no higher order ion densities. Charge conservation dictates that the electron density at order  $\epsilon^0$ follows
\begin{equation}
	n_0 = \sum_{Z} Z n_\mrm{ion}^{(Z)}\,\mbox{.}
	\label{eq:el_dens_expl}
\end{equation}
We introduce the ion source $S^{(Z)}$ as 
\begin{equation}
	S^{(Z)} = W^{(Z)} n_\mrm{ion}^{(Z-1)} - W^{(Z+1)} n_\mrm{ion}^{(Z)}\,\mbox{,}
\end{equation}
and the electron source
\begin{equation}
	S = \sum_{Z} Z S^{(Z)} \,\mbox{.}
\end{equation}
In general, the ionization rates and thus $S$ depend on the total electric field $\Evec$. In the multiple scale approach, one should formally use the best approximation to the total electric field, i.e., the sum of all known orders. However, it is usually sufficient to take just on the lowest order electric field $\Evec_1$.

\section{\label{app:iota} Transformation of the 1D current source $\bm{\iota_2}$ into the co-moving pulse frame}
We rewrite the current source Eq.~(\ref{eq:iota_2}) in 1D geometry. Using $\partial_x = \partial_y = 0$ and 
$J_{1,z}=0$ we get
\begin{equation}
\begin{split}
     &\iota_{2,z}  = - \frac{n_0}{2 
q_\mrm{e}}\partial_z{\left|\frac{\Jvec_1}{n_0}\right|}^2 - 
\frac{J_{1,x}}{q_\mrm{e}}\partial_z 
\int\limits_{-\infty}^t\frac{J_{1,x}}{n_0}\left(\nu_\mrm{ei} + \frac{\partial_{t'}n_0}{n_0}\right)\,dt' \\
     & ~ - \frac{J_{1,y}}{q_\mrm{e}}\partial_z 
\int\limits_{-\infty}^t\frac{J_{1,y}}{n_0}\left(\nu_\mrm{ei} + \frac{\partial_{t'}n_0}{n_0}\right)\,dt'-\frac{2q_\mrm{e}}{3m_\mrm{e}}\partial_z 
\left(n_0E_{\mrm{th}}\right)\,\mbox{.}
     \label{eq:iota_2_app1}
\end{split}
\end{equation}
We also find that $\iota_{2,x}=\iota_{2,y}=0$, so $\bm\iota_{2}$ is purely longitudinal in 1D. 
Next we transform 
$\iota_{2,z}$ into the co-moving pulse frame $(z,t)\mapsto(\xi=z, \tau=t 
- z/c)$. According to the approximation we made in Eq.~(\ref{eq:J_L}), the 
current $\Jvec_1$ can be calculated directly from the vacuum laser field 
$\Evec_\mrm{L}$. Because $\Evec_\mrm{L}$ and thus $\Jvec_1$ do not 
change their temporal shape upon propagation along $z$, they are invariant in 
the new variable $\xi$ and the $z$ derivative 
transforms as $\partial_z = \partial_\xi - \partial_\tau / c =  - 
\partial_\tau / c$ leading to
\begin{equation}
\begin{split}
     \bm\iota_2 & = \frac{n_0}{2 q_\mrm{e} 
c}\partial_\tau{\left|\frac{\Jvec_1}{n_0}\right|}^2 \evec_{z} + 
\frac{n_0}{q_\mrm{e} c}\left|\frac{\Jvec_1}{n_0}\right|^2\left(\nu_\mrm{ei} + \frac{\partial_{\tau}n_0}{n_0}\right)\evec_{z}   \\
& \quad + 
\frac{2q_\mrm{e}}{3m_\mrm{e} c}\partial_\tau 
\left(n_0E_{\mrm{th}}\right)\evec_{z} \,\mbox{.}
     \label{eq:iota_2_app2}
\end{split}
\end{equation}

\section{\label{app:iota_pond_par}Ponderomotive source in quasi-monochromatic paraxial approximation}
Here, we derive an approximate expression for the ponderomotive source term 
\begin{equation}
	\bm\iota_2^\mrm{pond}  = - \frac{n_0}{2 q_\mrm{e}}\nabla{\left|\frac{\Jvec_1}{n_0}\right|}^2 \,\mbox{.}
\end{equation}
The Gaussian 2D laser electric field is computed in the quasi-monochromatic paraxial approximation as
\begin{equation}
	\Evec_\mrm{L}^{2\mrm{D}}(x,z,t) \approx \Re \frac{E_\mrm{L}^0 \, e^{-\frac{x^2}{w_0^2 \left(1+\rmi\frac{z}{z_\mrm{R}}\right)}-\frac{\tau^2}{t_0^2}-\rmi\left(\omega_\mrm{L}\tau - \frac{\pi}{2} \right)}}{\sqrt{1+\rmi\frac{z}{z_\mrm{R}}}} \evec_y \,\mbox{,}
\end{equation}
with $\tau=t - z/c$, and the Rayleigh length $z_\mrm{R}=w_0^2\omega_\mrm{L}/2c$. 
The symbol $\Re$ denotes the real part of a complex quantity.
In general, the current $\Jvec_1$ has to be calculated by solving the Maxwell's equations coupled to Eqs.~(\ref{eq:cont_1}) and (\ref{eq:heating}). However, when using $\Jvec_1$ to calculate the source term $\iota_2$ in order to study the TC mechanism, it is sufficient to approximate 
\begin{align}
	\Jvec_1 (x,z,t)
	& \approx \frac{q_\mrm{e}^2}{m_\mrm{e}}\int\limits_{-\infty}^t n_0(x,z,t') {\Evec}_{\mrm{L}}^{2\mrm{D}}(x,z,t') \,dt' \\ 
	&\approx \frac{n_0 q_\mrm{e}^2}{m_\mrm{e}\omega_\mrm{L}}\Re \frac{\rmi E_\mrm{L}^0 \, e^{-\frac{x^2}{w_0^2 \left(1+\rmi\frac{z}{z_\mrm{R}}\right)}-\frac{\tau^2}{t_0^2}-\rmi\left(\omega_\mrm{L}\tau - \frac{\pi}{2} \right)}}{\sqrt{1+\rmi\frac{z}{z_\mrm{R}}}} \evec_y \nonumber \,\mbox{.}
\end{align} 
In the following computation of $\nabla{\left|\Jvec_1/n_0\right|}^2$ we will omit the $z$ dependent Gouy phase
as well as the transverse phase curvature. The former would give a $z$ dependent time shift for $\iota_2^\mrm{pond}$ of maximum half a laser period, while the latter is almost flat near focus where we look for a good approximation. Thus, both phases are without a greater importance for THz waves. Then, the ponderomotive source writes in terms of the optical intensity
\begin{equation}
	\bm\iota_2^\mrm{pond} \approx - \frac{n_0 q_\mrm{e}^3}{2m_\mrm{e}^2\omega_\mrm{L}^2\epsilon_0 c}
	\nabla\left\{I_\mrm{L}^{2\mrm{D}}\left[1 + \cos(2\omega_\mrm{L}\tau)\right]\right\}\,\mbox{,} \label{eq:I_Gauss_app}
\end{equation}	
with
\begin{align}
	I_\mrm{L}^{2\mrm{D}}(x,z,t) & = \frac{\epsilon_0 c \left(E_\mrm{L}^0\right)^2 }{2} \frac{w_0}{w(z)} \, e^{-\frac{2x^2}{w(z)}-\frac{2\tau^2}{t_0^2}}\,\mbox{,} \label{eq:I_Gauss_int_2d} \\
	w(z) & = w_0 \sqrt{1+\left(\frac{z}{z_\mrm{R}}\right)^2} \,\mbox{.} 
\end{align}	
A similar treatment applies to a Gaussian 3D laser electric field 
\begin{equation}
	\Evec_\mrm{L}^{3\mrm{D}}(\rvec,t) \approx \Re \frac{E_\mrm{L}^0 \, e^{-\frac{x^2+y^2}{w_0^2 \left(1+\rmi\frac{z}{z_\mrm{R}}\right)}-\frac{\tau^2}{t_0^2}-\rmi\left(\omega_\mrm{L}\tau - \frac{\pi}{2} \right)}}{{1+\rmi\frac{z}{z_\mrm{R}}}} \evec_y  \,\mbox{.}
\end{equation}
One just has to plug the expression for the 3D intensity
\begin{equation}
	I_\mrm{L}^{3\mrm{D}}(\rvec,t) = \frac{\epsilon_0 c \left(E_\mrm{L}^0\right)^2}{2} \left[\frac{w_0}{w(z)}\right]^2 \, e^{-\frac{2x^2+2y^2}{w(z)}-\frac{2\tau^2}{t_0^2}} \label{eq:I_Gauss_int_3d}
\end{equation}	
into the expression for the ponderomotive source Eq.~(\ref{eq:I_Gauss_app}).
Thus, the longitudinal ponderomotive source $\iota_{2,z}^\mrm{pond}$ for a 2D/3D Gaussian driving pulse, which we need to evaluate our model, reads
\begin{align}
	\iota_{2,z}^\mrm{pond} &= - \frac{q_\mrm{e}^3n_0 I_\mrm{L}}{2m_\mrm{e}^2\omega_\mrm{L}^2\epsilon_0 c}
	\left\{ \left[ \frac{z\left(1-D+\frac{4\rvec_{\perp}^2}{w^2(z)}\right)}{z_\mrm{R}^2+z^2}
	+\frac{4\tau}{c t_0^2} \right] \right. \nonumber \\
	& \quad \left. \times \left[1 + \cos(2\omega_\mrm{L}\tau)\right] 
	+ 2 \frac{\omega_\mrm{L}}{c}\sin(2\omega_\mrm{L}\tau)\right\} \,\mbox{,}
	\label{eq:iota_pond_app}
\end{align}
where $I_\mrm{L}$ is given by Eq.~(\ref{eq:I_Gauss_int_2d}) or (\ref{eq:I_Gauss_int_3d}), $D=2$ or 3 is the number of dimensions, and $\rvec_{\perp}^2=x^2$ or $x^2+y^2$, respectively. Here, the term $\propto z$ appears due to beam focusing, while the term $\propto \tau$ reflects the longitudinal ponderomotive source as it already exist in 1D. Both are equally important in 2D or 3D geometry. The product $n_0 I_d$ produces THz as well as SH frequencies due to the step-like increase in time of $n_0$. 

\section{\label{app:static}Non-radiating solutions of the wave equation}
We want to show that the general curl-free solution $\Evec_{2,\mrm{d}}$ to the wave equation (\ref{eq:E_wave}) in the collisionless case ($\nu_\mrm{ei}=0$), after the driving pulse has passed ($\bm\iota_2=0$, $\partial_t n_0=0$), is given by Eq.~(\ref{eq:stat_osc}) of the main text. Because $\nabla\times\Evec_{2,\mrm{d}}=0$, Eq.~(\ref{eq:E_wave}) reduces to an oscillator equation
\begin{equation}
	\partial_t^2 \Evec_{2,\mrm{d}} + \frac{q_\mrm{e}^2n_0}{m_\mrm{e} \epsilon_0} \Evec_{2,\mrm{d}} = 0\label{eq:osc_app}\,\mbox{,}
\end{equation}
and we can write the general solution as
\begin{equation}
	\Evec_{2,\mrm{d}}(\rvec,t) = \gvec(\rvec)\exp\left(\rmi\sqrt{\frac{q_\mrm{e}^2n_0}{m_\mrm{e} \epsilon_0}}t\right)\,\mbox{.}
	\label{eq:E_d_app1}
\end{equation}
For convenience, Eq.~(\ref{eq:E_d_app1}) is written in complex form, and $\gvec(\rvec)$ is a complex valued function fulfilling
\begin{equation}
	\nabla \times \gvec(\rvec)=0, \qquad \gvec(\rvec)\times \nabla n_0(\rvec) = 0\,\mbox{.}
\end{equation}
Thus, we can write $\gvec(\rvec) = \nabla h(\rvec)$ with some scalar complex valued function $h(\rvec)$. 
For spatially varying $n_0$, $\nabla h \times \nabla n_0 = 0$ further implies $h(\rvec) = f(n_0)$ with some complex valued function $f$, and we have
\begin{equation}
	\Evec_{2,\mrm{d}} = \exp\left(\rmi\sqrt{\frac{q_\mrm{e}^2n_0}{m_\mrm{e} \epsilon_0}}t\right) \nabla f(n_0)\,\mbox{.} \label{eq:stat_osc_comp}
\end{equation}
Taking the real part of Eq.~(\ref{eq:stat_osc_comp}), we identify $A(n_0)=\left|\partial_{n_0}f(n_0) \right|$ and 
$\phi(n_0)=\arg\left[\partial_{n_0}f(n_0) \right]$ in Eq.~(\ref{eq:stat_osc}).

\section{\label{app:rad_far}Far-field emission from a current}
As has been pointed out in Sec.~\ref{sec:Dist}, it is sufficient to know the current density $\Jvec$ in order to calculate the far-field emission. To give an explicit expression for the far field, we consider the wave equation for the magnetic field in Fourier space
\begin{equation}
	\Delta\hat{\Bvec} + \frac{\omega^2}{c^2}\hat{\Bvec}=-\mu_0\nabla\times\hat{\Jvec}\,\mbox{,}
	\label{eq:B_wave_w_app}
\end{equation}
where the temporal Fourier transform (here given for the magnetic field) is defined as
\begin{equation}
	\hat{\Bvec}(\omega) = \int\limits_{-\infty}^{\infty} \Bvec(t)\exp\left(\rmi\omega t\right)\,dt\,\mbox{.}
	\label{eq:FT}
\end{equation}
Throughout the paper the Fourier transform is denoted by ''~$\hat{}$~''.
Solutions of Eq.~(\ref{eq:B_wave_w_app}) can be written as~\cite{jackson}
\begin{equation}
	\hat{\Bvec}(\rvec) = \mu_0 \! \int\limits_{V_\mrm{plasma}}\! \nabla_{\rvec'}\times\hat{\Jvec}(\rvec') \, G^{3\mrm{D}}(\rvec-\rvec')\,d^3\rvec'
	\label{eq:B_wave_sol_app}
\end{equation}
with the Green function
\begin{equation}
	G^{3\mrm{D}}(\rvec) = \frac{\exp\left(\pm\rmi \frac{\omega}{c} |\rvec|\right)}{4\pi |\rvec|}\,\mbox{.}
\end{equation}
The $\pm$ indicates whether the incoming or the outgoing wave is considered. Here, we have to consider outgoing waves and use the ''~$-$~'' sign. Integration by parts in Eq.~(\ref{eq:B_wave_sol_app}) gives the far field approximation ($|\rvec|\gg|\rvec'|$)
\begin{equation}
	\hat{\Bvec}_\mrm{far}(\rvec) \approx -\rmi \mu_0 \frac{\omega}{c} \frac{\rvec}{|\rvec|} \times \!\int\limits_{V_\mrm{plasma}}
	\!\hat{\Jvec}(\rvec') \, G^{3\mrm{D}}(\rvec-\rvec')\,d^3\rvec'\label{eq:B_far_app}\,\mbox{.}
\end{equation}
The corresponding electric field in the far field, in particular outside the plasma volume, can then be computed from $c^2\nabla\times \hat{\Bvec}_\mrm{far} = -\rmi \omega \hat{\Evec}_\mrm{far}$ as
\begin{equation}
	\hat{\Evec}_\mrm{far}(\rvec) \approx -c \frac{\rvec}{|\rvec|} \times \hat{\Bvec}_\mrm{far}(\rvec) \label{eq:E_far_app}\,\mbox{.}
\end{equation}
Special care has to be taken when it comes to evaluating Eq.~(\ref{eq:B_far_app}) for 2D geometries with translational invariance in, e.g., $y$-direction. Then, the integration over $y$ can be performed analytically leading to the 2D Green function valid in the far field ($K=\frac{\omega}{c} \sqrt{x^2+z^2} \gg1$)
\begin{align}
	& G ^{2\mrm{D}}(x,z) = \int\limits_{-\infty}^{\infty}\frac{\exp\left(\rmi \frac{\omega}{c} |\rvec|\right)}{4\pi |\rvec|} \, dy
	= \int\limits_{0}^{\infty}\frac{\exp\left(\rmi \theta +\rmi K \right)}{2\pi \sqrt{\theta^2+2K\theta } } \, d\theta \nonumber \\
	& \approx \int\limits_{0}^{\infty}\frac{\exp\left(\rmi \theta +\rmi K \right)}{2\pi \sqrt{2K\theta  } } \, d\theta  = \frac{\exp\left(\rmi \frac{\omega}{c} \sqrt{x^2+z^2} + \rmi \frac{\pi}{4}\right)}{\sqrt{8\pi \frac{\omega}{c} \sqrt{x^2+z^2} }}\,\mbox{.}
\end{align}
Here, we used the substitution $\theta = \frac{\omega}{c} |\rvec| - K$.

\section{\label{app:rad_thin}Far-field emission from a transversely narrow plasma}
For a narrow plasma with transverse extension $R\geq|\rvec'_{\perp}|$ and length $L\geq|\rvec'|$ one can show that under certain conditions only the longitudinal current $J_z$ contributes significantly to radiation in the far-field ($|\rvec|\gg|\rvec'|$). These conditions, which are typically found for the TC mechanism, are:
\begin{itemize}
        \item[1.] The plasma is symmetric, 
        \begin{equation}
        n_0(\rvec'_{\perp},z')=n_0(-\rvec'_{\perp},z')\,\mbox{.}
        \end{equation}
	\item[2.] The plasma is narrow, in the sense that
	\begin{equation}
		\big||\rvec - \rvec'| - |\rvec_{\perp} + (z - z') \evec_z|\big|\leq|\rvec'_{\perp}|\leq R \ll 
		\frac{\pi c}{\omega}\,\mbox{,}
	\end{equation}
	and therefore
	\begin{equation}
	  G^{3\mrm{D}}(\rvec_\perp-\rvec_\perp',z-z') \approx G^{3\mrm{D}}(\rvec_\perp,z-z')\,\mbox{.}
	\end{equation}	
	\item[3.] The current is anti-symmetric [c.f.\ Eqs.~(\ref{eq:sym_a})]
	\begin{align}
		\Jvec_\perp(\rvec_\perp,z) & = -\Jvec_\perp(-\rvec_\perp,z) \\
		J_z(\rvec_\perp,z) & = J_z(-\rvec_\perp,z)\,\mbox{.}
	\end{align}
\end{itemize}
Then, the integral in Eq.~(\ref{eq:B_far_app})
\begin{equation}
	\int\limits_{V_\mrm{plasma}} \! \hat{\Jvec}_\perp(\rvec')
		G^{3\mrm{D}}(\rvec-\rvec')\,d^3\rvec' \approx 0
\end{equation}
and thus the contributions of $\hat{\Jvec}_\perp$ to the far field vanish. In 2D geometry, the same argumentation holds when replacing $\rvec'_{\perp}\rightarrow x'$ and $G^{3\mrm{D}}\rightarrow G^{2\mrm{D}}$.

\section{\label{app:spectrum}Far-field power spectrum}
We define the far-field power spectrum as 
\begin{equation}
	P^{3\mrm{D}}_\mrm{far}(r,\theta,\varphi,\omega) = \left[\hat{\Evec}_\mrm{far}(r,\theta,\varphi,\omega)\times\hat{\Hvec}_\mrm{far}^\star(r,\theta,\varphi,\omega)\right]\cdot\,\evec_r\,\mbox{.}
\end{equation}
Here, $\hat{\Hvec}_\mrm{far}=\hat{\Bvec}_\mrm{far}/\mu_0$ and $\hat{\Evec}_\mrm{far}$ are given in frequency space [see Eq.~(\ref{eq:FT})], and we switched to spherical coordinates ($r,\theta,\varphi$) for convenience. Then, the detection angle is given by ($\theta,\varphi$) and the detector distance by $r$. The vector $\evec_r$ is the unit vector in $\rvec$-direction and normal to the radiation sphere. For our 2D geometry with translational invariance in $y$-direction we use 
polar coordinates ($r,\varphi$) to parametrize the $(x,z)$ plane.

The integral of $P_\mrm{far}$ over $\omega$ gives the frequency integrated power spectrum. The integral of $P_\mrm{far}$ over ($\theta,\varphi$), i.e., the radiation sphere, gives the angle integrated power spectrum. Performing both integrals we get the total radiated energy. For the THz energy we use
\begin{equation}
	E^{3\mrm{D}}_\mrm{THz} = \int\limits_{0}^{\omega_\mrm{m}}\int\limits_{-\frac{\pi}{2}}^{\frac{\pi}{2}}\int\limits_{0}^{2\pi} P^{3\mrm{D}}_\mrm{far}(r,\theta,\varphi,\omega) \, r^2 \sin\theta \,d\varphi\, d\theta \, d\omega\,\mbox{,} \label{eq:ethz3d}
\end{equation}
where we count the THz signal up to $\omega_\mrm{m}$. 
In 2D geometry, the THz energy density is defined analogous as
\begin{equation}
	E^{2\mrm{D}}_\mrm{THz} = \int\limits_{0}^{\omega_\mrm{m}}\int\limits_{0}^{2\pi} P^{2\mrm{D}}_\mrm{far}(r,\varphi,\omega) \, r \,d\varphi\, d\omega\,\mbox{.} \label{eq:ethz2d}
\end{equation}
When exploiting Eqs.~(\ref{eq:B_far_app}) and (\ref{eq:E_far_app}) for the far field, it is possible to show that the integration over all angles cancels the $r$ dependency in Eqs.~(\ref{eq:ethz3d}) and (\ref{eq:ethz2d}), respectively. When we use fields obtained from PIC simulations to compute $E_\mrm{THz}$, we use the invariance in $r$ as an additional consistency check for our codes.

\section{\label{app:plasma_extend}Scaling of the plasma volume with the laser pulse energy}
We define the plasma volume $V_\mrm{plasma}$ of an elongated plasma with electron density $n_0$ after ionization by a Gaussian laser pulse as
\begin{equation}
	V_\mrm{plasma} = \{\rvec\,|\,n_0(\rvec,t=\infty)\geq n_\mrm{th}\}\label{eq:plasma_vol_app1}\,\mbox{,}
\end{equation}
where $n_\mrm{th}$ is a threshold electron density. For Gaussian quasi-monochromatic laser pulses, at each position  $\rvec$, the temporal dependency of the laser intensity $I_\mrm{L}$ is the same up to the amplitude. The larger the amplitude, the more electrons are ionized and thus we can rewrite Eq.~(\ref{eq:plasma_vol_app1}) in terms of the intensity
\begin{equation}
	V_\mrm{plasma} = \{\rvec\,|\,I_\mrm{L}(\rvec,\tau=0)\geq I_\mrm{th}\}\label{eq:plasma_vol_app2}\,\mbox{.}
\end{equation}
For a threshold electron density $n_\mrm{th} \ll n_0$ resp.\ threshold intensity $I_\mrm{th} \ll I_0$ we can expect a bone shaped plasma [see, e.g., Fig.~\ref{fig:plasma_Ar}(a)], and the maximum transverse size is achieved off focus at $z \gg z_\mrm{R}$. 
In order to estimate transverse and longitudinal extension $R$ resp.\ $L$ of the plasma, we use the intensity profiles
$I_\mrm{L}^{2\mrm{D}}$ and $I_\mrm{L}^{3\mrm{D}}$ given in Eqs.~(\ref{eq:I_Gauss_int_2d}) and (\ref{eq:I_Gauss_int_3d}). 
For $z \gg z_\mrm{R}$ and thus $w(z)\approx w_0 z / z_\mrm{R}$, we can see that 
\begin{align}
	2\mrm{D}:&\quad L \propto I_\mrm{L}^0 \propto E_\mrm{p}  & R &\propto I_\mrm{L}^0 \propto E_\mrm{p}\\
	3\mrm{D}:&\quad L \propto \sqrt{I_\mrm{L}^0} \propto \sqrt{E_\mrm{p}} & R  &\propto \sqrt{I_\mrm{L}^0} \propto \sqrt{E_\mrm{p}}\,\mbox{.}
\end{align}

\newpage

\bibliography{mybibfile}

\begin{thebibliography}{40}%
\makeatletter
\providecommand \@ifxundefined [1]{%
 \@ifx{#1\undefined}
}%
\providecommand \@ifnum [1]{%
 \ifnum #1\expandafter \@firstoftwo
 \else \expandafter \@secondoftwo
 \fi
}%
\providecommand \@ifx [1]{%
 \ifx #1\expandafter \@firstoftwo
 \else \expandafter \@secondoftwo
 \fi
}%
\providecommand \natexlab [1]{#1}%
\providecommand \enquote  [1]{``#1''}%
\providecommand \bibnamefont  [1]{#1}%
\providecommand \bibfnamefont [1]{#1}%
\providecommand \citenamefont [1]{#1}%
\providecommand \href@noop [0]{\@secondoftwo}%
\providecommand \href [0]{\begingroup \@sanitize@url \@href}%
\providecommand \@href[1]{\@@startlink{#1}\@@href}%
\providecommand \@@href[1]{\endgroup#1\@@endlink}%
\providecommand \@sanitize@url [0]{\catcode `\\12\catcode `\$12\catcode
  `\&12\catcode `\#12\catcode `\^12\catcode `\_12\catcode `\%12\relax}%
\providecommand \@@startlink[1]{}%
\providecommand \@@endlink[0]{}%
\providecommand \url  [0]{\begingroup\@sanitize@url \@url }%
\providecommand \@url [1]{\endgroup\@href {#1}{\urlprefix }}%
\providecommand \urlprefix  [0]{URL }%
\providecommand \Eprint [0]{\href }%
\providecommand \doibase [0]{http://dx.doi.org/}%
\providecommand \selectlanguage [0]{\@gobble}%
\providecommand \bibinfo  [0]{\@secondoftwo}%
\providecommand \bibfield  [0]{\@secondoftwo}%
\providecommand \translation [1]{[#1]}%
\providecommand \BibitemOpen [0]{}%
\providecommand \bibitemStop [0]{}%
\providecommand \bibitemNoStop [0]{.\EOS\space}%
\providecommand \EOS [0]{\spacefactor3000\relax}%
\providecommand \BibitemShut  [1]{\csname bibitem#1\endcsname}%
\let\auto@bib@innerbib\@empty
\bibitem [{\citenamefont {Chan}\ \emph {et~al.}(2007)\citenamefont {Chan},
  \citenamefont {Deibel},\ and\ \citenamefont
  {Mittleman}}]{0034-4885-70-8-R02}%
  \BibitemOpen
  \bibfield  {author} {\bibinfo {author} {\bibfnamefont {W.~L.}\ \bibnamefont
  {Chan}}, \bibinfo {author} {\bibfnamefont {J.}~\bibnamefont {Deibel}}, \ and\
  \bibinfo {author} {\bibfnamefont {D.~M.}\ \bibnamefont {Mittleman}},\
  }\href@noop {} {\bibfield  {journal} {\bibinfo  {journal} {Reports on
  Progress in Physics}\ }\textbf {\bibinfo {volume} {70}},\ \bibinfo {pages}
  {1325} (\bibinfo {year} {2007})}\BibitemShut {NoStop}%
\bibitem [{\citenamefont {Tonouchi}(2007)}]{Tonouchi}%
  \BibitemOpen
  \bibfield  {author} {\bibinfo {author} {\bibfnamefont {M.}~\bibnamefont
  {Tonouchi}},\ }\href@noop {} {\bibfield  {journal} {\bibinfo  {journal} {Nat.
  Photon.}\ }\textbf {\bibinfo {volume} {1}},\ \bibinfo {pages} {97} (\bibinfo
  {year} {2007})}\BibitemShut {NoStop}%
\bibitem [{\citenamefont {Liu}\ \emph {et~al.}(2010)\citenamefont {Liu},
  \citenamefont {Dai}, \citenamefont {Chin},\ and\ \citenamefont
  {Zhang}}]{Liu}%
  \BibitemOpen
  \bibfield  {author} {\bibinfo {author} {\bibfnamefont {J.}~\bibnamefont
  {Liu}}, \bibinfo {author} {\bibfnamefont {J.}~\bibnamefont {Dai}}, \bibinfo
  {author} {\bibfnamefont {S.~L.}\ \bibnamefont {Chin}}, \ and\ \bibinfo
  {author} {\bibfnamefont {X.-C.}\ \bibnamefont {Zhang}},\ }\href@noop {}
  {\bibfield  {journal} {\bibinfo  {journal} {Nat. Photon.}\ }\textbf {\bibinfo
  {volume} {4}},\ \bibinfo {pages} {627} (\bibinfo {year} {2010})}\BibitemShut
  {NoStop}%
\bibitem [{\citenamefont {Kampfrath}\ \emph {et~al.}(2013)\citenamefont
  {Kampfrath}, \citenamefont {Tanaka},\ and\ \citenamefont
  {Nelson}}]{Kampfrath}%
  \BibitemOpen
  \bibfield  {author} {\bibinfo {author} {\bibfnamefont {T.}~\bibnamefont
  {Kampfrath}}, \bibinfo {author} {\bibfnamefont {K.}~\bibnamefont {Tanaka}}, \
  and\ \bibinfo {author} {\bibfnamefont {K.~A.}\ \bibnamefont {Nelson}},\
  }\href@noop {} {\bibfield  {journal} {\bibinfo  {journal} {Nat. Photon.}\
  }\textbf {\bibinfo {volume} {7}},\ \bibinfo {pages} {680} (\bibinfo {year}
  {2013})}\BibitemShut {NoStop}%
\bibitem [{\citenamefont {Tuniz}\ \emph {et~al.}(2013)\citenamefont {Tuniz},
  \citenamefont {Kaltenecker}, \citenamefont {Fischer}, \citenamefont
  {Walther}, \citenamefont {Fleming}, \citenamefont {Argyros},\ and\
  \citenamefont {Kuhlmey}}]{Tuniz}%
  \BibitemOpen
  \bibfield  {author} {\bibinfo {author} {\bibfnamefont {A.}~\bibnamefont
  {Tuniz}}, \bibinfo {author} {\bibfnamefont {K.~J.}\ \bibnamefont
  {Kaltenecker}}, \bibinfo {author} {\bibfnamefont {B.~M.}\ \bibnamefont
  {Fischer}}, \bibinfo {author} {\bibfnamefont {M.}~\bibnamefont {Walther}},
  \bibinfo {author} {\bibfnamefont {S.~C.}\ \bibnamefont {Fleming}}, \bibinfo
  {author} {\bibfnamefont {A.}~\bibnamefont {Argyros}}, \ and\ \bibinfo
  {author} {\bibfnamefont {B.~T.}\ \bibnamefont {Kuhlmey}},\ }\href@noop {}
  {\bibfield  {journal} {\bibinfo  {journal} {Nat. Commun.}\ }\textbf {\bibinfo
  {volume} {4}},\ \bibinfo {pages} {2706} (\bibinfo {year} {2013})}\BibitemShut
  {NoStop}%
\bibitem [{\citenamefont {Stepanov}\ \emph {et~al.}(2008)\citenamefont
  {Stepanov}, \citenamefont {Bonacina}, \citenamefont {Chekalin},\ and\
  \citenamefont {Wolf}}]{Stepanov:08}%
  \BibitemOpen
  \bibfield  {author} {\bibinfo {author} {\bibfnamefont {A.~G.}\ \bibnamefont
  {Stepanov}}, \bibinfo {author} {\bibfnamefont {L.}~\bibnamefont {Bonacina}},
  \bibinfo {author} {\bibfnamefont {S.~V.}\ \bibnamefont {Chekalin}}, \ and\
  \bibinfo {author} {\bibfnamefont {J.-P.}\ \bibnamefont {Wolf}},\ }\href@noop
  {} {\bibfield  {journal} {\bibinfo  {journal} {Opt. Lett.}\ }\textbf
  {\bibinfo {volume} {33}},\ \bibinfo {pages} {2497} (\bibinfo {year}
  {2008})}\BibitemShut {NoStop}%
\bibitem [{\citenamefont {Vicario}\ \emph {et~al.}(2014)\citenamefont
  {Vicario}, \citenamefont {Monoszlai},\ and\ \citenamefont
  {Hauri}}]{PhysRevLett.112.213901}%
  \BibitemOpen
  \bibfield  {author} {\bibinfo {author} {\bibfnamefont {C.}~\bibnamefont
  {Vicario}}, \bibinfo {author} {\bibfnamefont {B.}~\bibnamefont {Monoszlai}},
  \ and\ \bibinfo {author} {\bibfnamefont {C.~P.}\ \bibnamefont {Hauri}},\
  }\href@noop {} {\bibfield  {journal} {\bibinfo  {journal} {Phys. Rev. Lett.}\
  }\textbf {\bibinfo {volume} {112}},\ \bibinfo {pages} {213901} (\bibinfo
  {year} {2014})}\BibitemShut {NoStop}%
\bibitem [{\citenamefont {Casalbuoni}\ \emph {et~al.}(2009)\citenamefont
  {Casalbuoni}, \citenamefont {Schmidt}, \citenamefont {Schm\"user},
  \citenamefont {Arsov},\ and\ \citenamefont {Wesch}}]{PhysRevSTAB.12.030705}%
  \BibitemOpen
  \bibfield  {author} {\bibinfo {author} {\bibfnamefont {S.}~\bibnamefont
  {Casalbuoni}}, \bibinfo {author} {\bibfnamefont {B.}~\bibnamefont {Schmidt}},
  \bibinfo {author} {\bibfnamefont {P.}~\bibnamefont {Schm\"user}}, \bibinfo
  {author} {\bibfnamefont {V.}~\bibnamefont {Arsov}}, \ and\ \bibinfo {author}
  {\bibfnamefont {S.}~\bibnamefont {Wesch}},\ }\href@noop {} {\bibfield
  {journal} {\bibinfo  {journal} {Phys. Rev. ST Accel. Beams}\ }\textbf
  {\bibinfo {volume} {12}},\ \bibinfo {pages} {030705} (\bibinfo {year}
  {2009})}\BibitemShut {NoStop}%
\bibitem [{\citenamefont {Wu}\ \emph {et~al.}(2013)\citenamefont {Wu},
  \citenamefont {Fisher}, \citenamefont {Goodfellow}, \citenamefont {Fuchs},
  \citenamefont {Daranciang}, \citenamefont {Hogan}, \citenamefont {Loos},\
  and\ \citenamefont {Lindenberg}}]{1.4790427}%
  \BibitemOpen
  \bibfield  {author} {\bibinfo {author} {\bibfnamefont {Z.}~\bibnamefont
  {Wu}}, \bibinfo {author} {\bibfnamefont {A.~S.}\ \bibnamefont {Fisher}},
  \bibinfo {author} {\bibfnamefont {J.}~\bibnamefont {Goodfellow}}, \bibinfo
  {author} {\bibfnamefont {M.}~\bibnamefont {Fuchs}}, \bibinfo {author}
  {\bibfnamefont {D.}~\bibnamefont {Daranciang}}, \bibinfo {author}
  {\bibfnamefont {M.}~\bibnamefont {Hogan}}, \bibinfo {author} {\bibfnamefont
  {H.}~\bibnamefont {Loos}}, \ and\ \bibinfo {author} {\bibfnamefont
  {A.}~\bibnamefont {Lindenberg}},\ }\href@noop {} {\bibfield  {journal}
  {\bibinfo  {journal} {Review of Scientific Instruments}\ }\textbf {\bibinfo
  {volume} {84}},\ \bibinfo {eid} {022701} (\bibinfo {year}
  {2013})}\BibitemShut {NoStop}%
\bibitem [{\citenamefont {Cook}\ and\ \citenamefont
  {Hochstrasser}(2000)}]{Cook}%
  \BibitemOpen
  \bibfield  {author} {\bibinfo {author} {\bibfnamefont {D.~J.}\ \bibnamefont
  {Cook}}\ and\ \bibinfo {author} {\bibfnamefont {R.~M.}\ \bibnamefont
  {Hochstrasser}},\ }\href@noop {} {\bibfield  {journal} {\bibinfo  {journal}
  {Opt. Lett.}\ }\textbf {\bibinfo {volume} {25}},\ \bibinfo {pages} {1210}
  (\bibinfo {year} {2000})}\BibitemShut {NoStop}%
\bibitem [{\citenamefont {Kim}\ \emph {et~al.}(2007)\citenamefont {Kim},
  \citenamefont {Glownia}, \citenamefont {Taylor},\ and\ \citenamefont
  {Rodriguez}}]{Kim2007}%
  \BibitemOpen
  \bibfield  {author} {\bibinfo {author} {\bibfnamefont {K.-Y.}\ \bibnamefont
  {Kim}}, \bibinfo {author} {\bibfnamefont {J.~H.}\ \bibnamefont {Glownia}},
  \bibinfo {author} {\bibfnamefont {A.~J.}\ \bibnamefont {Taylor}}, \ and\
  \bibinfo {author} {\bibfnamefont {G.}~\bibnamefont {Rodriguez}},\ }\href@noop
  {} {\bibfield  {journal} {\bibinfo  {journal} {Opt.\ Express}\ }\textbf
  {\bibinfo {volume} {15}},\ \bibinfo {pages} {4577} (\bibinfo {year}
  {2007})}\BibitemShut {NoStop}%
\bibitem [{\citenamefont {Sprangle}\ \emph {et~al.}(2004)\citenamefont
  {Sprangle}, \citenamefont {Pe\~nano}, \citenamefont {Hafizi},\ and\
  \citenamefont {Kapetanakos}}]{PhysRevE.69.066415}%
  \BibitemOpen
  \bibfield  {author} {\bibinfo {author} {\bibfnamefont {P.}~\bibnamefont
  {Sprangle}}, \bibinfo {author} {\bibfnamefont {J.~R.}\ \bibnamefont
  {Pe\~nano}}, \bibinfo {author} {\bibfnamefont {B.}~\bibnamefont {Hafizi}}, \
  and\ \bibinfo {author} {\bibfnamefont {C.~A.}\ \bibnamefont {Kapetanakos}},\
  }\href@noop {} {\bibfield  {journal} {\bibinfo  {journal} {Phys. Rev. E}\
  }\textbf {\bibinfo {volume} {69}},\ \bibinfo {pages} {066415} (\bibinfo
  {year} {2004})}\BibitemShut {NoStop}%
\bibitem [{\citenamefont {D'Amico}\ \emph {et~al.}(2007)\citenamefont
  {D'Amico}, \citenamefont {Houard}, \citenamefont {Franco}, \citenamefont
  {Prade}, \citenamefont {Mysyrowicz}, \citenamefont {Couairon},\ and\
  \citenamefont {Tikhonchuk}}]{PhysRevLett.98.235002}%
  \BibitemOpen
  \bibfield  {author} {\bibinfo {author} {\bibfnamefont {C.}~\bibnamefont
  {D'Amico}}, \bibinfo {author} {\bibfnamefont {A.}~\bibnamefont {Houard}},
  \bibinfo {author} {\bibfnamefont {M.}~\bibnamefont {Franco}}, \bibinfo
  {author} {\bibfnamefont {B.}~\bibnamefont {Prade}}, \bibinfo {author}
  {\bibfnamefont {A.}~\bibnamefont {Mysyrowicz}}, \bibinfo {author}
  {\bibfnamefont {A.}~\bibnamefont {Couairon}}, \ and\ \bibinfo {author}
  {\bibfnamefont {V.~T.}\ \bibnamefont {Tikhonchuk}},\ }\href@noop {}
  {\bibfield  {journal} {\bibinfo  {journal} {Phys. Rev. Lett.}\ }\textbf
  {\bibinfo {volume} {98}},\ \bibinfo {pages} {235002} (\bibinfo {year}
  {2007})}\BibitemShut {NoStop}%
\bibitem [{\citenamefont {Kim}\ \emph {et~al.}(2008)\citenamefont {Kim},
  \citenamefont {Taylor}, \citenamefont {Chin},\ and\ \citenamefont
  {Rodriguez}}]{Kim}%
  \BibitemOpen
  \bibfield  {author} {\bibinfo {author} {\bibfnamefont {K.~Y.}\ \bibnamefont
  {Kim}}, \bibinfo {author} {\bibfnamefont {A.~J.}\ \bibnamefont {Taylor}},
  \bibinfo {author} {\bibfnamefont {S.~L.}\ \bibnamefont {Chin}}, \ and\
  \bibinfo {author} {\bibfnamefont {G.}~\bibnamefont {Rodriguez}},\ }\href@noop
  {} {\bibfield  {journal} {\bibinfo  {journal} {Nat. Photon.}\ }\textbf
  {\bibinfo {volume} {2}},\ \bibinfo {pages} {605} (\bibinfo {year}
  {2008})}\BibitemShut {NoStop}%
\bibitem [{\citenamefont {Chen}\ \emph {et~al.}(2008)\citenamefont {Chen},
  \citenamefont {Pukhov}, \citenamefont {Peng},\ and\ \citenamefont
  {Willi}}]{PhysRevE.78.046406}%
  \BibitemOpen
  \bibfield  {author} {\bibinfo {author} {\bibfnamefont {M.}~\bibnamefont
  {Chen}}, \bibinfo {author} {\bibfnamefont {A.}~\bibnamefont {Pukhov}},
  \bibinfo {author} {\bibfnamefont {X.-Y.}\ \bibnamefont {Peng}}, \ and\
  \bibinfo {author} {\bibfnamefont {O.}~\bibnamefont {Willi}},\ }\href@noop {}
  {\bibfield  {journal} {\bibinfo  {journal} {Phys. Rev. E}\ }\textbf {\bibinfo
  {volume} {78}},\ \bibinfo {pages} {046406} (\bibinfo {year}
  {2008})}\BibitemShut {NoStop}%
\bibitem [{\citenamefont {Babushkin}\ \emph {et~al.}(2010)\citenamefont
  {Babushkin}, \citenamefont {Kuehn}, \citenamefont {K\"ohler}, \citenamefont
  {Skupin}, \citenamefont {Berg\'e}, \citenamefont {Reimann}, \citenamefont
  {Woerner}, \citenamefont {Herrmann},\ and\ \citenamefont
  {Elsaesser}}]{PhysRevLett.105.053903}%
  \BibitemOpen
  \bibfield  {author} {\bibinfo {author} {\bibfnamefont {I.}~\bibnamefont
  {Babushkin}}, \bibinfo {author} {\bibfnamefont {W.}~\bibnamefont {Kuehn}},
  \bibinfo {author} {\bibfnamefont {C.}~\bibnamefont {K\"ohler}}, \bibinfo
  {author} {\bibfnamefont {S.}~\bibnamefont {Skupin}}, \bibinfo {author}
  {\bibfnamefont {L.}~\bibnamefont {Berg\'e}}, \bibinfo {author} {\bibfnamefont
  {K.}~\bibnamefont {Reimann}}, \bibinfo {author} {\bibfnamefont
  {M.}~\bibnamefont {Woerner}}, \bibinfo {author} {\bibfnamefont
  {J.}~\bibnamefont {Herrmann}}, \ and\ \bibinfo {author} {\bibfnamefont
  {T.}~\bibnamefont {Elsaesser}},\ }\href@noop {} {\bibfield  {journal}
  {\bibinfo  {journal} {Phys. Rev. Lett.}\ }\textbf {\bibinfo {volume} {105}},\
  \bibinfo {pages} {053903} (\bibinfo {year} {2010})}\BibitemShut {NoStop}%
\bibitem [{\citenamefont {Balakin}\ \emph {et~al.}(2010)\citenamefont
  {Balakin}, \citenamefont {Borodin}, \citenamefont {Kotelnikov},\ and\
  \citenamefont {Shkurinov}}]{Balakin:10}%
  \BibitemOpen
  \bibfield  {author} {\bibinfo {author} {\bibfnamefont {A.~V.}\ \bibnamefont
  {Balakin}}, \bibinfo {author} {\bibfnamefont {A.~V.}\ \bibnamefont
  {Borodin}}, \bibinfo {author} {\bibfnamefont {I.~A.}\ \bibnamefont
  {Kotelnikov}}, \ and\ \bibinfo {author} {\bibfnamefont {A.~P.}\ \bibnamefont
  {Shkurinov}},\ }\href@noop {} {\bibfield  {journal} {\bibinfo  {journal} {J.
  Opt. Soc. Am. B}\ }\textbf {\bibinfo {volume} {27}},\ \bibinfo {pages} {16}
  (\bibinfo {year} {2010})}\BibitemShut {NoStop}%
\bibitem [{\citenamefont {Zharova}\ \emph {et~al.}(2010)\citenamefont
  {Zharova}, \citenamefont {Mironov},\ and\ \citenamefont
  {Fadeev}}]{PhysRevE.82.056409}%
  \BibitemOpen
  \bibfield  {author} {\bibinfo {author} {\bibfnamefont {N.~A.}\ \bibnamefont
  {Zharova}}, \bibinfo {author} {\bibfnamefont {V.~A.}\ \bibnamefont
  {Mironov}}, \ and\ \bibinfo {author} {\bibfnamefont {D.~A.}\ \bibnamefont
  {Fadeev}},\ }\href@noop {} {\bibfield  {journal} {\bibinfo  {journal} {Phys.
  Rev. E}\ }\textbf {\bibinfo {volume} {82}},\ \bibinfo {pages} {056409}
  (\bibinfo {year} {2010})}\BibitemShut {NoStop}%
\bibitem [{\citenamefont {Pe\~nano}\ \emph {et~al.}(2010)\citenamefont
  {Pe\~nano}, \citenamefont {Sprangle}, \citenamefont {Hafizi}, \citenamefont
  {Gordon},\ and\ \citenamefont {Serafim}}]{PhysRevE.81.026407}%
  \BibitemOpen
  \bibfield  {author} {\bibinfo {author} {\bibfnamefont {J.}~\bibnamefont
  {Pe\~nano}}, \bibinfo {author} {\bibfnamefont {P.}~\bibnamefont {Sprangle}},
  \bibinfo {author} {\bibfnamefont {B.}~\bibnamefont {Hafizi}}, \bibinfo
  {author} {\bibfnamefont {D.}~\bibnamefont {Gordon}}, \ and\ \bibinfo {author}
  {\bibfnamefont {P.}~\bibnamefont {Serafim}},\ }\href@noop {} {\bibfield
  {journal} {\bibinfo  {journal} {Phys. Rev. E}\ }\textbf {\bibinfo {volume}
  {81}},\ \bibinfo {pages} {026407} (\bibinfo {year} {2010})}\BibitemShut
  {NoStop}%
\bibitem [{\citenamefont {Babushkin}\ \emph {et~al.}(2011)\citenamefont
  {Babushkin}, \citenamefont {Skupin}, \citenamefont {Husakou}, \citenamefont
  {K\"ohler}, \citenamefont {Cabrera-Granado}, \citenamefont {Berg\'e},\ and\
  \citenamefont {Herrmann}}]{1367-2630-13-12-123029}%
  \BibitemOpen
  \bibfield  {author} {\bibinfo {author} {\bibfnamefont {I.}~\bibnamefont
  {Babushkin}}, \bibinfo {author} {\bibfnamefont {S.}~\bibnamefont {Skupin}},
  \bibinfo {author} {\bibfnamefont {A.}~\bibnamefont {Husakou}}, \bibinfo
  {author} {\bibfnamefont {C.}~\bibnamefont {K\"ohler}}, \bibinfo {author}
  {\bibfnamefont {E.}~\bibnamefont {Cabrera-Granado}}, \bibinfo {author}
  {\bibfnamefont {L.}~\bibnamefont {Berg\'e}}, \ and\ \bibinfo {author}
  {\bibfnamefont {J.}~\bibnamefont {Herrmann}},\ }\href@noop {} {\bibfield
  {journal} {\bibinfo  {journal} {New Journal of Physics}\ }\textbf {\bibinfo
  {volume} {13}},\ \bibinfo {pages} {123029} (\bibinfo {year}
  {2011})}\BibitemShut {NoStop}%
\bibitem [{\citenamefont {Debayle}\ \emph {et~al.}(2014)\citenamefont
  {Debayle}, \citenamefont {Gremillet}, \citenamefont {Berg\'{e}},\ and\
  \citenamefont {K\"{o}hler}}]{Debayle:14}%
  \BibitemOpen
  \bibfield  {author} {\bibinfo {author} {\bibfnamefont {A.}~\bibnamefont
  {Debayle}}, \bibinfo {author} {\bibfnamefont {L.}~\bibnamefont {Gremillet}},
  \bibinfo {author} {\bibfnamefont {L.}~\bibnamefont {Berg\'{e}}}, \ and\
  \bibinfo {author} {\bibfnamefont {C.}~\bibnamefont {K\"{o}hler}},\
  }\href@noop {} {\bibfield  {journal} {\bibinfo  {journal} {Opt. Express}\
  }\textbf {\bibinfo {volume} {22}},\ \bibinfo {pages} {13691} (\bibinfo {year}
  {2014})}\BibitemShut {NoStop}%
\bibitem [{\citenamefont {Buccheri}\ and\ \citenamefont
  {Zhang}(2015)}]{Buccheri:15}%
  \BibitemOpen
  \bibfield  {author} {\bibinfo {author} {\bibfnamefont {F.}~\bibnamefont
  {Buccheri}}\ and\ \bibinfo {author} {\bibfnamefont {X.-C.}\ \bibnamefont
  {Zhang}},\ }\href@noop {} {\bibfield  {journal} {\bibinfo  {journal}
  {Optica}\ }\textbf {\bibinfo {volume} {2}},\ \bibinfo {pages} {366} (\bibinfo
  {year} {2015})}\BibitemShut {NoStop}%
\bibitem [{\citenamefont {Andreeva}\ \emph {et~al.}(2016)\citenamefont
  {Andreeva}, \citenamefont {Kosareva}, \citenamefont {Panov}, \citenamefont
  {Shipilo}, \citenamefont {Solyankin}, \citenamefont {Esaulkov}, \citenamefont
  {Gonz\'alez~de Alaiza~Mart\'{\i}nez}, \citenamefont {Shkurinov},
  \citenamefont {Makarov}, \citenamefont {Berg\'e},\ and\ \citenamefont
  {Chin}}]{PhysRevLett.116.063902}%
  \BibitemOpen
  \bibfield  {author} {\bibinfo {author} {\bibfnamefont {V.~A.}\ \bibnamefont
  {Andreeva}}, \bibinfo {author} {\bibfnamefont {O.~G.}\ \bibnamefont
  {Kosareva}}, \bibinfo {author} {\bibfnamefont {N.~A.}\ \bibnamefont {Panov}},
  \bibinfo {author} {\bibfnamefont {D.~E.}\ \bibnamefont {Shipilo}}, \bibinfo
  {author} {\bibfnamefont {P.~M.}\ \bibnamefont {Solyankin}}, \bibinfo {author}
  {\bibfnamefont {M.~N.}\ \bibnamefont {Esaulkov}}, \bibinfo {author}
  {\bibfnamefont {P.}~\bibnamefont {Gonz\'alez~de Alaiza~Mart\'{\i}nez}},
  \bibinfo {author} {\bibfnamefont {A.~P.}\ \bibnamefont {Shkurinov}}, \bibinfo
  {author} {\bibfnamefont {V.~A.}\ \bibnamefont {Makarov}}, \bibinfo {author}
  {\bibfnamefont {L.}~\bibnamefont {Berg\'e}}, \ and\ \bibinfo {author}
  {\bibfnamefont {S.~L.}\ \bibnamefont {Chin}},\ }\href@noop {} {\bibfield
  {journal} {\bibinfo  {journal} {Phys. Rev. Lett.}\ }\textbf {\bibinfo
  {volume} {116}},\ \bibinfo {pages} {063902} (\bibinfo {year}
  {2016})}\BibitemShut {NoStop}%
\bibitem [{\citenamefont {Koulouklidis}\ \emph {et~al.}(2016)\citenamefont
  {Koulouklidis}, \citenamefont {Fedorov},\ and\ \citenamefont
  {Tzortzakis}}]{PhysRevA.93.033844}%
  \BibitemOpen
  \bibfield  {author} {\bibinfo {author} {\bibfnamefont {A.~D.}\ \bibnamefont
  {Koulouklidis}}, \bibinfo {author} {\bibfnamefont {V.~Y.}\ \bibnamefont
  {Fedorov}}, \ and\ \bibinfo {author} {\bibfnamefont {S.}~\bibnamefont
  {Tzortzakis}},\ }\href@noop {} {\bibfield  {journal} {\bibinfo  {journal}
  {Phys. Rev. A}\ }\textbf {\bibinfo {volume} {93}},\ \bibinfo {pages} {033844}
  (\bibinfo {year} {2016})}\BibitemShut {NoStop}%
\bibitem [{\citenamefont {Gonz\'alez~de Alaiza~Mart\'inez}\ \emph
  {et~al.}(2016)\citenamefont {Gonz\'alez~de Alaiza~Mart\'inez}, \citenamefont
  {Davoine}, \citenamefont {Debayle}, \citenamefont {Gremillet},\ and\
  \citenamefont {Berg\'e}}]{PedrosArticle}%
  \BibitemOpen
  \bibfield  {author} {\bibinfo {author} {\bibfnamefont {P.}~\bibnamefont
  {Gonz\'alez~de Alaiza~Mart\'inez}}, \bibinfo {author} {\bibfnamefont
  {X.}~\bibnamefont {Davoine}}, \bibinfo {author} {\bibfnamefont
  {A.}~\bibnamefont {Debayle}}, \bibinfo {author} {\bibfnamefont
  {L.}~\bibnamefont {Gremillet}}, \ and\ \bibinfo {author} {\bibfnamefont
  {L.}~\bibnamefont {Berg\'e}},\ }\href@noop {} {\bibfield  {journal} {\bibinfo
   {journal} {Scientific Reports}\ }\textbf {\bibinfo {volume} {6}},\ \bibinfo
  {pages} {26743} (\bibinfo {year} {2016})}\BibitemShut {NoStop}%
\bibitem [{\citenamefont {D'Amico}\ \emph {et~al.}(2008)\citenamefont
  {D'Amico}, \citenamefont {Houard}, \citenamefont {Akturk}, \citenamefont
  {Liu}, \citenamefont {Le~Bloas}, \citenamefont {Franco}, \citenamefont
  {Prade}, \citenamefont {Couairon}, \citenamefont {Tikhonchuk},\ and\
  \citenamefont {Mysyrowicz}}]{ISI:000253084000007}%
  \BibitemOpen
  \bibfield  {author} {\bibinfo {author} {\bibfnamefont {C.}~\bibnamefont
  {D'Amico}}, \bibinfo {author} {\bibfnamefont {A.}~\bibnamefont {Houard}},
  \bibinfo {author} {\bibfnamefont {S.}~\bibnamefont {Akturk}}, \bibinfo
  {author} {\bibfnamefont {Y.}~\bibnamefont {Liu}}, \bibinfo {author}
  {\bibfnamefont {J.}~\bibnamefont {Le~Bloas}}, \bibinfo {author}
  {\bibfnamefont {M.}~\bibnamefont {Franco}}, \bibinfo {author} {\bibfnamefont
  {B.}~\bibnamefont {Prade}}, \bibinfo {author} {\bibfnamefont
  {A.}~\bibnamefont {Couairon}}, \bibinfo {author} {\bibfnamefont {V.~T.}\
  \bibnamefont {Tikhonchuk}}, \ and\ \bibinfo {author} {\bibfnamefont
  {A.}~\bibnamefont {Mysyrowicz}},\ }\href@noop {} {\bibfield  {journal}
  {\bibinfo  {journal} {New.\ J.\ Phys.}\ }\textbf {\bibinfo {volume} {{10}}}
  (\bibinfo {year} {{2008}})}\BibitemShut {NoStop}%
\bibitem [{NIS()}]{NIST}%
  \BibitemOpen
  \href@noop {} {}\bibinfo {howpublished} {NIST Atomic Spectra
  Database}\BibitemShut {NoStop}%
\bibitem [{\citenamefont {Nuter}\ and\ \citenamefont
  {Tikhonchuk}(2013)}]{PhysRevE.87.043109}%
  \BibitemOpen
  \bibfield  {author} {\bibinfo {author} {\bibfnamefont {R.}~\bibnamefont
  {Nuter}}\ and\ \bibinfo {author} {\bibfnamefont {V.}~\bibnamefont
  {Tikhonchuk}},\ }\href@noop {} {\bibfield  {journal} {\bibinfo  {journal}
  {Phys. Rev. E}\ }\textbf {\bibinfo {volume} {87}},\ \bibinfo {pages} {043109}
  (\bibinfo {year} {2013})}\BibitemShut {NoStop}%
\bibitem [{\citenamefont {Nuter}\ \emph {et~al.}(2011)\citenamefont {Nuter},
  \citenamefont {Gremillet}, \citenamefont {Lefebvre}, \citenamefont
  {L\'{e}vy}, \citenamefont {Ceccotti},\ and\ \citenamefont {Martin}}]{Nut11}%
  \BibitemOpen
  \bibfield  {author} {\bibinfo {author} {\bibfnamefont {R.}~\bibnamefont
  {Nuter}}, \bibinfo {author} {\bibfnamefont {L.}~\bibnamefont {Gremillet}},
  \bibinfo {author} {\bibfnamefont {E.}~\bibnamefont {Lefebvre}}, \bibinfo
  {author} {\bibfnamefont {A.}~\bibnamefont {L\'{e}vy}}, \bibinfo {author}
  {\bibfnamefont {T.}~\bibnamefont {Ceccotti}}, \ and\ \bibinfo {author}
  {\bibfnamefont {P.}~\bibnamefont {Martin}},\ }\href@noop {} {\bibfield
  {journal} {\bibinfo  {journal} {Physics of Plasmas}\ }\textbf {\bibinfo
  {volume} {18}},\ \bibinfo {eid} {033107} (\bibinfo {year}
  {2011})}\BibitemShut {NoStop}%
\bibitem [{\citenamefont {Huba}(2013)}]{Huba2013}%
  \BibitemOpen
  \bibfield  {author} {\bibinfo {author} {\bibfnamefont {J.~D.}\ \bibnamefont
  {Huba}},\ }\href@noop {} {\emph {\bibinfo {title} {Plasma Physics}}}\
  (\bibinfo  {publisher} {Naval Research Laboratory},\ \bibinfo {address}
  {Washington, DC},\ \bibinfo {year} {2013})\BibitemShut {NoStop}%
\bibitem [{\citenamefont {Jackson}(1999)}]{jackson}%
  \BibitemOpen
  \bibfield  {author} {\bibinfo {author} {\bibfnamefont {J.~D.}\ \bibnamefont
  {Jackson}},\ }\href@noop {} {\emph {\bibinfo {title} {Classical
  electrodynamics}}},\ \bibinfo {edition} {3rd}\ ed.\ (\bibinfo  {publisher}
  {Wiley},\ \bibinfo {address} {New York, {NY}},\ \bibinfo {year}
  {1999})\BibitemShut {NoStop}%
\bibitem [{\citenamefont {Cabrera-Granado}\ \emph {et~al.}(2015)\citenamefont
  {Cabrera-Granado}, \citenamefont {Chen}, \citenamefont {Babushkin},
  \citenamefont {Berg{\'{e}}},\ and\ \citenamefont
  {Skupin}}]{Cabrera-Granado2015}%
  \BibitemOpen
  \bibfield  {author} {\bibinfo {author} {\bibfnamefont {E.}~\bibnamefont
  {Cabrera-Granado}}, \bibinfo {author} {\bibfnamefont {Y.}~\bibnamefont
  {Chen}}, \bibinfo {author} {\bibfnamefont {I.}~\bibnamefont {Babushkin}},
  \bibinfo {author} {\bibfnamefont {L.}~\bibnamefont {Berg{\'{e}}}}, \ and\
  \bibinfo {author} {\bibfnamefont {S.}~\bibnamefont {Skupin}},\ }\href@noop {}
  {\bibfield  {journal} {\bibinfo  {journal} {New J.\ Phys.}\ }\textbf
  {\bibinfo {volume} {17}},\ \bibinfo {pages} {023060} (\bibinfo {year}
  {2015})}\BibitemShut {NoStop}%
\bibitem [{\citenamefont {Rae}\ and\ \citenamefont
  {Burnett}(1992)}]{PhysRevA.46.2077}%
  \BibitemOpen
  \bibfield  {author} {\bibinfo {author} {\bibfnamefont {S.~C.}\ \bibnamefont
  {Rae}}\ and\ \bibinfo {author} {\bibfnamefont {K.}~\bibnamefont {Burnett}},\
  }\href@noop {} {\bibfield  {journal} {\bibinfo  {journal} {Phys. Rev. A}\
  }\textbf {\bibinfo {volume} {46}},\ \bibinfo {pages} {2077} (\bibinfo {year}
  {1992})}\BibitemShut {NoStop}%
\bibitem [{\citenamefont {P\'{e}rez}\ \emph {et~al.}(2012)\citenamefont
  {P\'{e}rez}, \citenamefont {Gremillet}, \citenamefont {Decoster},
  \citenamefont {Drouin},\ and\ \citenamefont {Lefebvre}}]{Collisions}%
  \BibitemOpen
  \bibfield  {author} {\bibinfo {author} {\bibfnamefont {F.}~\bibnamefont
  {P\'{e}rez}}, \bibinfo {author} {\bibfnamefont {L.}~\bibnamefont
  {Gremillet}}, \bibinfo {author} {\bibfnamefont {A.}~\bibnamefont {Decoster}},
  \bibinfo {author} {\bibfnamefont {M.}~\bibnamefont {Drouin}}, \ and\ \bibinfo
  {author} {\bibfnamefont {E.}~\bibnamefont {Lefebvre}},\ }\href@noop {}
  {\bibfield  {journal} {\bibinfo  {journal} {Physics of Plasmas}\ }\textbf
  {\bibinfo {volume} {19}},\ \bibinfo {eid} {083104} (\bibinfo {year}
  {2012})}\BibitemShut {NoStop}%
\bibitem [{Note1()}]{Note1}%
  \BibitemOpen
  \bibinfo {note} {For our 50-fs example pulse the final electron density is
  $n_0(t\rightarrow \infty ) \approx n_\protect \mathrm {a}$}\BibitemShut
  {NoStop}%
\bibitem [{\citenamefont {Tikhonchuk}(2002)}]{PhysRevLett.89.209301}%
  \BibitemOpen
  \bibfield  {author} {\bibinfo {author} {\bibfnamefont {V.~T.}\ \bibnamefont
  {Tikhonchuk}},\ }\href@noop {} {\bibfield  {journal} {\bibinfo  {journal}
  {Phys. Rev. Lett.}\ }\textbf {\bibinfo {volume} {89}},\ \bibinfo {pages}
  {209301} (\bibinfo {year} {2002})}\BibitemShut {NoStop}%
\bibitem [{Note2()}]{Note2}%
  \BibitemOpen
  \bibinfo {note} {If the absolute of $\protect \mathbf {E}^\protect \mathrm
  {PIC}$ or $\nabla n_\protect \mathrm {e}^\protect \mathrm {PIC}$ is smaller
  than 1\% of its average value in the whole box, we set the value to unity
  since the angle between zero-vectors cannot be defined.}\BibitemShut {Stop}%
\bibitem [{\citenamefont {Hasan}\ \emph {et~al.}(2013)\citenamefont {Hasan},
  \citenamefont {Etrich}, \citenamefont {Filter}, \citenamefont {Rockstuhl},\
  and\ \citenamefont {Lederer}}]{PhysRevB.88.205125}%
  \BibitemOpen
  \bibfield  {author} {\bibinfo {author} {\bibfnamefont {S.~B.}\ \bibnamefont
  {Hasan}}, \bibinfo {author} {\bibfnamefont {C.}~\bibnamefont {Etrich}},
  \bibinfo {author} {\bibfnamefont {R.}~\bibnamefont {Filter}}, \bibinfo
  {author} {\bibfnamefont {C.}~\bibnamefont {Rockstuhl}}, \ and\ \bibinfo
  {author} {\bibfnamefont {F.}~\bibnamefont {Lederer}},\ }\href@noop {}
  {\bibfield  {journal} {\bibinfo  {journal} {Phys. Rev. B}\ }\textbf {\bibinfo
  {volume} {88}},\ \bibinfo {pages} {205125} (\bibinfo {year}
  {2013})}\BibitemShut {NoStop}%
\bibitem [{\citenamefont {Ammosov}\ \emph {et~al.}(1986)\citenamefont
  {Ammosov}, \citenamefont {Delone},\ and\ \citenamefont
  {Krainov}}]{Ammosov-1986-Tunnel}%
  \BibitemOpen
  \bibfield  {author} {\bibinfo {author} {\bibfnamefont {M.}~\bibnamefont
  {Ammosov}}, \bibinfo {author} {\bibfnamefont {N.}~\bibnamefont {Delone}}, \
  and\ \bibinfo {author} {\bibfnamefont {V.}~\bibnamefont {Krainov}},\
  }\href@noop {} {\bibfield  {journal} {\bibinfo  {journal} {Sov. Phys. JETP}\
  }\textbf {\bibinfo {volume} {64}},\ \bibinfo {pages} {1191} (\bibinfo {year}
  {1986})}\BibitemShut {NoStop}%
\bibitem [{\citenamefont {Yudin}\ and\ \citenamefont
  {Ivanov}(2001)}]{PhysRevA.64.013409}%
  \BibitemOpen
  \bibfield  {author} {\bibinfo {author} {\bibfnamefont {G.~L.}\ \bibnamefont
  {Yudin}}\ and\ \bibinfo {author} {\bibfnamefont {M.~Y.}\ \bibnamefont
  {Ivanov}},\ }\href@noop {} {\bibfield  {journal} {\bibinfo  {journal} {Phys.
  Rev. A}\ }\textbf {\bibinfo {volume} {64}},\ \bibinfo {pages} {013409}
  (\bibinfo {year} {2001})}\BibitemShut {NoStop}%
\end{thebibliography}%

\end{document}